\documentclass[a4paper,11pt,oneside]{amsart}

\usepackage[english]{babel}

\usepackage{fullpage}

\usepackage[utf8]{inputenc} 
\usepackage{lmodern} 
\usepackage[T1]{fontenc}

\usepackage{amssymb}
\usepackage{mathtools} 
	\numberwithin{equation}{section}

\usepackage{verbatim} 

\usepackage[dvipsnames]{xcolor} 

\usepackage{graphicx}

\usepackage{tikz}
	\usetikzlibrary{decorations.pathreplacing}
	\usetikzlibrary{decorations.pathmorphing} 
	\usetikzlibrary{matrix,arrows} 
	\usetikzlibrary{patterns} 

\usepackage{booktabs}
\usepackage{multicol}

\DeclarePairedDelimiter{\ket}{\lvert}{\rangle}
\DeclarePairedDelimiterX{\ketbra}[2]{\lvert}{\rvert}{#1\rangle \langle#2}
\DeclarePairedDelimiterX{\braket}[2]{\langle}{\rangle}{#1\vert#2}

\DeclarePairedDelimiterX{\cbraket}[2]{\langle\!\langle}{\rangle}{#1\vert#2}
\DeclarePairedDelimiterX{\bracket}[2]{\langle}{\rangle\!\rangle}{#1\vert#2}
\DeclarePairedDelimiterX{\cbracket}[2]{\langle\!\langle}{\rangle\!\rangle}{#1\vert#2}

\usepackage{bm} 

\newcommand{\I}{\mathrm{i}}
\newcommand{\E}{\mathrm{e}}

\newcommand{\bd}{\begin{displaymath}}
	\newcommand{\ed}{\end{displaymath}}
\newcommand{\be}{\begin{equation}}
	\newcommand{\ee}{\end{equation}}
\newcommand{\bea}{\begin{eqnarray}}
	\newcommand{\eea}{\end{eqnarray}}

\newcommand{\Z}{\mathbb{Z}}

\DeclareMathOperator*{\ordprod}{\prod\limits^{\vbox to -.5ex{\kern-0.5ex\hbox{$\leftharpoonup$}\vss}}}
\DeclareMathOperator*{\ordprodopp}{\prod\limits^{\vbox to -.5ex{\kern-0.5ex\hbox{$\rightharpoonup$}\vss}}}

\makeatletter
\DeclareRobustCommand\widecheck[1]{{\mathpalette\@widecheck{#1}}}
\def\@widecheck#1#2{%
	\setbox\z@\hbox{\m@th$#1#2$}%
	\setbox\tw@\hbox{\m@th$#1%
		\widehat{%
			\vrule\@width\z@\@height\ht\z@
			\vrule\@height\z@\@width\wd\z@}$}%
	\dp\tw@-\ht\z@
	\@tempdima\ht\z@ \advance\@tempdima2\ht\tw@ \divide\@tempdima\thr@@
	\setbox\tw@\hbox{%
		\raise\@tempdima\hbox{\scalebox{1}[-1]{\lower\@tempdima\box
				\tw@}}}%
	{\ooalign{\box\tw@ \cr \box\z@}}}
\makeatother

\usepackage{marginnote}
	
\usepackage[
	ocgcolorlinks, 
	linkcolor=BlueViolet, citecolor=OliveGreen, urlcolor=RawSienna, filecolor=Sepia,
	hyperfootnotes=false,
	]{hyperref} 

\usepackage{bbm} 
	\newcommand{\id}{\mathbbm{1}}

\DeclareMathOperator*{\Res}{Res}
\DeclareMathOperator*{\dn}{dn}
\DeclareMathOperator*{\cn}{cn}
\DeclareMathOperator*{\sn}{sn}

\newcommand{\EE}{F}
\renewcommand{\SS}{T}
\newcommand{\VE}{\reflectbox{\textit{I}}\mspace{-6mu}\EE}
\newcommand{\nncrossing}{\gamma'}
\newcommand{\longbar}[1]{\,\overline{\!#1\!}\,}
\newcommand{\diag}{{/\mspace{-5mu}/}}

\DeclareFontEncoding{LGR}{}{}
\DeclareSymbolFont{sfgreek}{LGR}{cmss}{m}{n}
\SetSymbolFont{sfgreek}{bold}{LGR}{cmss}{bx}{n}
\DeclareMathSymbol{\salpha}{\mathord}{sfgreek}{`a}
\DeclareMathSymbol{\sbeta}{\mathord}{sfgreek}{`b}
\DeclareMathSymbol{\sgamma}{\mathord}{sfgreek}{`g}
\DeclareMathSymbol{\sdelta}{\mathord}{sfgreek}{`d}
\DeclareMathSymbol{\sepsilon}{\mathord}{sfgreek}{`e}
\DeclareMathSymbol{\szeta}{\mathord}{sfgreek}{`z}
\DeclareMathSymbol{\seta}{\mathord}{sfgreek}{`h}
\DeclareMathSymbol{\stheta}{\mathord}{sfgreek}{`j}
\DeclareMathSymbol{\siota}{\mathord}{sfgreek}{`i}
\DeclareMathSymbol{\skappa}{\mathord}{sfgreek}{`k}
\DeclareMathSymbol{\slambda}{\mathord}{sfgreek}{`l}
\DeclareMathSymbol{\smu}{\mathord}{sfgreek}{`m}
\DeclareMathSymbol{\snu}{\mathord}{sfgreek}{`n}
\DeclareMathSymbol{\sxi}{\mathord}{sfgreek}{`x}
\DeclareMathSymbol{\somicron}{\mathord}{sfgreek}{`o}
\DeclareMathSymbol{\spi}{\mathord}{sfgreek}{`p}
\DeclareMathSymbol{\srho}{\mathord}{sfgreek}{`r}
\DeclareMathSymbol{\ssigma}{\mathord}{sfgreek}{`s}
\DeclareMathSymbol{\stau}{\mathord}{sfgreek}{`t}
\DeclareMathSymbol{\supsilon}{\mathord}{sfgreek}{`u}
\DeclareMathSymbol{\sphi}{\mathord}{sfgreek}{`f}
\DeclareMathSymbol{\schi}{\mathord}{sfgreek}{`q}
\DeclareMathSymbol{\spsi}{\mathord}{sfgreek}{`y}
\DeclareMathSymbol{\somega}{\mathord}{sfgreek}{`w}

\usepackage[alphabetic,short-journals,initials]{amsrefs}

\begin{document}
	
	\title{Landscapes of integrable long-range spin chains}
	
	\author{Rob Klabbers\textsuperscript{$a$}, Jules Lamers\textsuperscript{$b$\,$c$}}
	\address{%
	\textsuperscript{$a$}\,Humboldt-Universität zu Berlin,~Zum Großen Windkanal 2, 12489 Berlin, Germany \\
	\textsuperscript{$b$}\,Deutsches Elektronen-Synchrotron DESY, Notkestraße 85, 22607 Hamburg, Germany \\
	\textsuperscript{$c$}\,Present address: School of Mathematics and Statistics, University of Glasgow, University Place, Glasgow G12 8QQ, UK} 
	
\begin{abstract} 
	\noindent We clarify how the elliptic integrable spin chain recently found by Matushko and Zotov (MZ) relates to various other known long-range spin chains. 
	
	We evaluate various limits. More precisely, we tweak the MZ chain to allow for a short-range limit, and show it is the \textsc{xx} model with $q$-deformed \emph{anti}periodic boundary conditions.
	Taking $q\to1$ gives the elliptic spin chain of Sechin and Zotov (SZ), whose trigonometric case is due to Fukui and Kawakami. 
	It, too, can be adjusted to admit a short-range limit, which we demonstrate to be the antiperiodic \textsc{xx} model.
	By identifying the translation operator of the MZ$'$ chain, which is nontrivial, we show that antiperiodicity is a persistent feature.
	
	We compare the resulting (vertex-type) landscape of the MZ chain with the (face-type) landscape containing the Heisenberg \textsc{xxx} and Haldane--Shastry chains. We find that the landscapes
		only share a single point: the rational Haldane--Shastry chain. 
	Using wrapping we show that the SZ chain is the anti\-periodic version of the Inozemtsev chain in a precise sense,
	and expand both chains around their nearest-neighbour limits to facilitate their interpretations as long-range deformations.
\end{abstract}

\begin{flushright}
	\scriptsize{DESY-24-062\,\raisebox{1pt}{\textbullet}\,HU-EP-24/13\,\raisebox{1pt}{\textbullet}\,HU-Mathematik 2024-01}
\end{flushright}


\maketitle

\vspace{-1\baselineskip}

\setcounter{tocdepth}{2}
\tableofcontents
\vspace{-3\baselineskip}

\section{Introduction} 
\label{sec:intro}

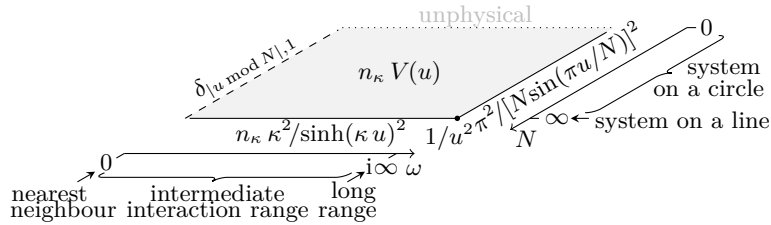
\begin{figure}[h]
	\!\!\begin{tikzpicture}[x={(-0.866cm,-0.5cm)}, y={(1cm,0cm)}, z={(0cm,1cm)}, scale=.6, font=\footnotesize]
		\fill [gray!10] (0,0,0) -- (4,0,0) -- (4,6,0) -- (0,6,0) -- cycle;
		\draw (4,.1,0) -- (4,6,0) -- (.2,6,0);
		\draw [dashed] (4,0,0) -- (0.1,0,0);
		\draw [dotted] (0,0,0) -- (0,6,0);
		\node at (-.5,3-.5,0) {\textcolor{gray!50}{unphysical}};
		\node at (2,3,0) {$n_\kappa \, V\mspace{-1mu}(u)$};
		\node at (2-.3,0,0) [xshift=-.4cm] {\rotatebox{30}{$\delta_{|u \,\mathrm{mod}\,N|, 1}$}};
		\node at (2,6,0) [xshift=.25cm, yshift=-.1cm] {\rotatebox{30}{$\ \pi^2 \!/[N \! \sin(\pi u/N)]^2$}};
		\node at (4,3,0) [yshift=-.2cm]{$n_\kappa \, \kappa^2 \!/\!\sinh(\kappa\,u)^2$};
		\fill[black] (4,6,0) circle (.06cm); 
		\node [xshift=-.1cm,yshift=-.25cm] at (4,6,0) {$1/u^2$};
		\draw [->] (0,7+.6,0) -- (4.5,7+.6,0);
		\node at (4.9,7.6+.6+.1,0) {$N$};
		\draw (0,7+.6,0) -- (0,7+.2+.6,0);
		\node [xshift = -.1cm] at (0,7.6+.6,0) {$0$};
		\draw (4,7+.6,0) -- (4,7+.2+.6,0);
		\node at (4,7.6+.6,0) {$\infty$};
		\draw [->] (5+.55,0,0) -- (5+.55,6.4,0);
		\node at (5.7+.55,6.9+.1,0) {$\omega$};
		\draw (5+.55,0,0) -- (5.2+.55,0,0);
		\node [shift ={(.05cm,.05cm)}] at (5.55+.55,0,0) {$0$};
		\draw (5+.55,6,0) -- (5.2+.55,6,0);
		\node [xshift = .1cm] at (5.55+.55,6,0) {$\I\mspace{1mu}\infty$};
		\draw [rounded corners=3pt] (0+.25+.1,8.8-.2,0) -- (0+.25+.1,8.8,0) -- (2,8.8,0) -- (2,8.8+.2,0) (2,8.8+.2,0) -- (2,8.8,0) -- (4-.25,8.8,0) -- (4-.25,8.8-.2,0);
		\node at (2,8.8+.2,0) [right, shift={(.07cm,.05cm)}] {system};
		\node at (2,8.8+.2,0) [right, shift={(-.38cm,-.2cm)}] {on a circle};
		\draw [->, >=stealth] (4,8.8+.1,0) -- (4,8.8-.3,0);
		\node at (4,8.8+.1,0) [right, shift={(-.05cm,-.05cm)}] {system on a line};
		\draw [->, >=stealth] (6.6+.2,0+.05,0) -- (6.6-.25,0+.05,0);
		\node at (6.6+1,0+.05,0) [shift={(.05cm,.1cm)}] {nearest\vphantom{l}};
		\node at (6.6+1,0+.05,0) [shift={(.23cm,-.15cm)}] {neighbour};
		\draw [rounded corners=3pt] (6.6-.2,0+.3,0) -- (6.6,0+.25,0) -- (6.6,3,0) -- (6.6+.1,3,0) (6.6+.1,3,0) -- (6.6,3,0) -- (6.6,6-.25,0) -- (6.6-.2,6-.25,0);
		\node at (6.6+1,3,0) [shift={(.5cm,.1cm)}] {intermediate};
		\node at (6.6+1,3,0) [shift={(.5cm,-.15cm)}] {\vphantom{l}interaction range};
		\draw [->, >=stealth] (6.6+.2,6,0) -- (6.6-.25,6,0);
		\node at (6.6+1,6,0) [shift={(.5cm,.1cm)}] {long};
		\node at (6.6+1,6,0) [shift={(.42cm,-.15cm)}] {\vphantom{l}range};
	\end{tikzpicture}
	\caption{Landscape given by \eqref{eq:V_intro}, parametrised by periods $(N,\omega)$, for us 
	$\in \mathbb{N}_{\geqslant 2} \times \I\,\mathbb{R}_{>0}$
	and $\omega = \I \pi/\kappa$.
	The limit $\omega \to \I \, 0^+$ requires a normalising factor $n_\kappa$, e.g.\ $n_\kappa = \sinh^2(\kappa)/\kappa^2$, and that $u \in \mathbb{R}$ with $|u|\geqslant 1$.}
	\label{fg:landscape_potential}
\end{figure}

\noindent 
The elliptic world provides a vast panorama comprising rational, trigonometric and hyperbolic regions. For example, the Weierstra{\ss} elliptic function
\begin{equation} \label{eq:V_intro}
	V\mspace{-1mu}(u) \coloneqq \wp(u) + \text{cst} \, ,\qquad 
	V\mspace{-1mu}(u+N) = V\mspace{-1mu}(u+\omega) = V\mspace{-1mu}(u) \, ,
\end{equation}
unifies $1/\!\sin^2$ (for $\omega \to \I \mspace{2mu} \infty$), $1/\!\sinh^2$ (as $N\to\infty$), and their common limit $1/u^2$; see Fig.~\ref{fg:landscape_potential} and \textsection\ref{sec:preliminaries} for details. 
These are precisely the four cases for which the (quantum) Calogero--Sutherland system is (quantum) integrable, with \eqref{eq:V_intro} or its limits as potential~\cite{sutherland1971exact}. 
These cases persist to be integrable for the `relativistic' generalisation, which is a $q$-deformation, to the (quantum) Ruijsenaars--Macdonald system~\cite{Rui_87}.

We will be interested in integrable spin chains. The prototype is the nearest-neighbour spin-1/2 \emph{Heisenberg chain},
\begin{equation} \label{eq:Heis}
	H_\text{Heis} = -\frac{1}{2} \sum_{i=1}^{N} \bigl( \sigma^x_i \, \sigma^x_{i+1} + \Gamma \, \sigma^y_i \, \sigma^y_{i+1}  + \Delta \, (\sigma^z_i \, \sigma^z_{i+1} -1) \bigr) \, ,
\end{equation}
with $\sigma^\alpha_i$ the Pauli matrices acting on the $i$th factor of $(\mathbb{C}^2)^{\otimes N}$, and periodic boundary conditions $\sigma^\alpha_{N+1} \equiv \sigma^\alpha_1$. 
It has three levels, with increasing complexity: 
\medskip
\begin{center}
	\begin{tabular}[h]{lll}
	\toprule
	$\!\!$\rlap{Heisenberg chain} & & \clap{spin symmetry} \\
	\midrule
	$\!\!$\textsc{xxx} ($\Gamma=\Delta=1$) & isotropic & $\mathit{SU}\mspace{-1mu}(2)$: $\mathfrak{sl}_2$-invariant \\
	$\!\!$\textsc{xxz} $\mspace{2mu}$($\Gamma=1$) & partially isotropic & $S^z$: weight-preserving$\!\!\!\!$ \\
	$\!\!$\textsc{xyz} & anisotropic & spin flip: Weyl group \\
	\bottomrule
	\end{tabular}
\end{center}
\medskip
Here $\mathit{SU}\mspace{-1mu}(2)$ acts by $S^\alpha = \sum_i \sigma_i^\alpha/2$.
The integrability of these levels of the Heisenberg chain is based on rational, trigonometric and elliptic \textit{R}-matrices, respectively. The latter have two versions, 
\begin{itemize}
	\item[\textit{i.}] Baxter's \textit{R}-matrix of (eight-)\emph{vertex} type~\cite{baxter1972one}, 
	\item[\textit{ii.}] Felder's dynamical elliptic \textit{R}-matrix of \emph{face} type~\cite{felder1994elliptic},
\end{itemize}
that are linked by the face-vertex transformation~\cite{baxter1973eight123}. For the \textsc{xyz} chain, which lives on the vertex side \cite{sutherland1970two}, (\textit{ii}) serves as a technical tool~\cite{baxter1973eight123,TF79,FELDER1996485}. In short: if $N$ is even then, on the zero-weight space (`equator'), the transfer matrices corresponding to (\textit{i}) and (\textit{ii}) are conjugate. Bethe-ansatz techniques on the face side thus yields the thermodynamics of the \textsc{xyz} chain in the phase where the equator dominates the spectrum, which determines all other phases by dualities.

Integrable spin chains with \emph{long-range} interactions provide a much broader vista. From this perspective \eqref{eq:Heis}, and its variants, is but one extreme. The function~\eqref{eq:V_intro} reappears as the pair potential of the isotropic \emph{Inozemtsev chain} \cite{Inozemtsev:1989yq}
\begin{equation} \label{eq:Ino}
	H_\text{Ino} = -\frac{1}{2} \sum_{i<j}^N V\mspace{-1mu}(i-j) \,
	\bigl( \sigma^x_i \, \sigma^x_j + \sigma^y_i \, \sigma^y_j + \sigma^z_i \, \sigma^z_j - 1 \bigr) \, .
\end{equation}
Note that $-\tfrac{1}{2} \bigl(\sigma^x_i \, \sigma^x_j + \sigma^y_i \, \sigma^y_j + \sigma^z_i \, \sigma^z_j - 1 \bigr) = 1- P_{ij}$ in terms of the spin permutation. The Inozemtsev chain is exactly solvable via a connection to the elliptic Calogero--Sutherland system~\cite{Inozemtsev_2000}, cf.~\cite{klabbers2022coordinate}. 
When $\omega \to \I\,\infty$ we get the trigonometric potential 
\begin{equation} \label{eq:V_tri_intro}
	\lim_{\kappa\to 0} V\mspace{-1mu}(u) = V_\text{tri}(u) \coloneqq \frac{(\pi/N)^2}{\sin^2 \bigl(\tfrac{\pi}{N} \mspace{1mu} u\bigr)} \, ,
\end{equation}
and \eqref{eq:Ino} reduces to the \emph{Haldane--Shastry chain}~\cites{haldane1988exact, shastry1988exact}. Its spin symmetry
is enhanced to the Yangian of $\mathfrak{gl}_2$~\cite{HH+_92,bernard1993yang} and it has explicit eigenvectors featuring Jack polynomials~\cite{Hal_91a,bernard1993yang}. This extreme of the Inozemtsev chain is the critical case for long-range order. The spin-chain lattice spacing provides a `\textsc{uv} cutoff' that furthermore allows one to make sense of the opposite limit $\omega \to \I \, 0^+$. Here, upon rescaling and with a suitably chosen constant in \eqref{eq:V_intro}, \eqref{eq:Ino} reduces to the Heisenberg \textsc{xxx} chain. The Inozemtsev chain thus interpolates between the nearest-neighbour Heisenberg \textsc{xxx} chain and the truly long-range Haldane--Shastry chains in a way that is exactly solvable throughout.%
\footnote{\ Note our double use of the adjective `long range':
\begin{itemize}
	\item for the general \emph{class} of spin chains such as \eqref{eq:Ino} (as well as their deformed generalisations);
	\item for their (trigonometric) \emph{limit} $\omega\to\I\,\infty$ of maximal interaction range\,---\,as opposed to the `short-range limit' $\omega\to\I\,0^+$.
\end{itemize}
We trust that this will not cause any confusion.}
\bigskip

In recent years it has become clear that, like \eqref{eq:Heis}, long-range spin chains admit integrable generalisations with less spin symmetry.\footnote{\ In \cite{haldane1988exact} it was suggested that an anisotropy $\Delta=m\,(m+1)/2$ can be included in the Haldane--Shastry chain \eqref{eq:Ino}--\eqref{eq:V_tri_intro} just like in \eqref{eq:Heis}. However, as numerical diagonalisation for low $N$ attests, this is too naive: it \emph{destroys} the special properties, e.g.\ high degeneracies due to the enhanced spin symmetry, rather than \emph{deforming} them. The correct mathematical generalisation has a more complicated hamiltonian, see below, such that the remarkable spectrum persists.} The \textsc{xxz}-type Haldane--Shastry and Inozemtsev chain are now known~\cite{Ugl_95u,Lam_18,lamers2022spin,KL_23}. Long-range integrability requires very specific `chiral' interactions: schematically,
\begin{equation} \label{eq:S^LR_diagr}
	\tikz[baseline={([yshift=-.5*11pt*0.15]current bounding box.center)},xscale=0.3,yscale=0.15,font=\footnotesize]{
		\draw[->] (9.5,0) node[below]{$N$} -- (9.5,10); 
		\draw[rounded corners=2pt,->] (8,0) node[below]{$j$} 
		-- (8,1.5) -- (5,4.5) -- (5,5.5) -- (8,8.5) -- (8,10);
		\draw[rounded corners=2pt,->] (7,0) -- (7,1.5) -- (8,2.5) -- (8,7.5) -- (7,8.5) -- (7,10); 
		\draw[rounded corners=2pt,->] (6,0) -- (6,2.5) -- (7,3.5) -- (7,6.5) -- (6,7.5) -- (6,10);
		\draw[rounded corners=2pt,->] (5,0) -- (5,3.5) -- (6,4.5) -- (6,5.5) -- (5,6.5) -- (5,10); 
		\draw[->] (4,0) node[below]{$i$} 
		-- (4,10);
		\draw[->] (2.5,0) node[below]{$1$} -- (2.5,10);
		\draw[style={decorate, decoration={zigzag,amplitude=.5mm,segment length=1mm}}] (4,5) -- (5,5);
		\foreach \x in {-1,...,1} \draw (3.25+.3*\x,5) node{$\cdot\mathstrut$};
		\foreach \x in {-1,...,1} \draw (8.75+.3*\x,5) node{$\cdot\mathstrut$};	
	} \! , 
	\quad
	\tikz[baseline={([yshift=-.5*11pt*0.15]current bounding box.center)},xscale=-0.3,yscale=0.15,font=\footnotesize]{
		\draw[->] (9.5,0) node[below]{$1$} -- (9.5,10);
		\draw[rounded corners=2pt,->] (8,0) node[below]{$i$} -- (8,1.5) -- (5,4.5) -- (5,5.5) -- (8,8.5) -- (8,10);
		\draw[rounded corners=2pt,->] (7,0) -- (7,1.5) -- (8,2.5) -- (8,7.5) -- (7,8.5) -- (7,10);
		\draw[rounded corners=2pt,->] (6,0) -- (6,2.5) -- (7,3.5) -- (7,6.5) -- (6,7.5) -- (6,10);
		\draw[rounded corners=2pt,->] (5,0) -- (5,3.5) -- (6,4.5) -- (6,5.5) -- (5,6.5) -- (5,10);
		\draw[->] (4,0) node[below]{$j$} -- (4,10);
		\draw[->] (2.5,0) node[below]{$N$} -- (2.5,10);
		\draw[style={decorate, decoration={zigzag,amplitude=.5mm,segment length=1mm}}] (4,5) -- (5,5);
		\foreach \x in {-1,...,1} \draw (3.25+.3*\x,5) node{$\cdot\mathstrut$};
		\foreach \x in {-1,...,1} \draw (8.75+.3*\x,5) node{$\cdot\mathstrut$};	
	} \! ,
	\quad
	\begin{aligned} 
		\tikz[baseline={([yshift=-2*11pt*.15]current bounding box.center)},xscale=.25,yscale=.15]{
			\draw[rounded corners=2pt,->] (1,-.1) -- (1,.5) -- (0,1.5) -- (0,2.3);
			\draw[rounded corners=2pt,->] (0,-.1) -- (0,.5) -- (1,1.5) -- (1,2.3);
		} = {} & \text{$R$-matrix} \, , \\
		\tikz[baseline={([yshift=-2*11pt*0.15]current bounding box.center)},xscale=0.25,yscale=0.15]{
			\draw[->] (1,0) -- (1,2.4);
			\draw[->] (2,0) -- (2,2.4);
			\draw[style={decorate, decoration={zigzag,amplitude=.5mm,segment length=1mm}}] (1,1) -- (2,1);
		} = {} & \text{nearest-neighbour} \\[-.25\baselineskip]
		& \text{spin exchange} \, .
	\end{aligned}
\end{equation}
The physical picture is as follows: given a pair of sites $i<j$ take either spin, use \textit{R}-matrices to transport it next to the other spin, let it interact with its neighbour, and then transport it back to where it started. There are two hamiltonians, $H^\textsc{l}$ and $H^\textsc{r}$, which are sums over $i<j$ of the left/right interactions in \eqref{eq:S^LR_diagr} with a potential that $q$-deforms~\eqref{eq:V_intro} (or its trigonometric limit). When $q\to 1$, 
$\tikz[baseline={([yshift=-2*11pt*.15]current bounding box.center)},xscale=.25,yscale=.15]{
	\draw[rounded corners=2pt,->] (1,-.1) -- (1,.5) -- (0,1.5) -- (0,2.3);
	\draw[rounded corners=2pt,->] (0,-.1) -- (0,.5) -- (1,1.5) -- (1,2.3);
} \to P$ 
and $\tikz[baseline={([yshift=-2*11pt*0.15]current bounding box.center)},xscale=0.3,yscale=0.15]{
	\draw[->] (1,0) -- (1,2.4);
	\draw[->] (2,0) -- (2,2.4);
	\draw[style={decorate, decoration={zigzag,amplitude=.5mm,segment length=1mm}}] (1,1) -- (2,1);
} \to 1-P$, 
so that both interactions in \eqref{eq:S^LR_diagr} become $1-P_{ij}$, and we retrieve~\eqref{eq:Ino}. For the $q$-deformed Inozemtsev chain, 
$\tikz[baseline={([yshift=-2*11pt*.15]current bounding box.center)},xscale=.25,yscale=.15]{
	\draw[rounded corners=2pt,->] (1,-.1) -- (1,.5) -- (0,1.5) -- (0,2.3);
	\draw[rounded corners=2pt,->] (0,-.1) -- (0,.5) -- (1,1.5) -- (1,2.3);
}$ 
is the elliptic \textit{R}-matrix of face type, guaranteeing \textsc{xxz}-type spin symmetry. 
We will return to it in \textsection\ref{sec:comparison}. See Fig.~\ref{fg:landscape_face} for the resulting spin-chain landscape.
Until then, we will focus on the \emph{vertex} side.

\subsection{Undeformed level} 

\noindent 
Write $\sigma^0 \coloneqq \id$ for the identity on $\mathbb{C}^2$. The \emph{Sechin--Zotov chain} \cite{sechin2018r} is given by
\begin{equation} \label{eq:SZ_intro}
	H_{\textsc{sz}'} = \sum_{i<j}^N \sum_{\alpha = 0}^z 
	\frac{1}{4}\,V\mspace{-2mu}\biggl(\frac{i-j+ \omega_\alpha}{2}\biggr) \bigl( 1-P_{ij} \, \sigma^\alpha_i \mspace{2mu} \sigma^\alpha_j \bigr) \, ,
\end{equation}
where the inner sum is over $\alpha\in\{0,x,y,z\}$ and the pair potential receives shifts over the corners $\omega_\alpha$ of the fundamental domain of \eqref{eq:V_intro}, see \textsection\ref{sec:SZ_limits}.
Its spin symmetry is restricted to invariance under the Weyl group (global spin flip). 
The long-range limit $\omega \to \I\,\infty$ of \eqref{eq:SZ_intro} is a special case of a family of twisted Haldane--Shastry chains \cite{fukui_exact_1996} that we will call the \emph{Fukui--Kawakami chain},
\begin{equation} \label{eq:FK}
	H_{\textsc{fk}} = -\frac{1}{2} \sum_{i<j}^N V_\text{tri}(i-j) \Bigl( \cos\bigl( \tfrac{\pi}{N} (i-j)\bigr) \bigl(\sigma^x_i \sigma^x_j + \sigma^y_i \sigma^y_j \bigr) + \sigma^z_i \sigma^z_j - 1 \Bigr) \, .
\end{equation}
Its spin symmetry is enhanced to not just $S^z$-invariance, but to the Yangian of $\mathfrak{gl}_1$ \cite{fukui_exact_1996}. 

Our first set of results concern the Sechin--Zotov chain. We will show that, like for the Inozemtsev chain, the constant in the potential~\eqref{eq:V_intro} can be chosen such that the resulting variant of \eqref{eq:SZ_intro}, which we call the \emph{SZ\/$'\!$ chain}, admits a short-range limit $\omega \to \I \, 0^+$ upon suitable rescaling. Up to a global spin rotation its long-range limit is still \eqref{eq:FK}. Remarkably, the short-range limit of the SZ$'$ chain is nothing but the free-fermion ($\Delta=0$) Heisenberg \textsc{xxz} chain with antiperiodic (`Möbius') boundaries,
\begin{equation} \label{eq:antiperiodic_xx} 
	H_{\textsc{xx}'} = \sum_{i=1}^{N-1} \bigl( \sigma^x_i \, \sigma^x_{i+1} + \sigma^y_i \, \sigma^y_{i+1} \bigr) \: + \bigl( \sigma^x_N \, \sigma^x_1 - \sigma^y_N \, \sigma^y_1 \bigr) \, .
\end{equation}
The hamiltonian~\eqref{eq:antiperiodic_xx} also arises from a six-vertex transfer matrix with antiperiodic twist, and has been solved exactly in various ways, see e.g.\ \cite{batchelor_exact_1995, niccoli2013antiperiodic} and references therein. 
The SZ$'$ chain can thus be viewed as an integrable long-range deformation of the antiperiodic \textsc{xx} model. We moreover show how the SZ$'$ and FK chains should fit in the framework of `order-by-order integrability' from \cite{gombor2021integrable,gombor2022wrapping,de_leeuw_lifting_2023}.

\subsection{Deformed level} 

\noindent 
Recently, Matushko and Zotov \cite{MZ_23b} discovered a long-range spin chain that $q$-deforms \eqref{eq:SZ_intro}, see \textsection7 in \cite{MZ_23b}. Its (chiral) hamiltonians resemble those of the $q$-deformed Inozemtsev chain, except that it is based on elliptic \textit{R}-matrices of vertex-type. 
Our second set of results pertain to this model. 
In our (additive) notation, the crossing parameter $\eta$ sets the deformation parameter $q \sim \E^{\I \mspace{1mu} \eta}$.
We will show that the interactions of the Matushko--Zotov chain can be written in the form \eqref{eq:S^LR_diagr} and it has a potential that \textit{q}-deforms $V\mspace{-1mu}(u/2)/4$, which receives shifts depending on the spin direction as in \eqref{eq:SZ_intro}. We furthermore introduce the appropriate deformed translation operator.

Moreover, we will demonstrate that the pertinent elliptic functions can once more be chosen so as to extend to the full parameter space. We will similarly call this variant of \cite{MZ_23b} the \textit{MZ\/$'\!$ chain}. 
The long-range limit is, up to a spin rotation, the trigonometric MZ chain. 
The short-range limit of the MZ$'$ chain is a generalisation of \eqref{eq:antiperiodic_xx} with the same `bulk' terms but \textit{q}-deformed antiperiodic boundary terms.

Finally, we will compare the (vertex-type) landscape spanned by the MZ$'$ chain and the (face-type) landscape of the \textit{q}-deformed Inozemtsev chain. 
As the various limits attest, these two landscapes only share one single point (more about which in \textsection~\ref{sec:wrapping}): the face-vertex transformation does \emph{not} extend to the long-range spin chains. We will illustrate this with a practical comparison of various properties of the two sides.

While the case of spin 1/2 suffices for our purposes, everything readily extends to higher rank: just replace all \textit{R}-matrices by their known higher-rank counterparts.

\subsection{Outline} 

\noindent We will start at the \textit{q}-deformed level. The necessary elliptic preliminaries are collected in \textsection\ref{sec:preliminaries}. 
In \textsection\ref{sec:MZ_hams} we give the hamiltonians of the MZ$'$ chain and rewrite them in terms of interactions like in~\eqref{eq:S^LR_diagr}. 
The deformed translation operator is given in \textsection\ref{sec:MZ_transl}.
The core of the paper is \textsection\ref{sec:limits}.
In \textsection\ref{sec:MZ_limits} we evaluate the limits of the MZ$'$ chain, including the new short-range limit.
Arriving at the undeformed level, in \textsection\ref{sec:SZ_limits} we evaluate all limits of the SZ$'$ chain, including \eqref{eq:antiperiodic_xx}.
In \textsection\ref{sec:comparison} we compare the resulting (vertex-type) landscape with the (face-type) landscape of the \textit{q}-deformed Inozemtsev chain \cite{KL_23}. We furthermore show how the SZ$'$ chain is, in a precise sense, the antiperiodic counterpart of the Inozemtsev chain.
We conclude in \textsection\ref{sec:concl}.

There are several appendices. These contain an overview of all elliptic functions (\textsection\ref{app:ell_functions}) and  \textit{R}-matrices (\textsection\ref{app:R_matrices}) that we will need, along with their limits and relations, including the face-vertex transformation given in \textsection\ref{app:face_vertex}. The key identity \eqref{eq:RR_decomp} is derived in \textsection\ref{app:RR_decomp} and related to the Heisenberg \textsc{xyz} hamiltonian. The differences with the setup of \cite{MZ_23a,sechin2018r} are summarised in \textsection\ref{app:MZ_chain}.

\section{Preliminaries} 
\label{sec:preliminaries}

\noindent 
The long-range models that we are interested in feature elliptic functions. Appendix~\ref{app:ell_functions} contains an overview of the functions and properties that we will need. All of them can be constructed from a single building block: the odd Jacobi theta function. Given $N \in \mathbb{N}_{\geqslant 2}$ and $\omega = \I \pi/\kappa$ with $\kappa>0$, we take it to be\,%
\begin{equation} \label{eq:def_theta2}
	\begin{aligned} 
	\theta(u) \coloneqq {} & \frac{\sinh(\kappa \, u)}{\kappa} \prod_{n=1}^{\infty}\! \frac{\sinh [\kappa \, (n\,N + u)] \sinh[\kappa \, (n\,N - u)]}{\sinh^2 (\kappa\,n\,N)} \\
	= {} & \frac{\sinh(\kappa \, u)}{\kappa} + O\bigl(p^2\bigr) \, , \qquad p = \E^{-N\kappa} \, .
	\end{aligned}
\end{equation}
This is the unique odd entire function with $\theta'(0) = 1$ and double quasiperiodicity $\theta(u+\omega) = -\theta(u)$, $\theta(u+N) = - \E^{\kappa(2u+N)} \, \theta(u)$.
It is the standard odd Jacobi theta function with lattice parameter $\tau = \I \,N \mspace{1mu} \kappa/\pi$ and rescaled argument, see \textsection\ref{app:specialisations}.

For example, \eqref{eq:def_theta2} gives rise to 
the `prepotential'
\begin{equation} \label{eq:rho_2}
	\rho(u) \coloneqq \partial_u \log \theta(u) = \frac{\theta'(u)}{\theta(u)} = \zeta(u) + \frac{\eta_z}{\omega} \, u \, ,
\end{equation}
which is odd, has a simple pole (of residue $1$) at the origin, is $\omega$-periodic, and $N$-quasi\-periodic, $\rho(u+N) = \rho(u) + 2\kappa$. The final equality shows that $\rho$ differs from Weierstrass $\zeta$-function with quasiperiods $N,\omega$ by a linear term proportional to the constant $\eta_z \coloneqq 2 \,\zeta(-\omega/2)$, cf.\ \textsection\ref{app:ell_functions}. In turn, \eqref{eq:rho_2} yields the shifted Weierstra\ss\ potential from~\eqref{eq:V_intro} as\,%
\begin{equation} \label{eq:V}
	V\mspace{-1mu}(u) \coloneqq -\rho'(u) = \wp(u) - \frac{\eta_z}{\omega} = \wp(u) + \frac{\I \mspace{1mu} \kappa}{\pi}\,\eta_z \, ,
\end{equation}
This function is even, has a double pole at the origin, and is doubly periodic with periods $N$ and $\omega$. The landscape of functions contained in the potential is shown in Figure~\ref{fg:landscape_potential}. For details about the limits see \cite{Inozemtsev_1995}, \textsection2.2 of \cite{klabbers2022coordinate} and \textsection{E} of \cite{KL_23}.

\subsection{Baxter's \textit{R}-matrix} 
\label{sec:R8v}

\noindent
Baxter's eight-vertex \textit{R}-matrix \cite{baxter1972one} on $\mathbb{C}^2 \otimes \mathbb{C}^2$ has the form
\begin{equation} \label{eq:R8v}
	R(u;\eta) \coloneqq
	\begin{pmatrix}
		a(u;\eta) & \color{gray!80}{0} & \color{gray!80}{0} & d(u;\eta) \\
		\color{gray!80}{0} & b(u;\eta) & c(u;\eta) & \color{gray!80}{0} \\
		\color{gray!80}{0} & c(u;\eta) & b(u;\eta) & \color{gray!80}{0} \\
		d(u;\eta) & \color{gray!80}{0} & \color{gray!80}{0} & a(u;\eta)
	\end{pmatrix} \, .
\end{equation}
To specify the dependence of the vertex weights~$a(u;\eta),\dots,d(u;\eta)$ on the spectral parameter~$u$ and crossing parameter~$\eta$ we use the Kronecker elliptic function
\begin{equation} \label{eq:kronecker}
	\phi(u,v) \coloneqq\frac{\theta(u+v)}{\,\theta(u)\, \theta(v)} \, .
\end{equation}
It is related to \eqref{eq:V} by $\phi(u,v)\,\phi(u,-v) = \wp(u) - \wp(v)$. Next set
\begin{equation} \label{eq:g}
	g_\alpha(u;\eta) \coloneqq 
	\begin{cases} 
		\displaystyle \hphantom{\E^{-\kappa\,u} \,} \frac{\phi\bigl(u,(\eta+\omega_\alpha)/2\bigr)}{\phi(u,\eta)} \quad & \alpha = 0,z \, ,  \\[1.6ex]
		\displaystyle \E^{-\mspace{-1mu}\kappa\,u} \, \frac{\phi\bigl(u,(\eta+\omega_\alpha)/2\bigr)}{\phi(u,\eta)}\quad & \alpha = x,y \, ,
	\end{cases} 
\end{equation}
where the $\omega_\alpha$ that belong to our theta function \eqref{eq:def_theta2}, cf.~\textsection\ref{app:ell_functions}, are
\begin{equation} \label{eq:omega_alpha}
	\vec{\omega} \coloneqq (\omega_0,\omega_x,\omega_y,\omega_z) \coloneqq (0,N,N-\omega,-\omega)\, . 
\end{equation}
Recall that `$\sum_{\alpha = 0}^z$' means a sum over $\alpha \in \{0,x,y,z\}$. 
Then
\begin{equation} \label{eq:R8v_via_g}
	R(u;\eta) = \frac{1}{2} \sum_{\alpha = 0}^z g_\alpha(u;\eta) \, \sigma^\alpha \otimes \sigma^\alpha \, ,
\end{equation} 
so that the eight-vertex weights are
\begin{equation} \label{eq:8v_wts}
	\begin{aligned}
	a(u;\eta) & = \frac{g_0(u;\eta) + g_z(u;\eta)}{2} \, , \ \ && c(u;\eta) = \frac{g_x(u;\eta) + g_y(u;\eta)}{2} \, , \\
	b(u;\eta) & = \frac{g_0(u;\eta) - g_z(u;\eta)}{2} \, , && d(u;\eta) = \frac{g_x(u;\eta) - g_y(u;\eta)}{2} \, .
	\end{aligned}
\end{equation}
\smallskip

\textit{Anisotropic Heisenberg chain.} 
Traditional nearest-neighbour integrability tells us to introduce an auxiliary space, define a monodromy matrix, and trace over the auxiliary space to get the (homogeneous) transfer matrix $t(u;\eta) \coloneqq \mathrm{tr}_0 \bigl[ R_{0N}(u;\eta) \cdots R_{01}(u;\eta) \bigr]$. The Yang--Baxter equation and generic invertibility of the \textit{R}-matrix imply $\bigl[t(u;\eta),t(v;\eta)\bigr]=0$. Hence any expansion of the transfer matrix produces a family of commuting operators. Due to the `initial condition' $R(0;\eta) = P$ one expands at $u=0$. Then 
\begin{equation} \label{eq:homogeneous_translation}
	t(0;\eta) = P_{12} \cdots P_{N-1,N} 
\end{equation}
is the periodic lattice translation (right shift operator). The logarithmic derivative gives the Heisenberg \textsc{xyz} chain~\eqref{eq:Heis} \cite{sutherland1970two},
\begin{equation} \label{eq:log_der_transf}
\begin{aligned}
	t(0;\eta)^{-1} \, t'(0;\eta) = \sum_{i=1}^{\smash{N}} & P_{i,i+1} \, R'_{i,i+1}(0;\eta) \, , \\
	&P_{i,i+1} \, R'_{i,i+1}(0;\eta) = J \, \bigl( \sigma^x_i \, \sigma^x_{i+1} + \Gamma \, \sigma^y_i \, \sigma^y_{i+1} + \Delta \, \sigma^z_i \, \sigma^z_{i+1} \bigr) + C \, , 
	\end{aligned}
\end{equation}
with  $J,C$ constants, $\Gamma = \mathrm{dn}(\eta')$, and $\Delta = \mathrm{cn}(\eta')$ for $\eta' \coloneqq 2 K \eta/\omega$, see the end of \textsection\ref{app:RR_decomp}. 

For $\Gamma\neq 1$ spin excitations (magnons) can be created or annihilated in pairs, so standard Bethe-ansatz approaches requiring weight conservation do not apply. In 1973, Baxter invented two ways around this problem: the face-vertex transformation and the \textit{TQ}-equation. For us the former is relevant, see \textsection\ref{sec:FV} and \textsection\ref{app:face_vertex}.\! 
\medskip

\textit{Braid-like form.}
We prefer\,%
\footnote{\ Our reason for using \eqref{eq:Rch8v} is twofold. First, it can be viewed as a deformed spin permutation: $\check{R} \to P$ as $\eta \to 0$. Thus each ingredient of the hamiltonians will deform something nontrivial. Second, while the hamiltonians can be written via either $\check{R}$ (\textsection\ref{sec:MZ_hams}) or $R$ (as in \cite{MZ_23b}), the deformed translation operator (\textsection\ref{sec:MZ_transl}) naturally features $\check{R}$. \label{fn:Rch}}
to use the `braid-like' \textit{R}-matrix
\begin{equation} \label{eq:Rch8v}
	\check{R}(u;\eta) \coloneqq P \, R(u;\eta) = \frac{1}{2} \, P \sum_{\alpha = 0}^z g_\alpha(u;\eta) \, \sigma^\alpha \otimes \sigma^\alpha \, .
\end{equation} 
Note that its derivative already appeared in \eqref{eq:log_der_transf}. 
It obeys the `initial' condition $\check{R}(0;\eta) = \id \otimes \id$, is normalised such that $\check{R}(-u;\eta) \check{R}(u;\eta) = \id \otimes \id$, and satisfies the quantum Yang--Baxter equation (on $\mathbb{C}^2 \otimes \mathbb{C}^2 \otimes \mathbb{C}^2$) in the braid-like form
\begin{equation} \label{eq:YBE}
	\begin{aligned}
	\check{R}_{12}(u;\eta) \, & \check{R}_{23}(u+v;\eta)\, \check{R}_{12}(v;\eta) = \check{R}_{23}(v;\eta) \, \check{R}_{12}(v+u;\eta) \, \check{R}_{23}(u;\eta) \, .
	\end{aligned}
\end{equation}
Here $\check{R}_{12}(u;\eta) = \check{R}(u;\eta) \otimes \id$ and $\check{R}_{23}(u;\eta) = \id \otimes \check{R}(u;\eta)$: the subscripts of an \textit{R}-matrix indicate on which two tensor factors it acts nontrivially.
\medskip

\textit{Six-vertex limit.}
The trigonometric limit $\kappa \to 0$ of \eqref{eq:8v_wts} is not quite the usual (symmetric) six-vertex \textit{R}-matrix with weights
\begin{equation} \label{eq:6v_wts}
	\begin{aligned}
		a^\text{6v}(u;\gamma) & = 1 \, , && c^{\text{6v}}(u;\gamma) = \frac{\sin(\pi \gamma)}{\sin\bigl(\tfrac{\pi}{N} \mspace{1mu} u + \pi \gamma \bigr)} \, ,\! \\
		b^{\text{6v}}(u;\gamma) & = \frac{\sin\bigl(\tfrac{\pi}{N} \mspace{1mu} u\bigr)}{\sin\bigl(\tfrac{\pi}{N} \mspace{1mu} u + \pi \gamma \bigr)} \, , \quad && d^{\mspace{1mu}\text{6v}}(u;\gamma) = 0 \, ,
	\end{aligned}
\end{equation}
where we set $\eta = N\,\gamma$ so that $q=\E^{\I\pi\gamma}$ is independent of $N$. 
Rather, the trigonometric limit of \eqref{eq:Rch8v} differs by a spin rotation,
\begin{equation} \label{eq:R8v_trig}
	\lim_{\kappa \to 0} \check{R}(u;N\gamma) =  
	U^{\otimes 2} \, \check{R}^{\text{6v}}(u;\gamma) \, \bigl( U^{\otimes 2} \bigr)^{-1} \, ,
\end{equation}
where the (Hadamard-type) matrix
\begin{equation} \label{eq:U_mat}
	U \coloneqq \frac{1}{\sqrt{2}} 
	\begin{pmatrix}
	1 & -1 \\ 1 & 1 
	\end{pmatrix} , \quad\
	\begin{aligned}
		& U \, \sigma^0 \, U^{-1} = \sigma^0 \, , 
		&& U \, \sigma^x \, U^{-1} = -\sigma^z \, , \\ 
		& U \, \sigma^y \, U^{-1} = \sigma^y \, , 
		&& U \, \sigma^z \, U^{-1} = \ \sigma^x \, ,
	\end{aligned}
\end{equation}
swaps $\sigma^x \leftrightarrow \sigma^z$ up to a sign that drops out in $U^{\otimes 2}$. 
The explicit form of \eqref{eq:R8v_trig}, which still has eight vertices, is given in \textsection\ref{app:R_matrices}.

\section{The MZ$'$ spin chain}

\subsection{Chiral hamiltonians} 
\label{sec:MZ_hams}

\noindent
The \emph{MZ\/$'\!$ spin chain} is given by the `chiral' hamiltonians
\begin{subequations} \label{eq:MZ_ham_pre}
	\begin{gather}
	\begin{aligned}
	H^\textsc{l}_{\textsc{mz}'} = \sum_{i<j}^N  \Bigg(	\ordprod_{(j>)k(>i)} \!\!\!\!\!\!\! \check{R}_{k,k+1}\bigl(j-k;\eta\bigr) \Bigg)	\check{R}_{i,i+1}(j-i;\eta) \, \check{R}_{i,i+1}'(i-j;\eta) \Bigg( \ordprodopp_{(i<)k(<j)} \!\!\!\!\!\!\! \check{R}_{k,k+1}(k-j;\eta) \Bigg) \, , 
	\end{aligned}
\shortintertext{and}
	\begin{aligned}
	H^\textsc{r}_{\textsc{mz}'} = \sum_{i<j}^N \Bigg( \ordprodopp_{(i<)k(<j)} \!\!\!\!\!\!\! \check{R}_{k-1,k}(k-i;\eta) \Bigg) \check{R}_{j-1,j}(j-i;\eta) \, \check{R}_{j-1,j}'(i-j;\eta) \Bigg(  \ordprod_{(j>)k(>i)} \!\!\!\!\!\!\! \check{R}_{k-1,k}(i-k;\eta) \Bigg) \, ,
	\end{aligned}
	\end{gather}
\end{subequations}
where `$\leftharpoonup$' or `$\rightharpoonup$' on a product indicates the order of the factors.
Note that $\check{R}(-u;\eta) \, \check{R}'(u,\eta)$ already appears at $u=0$ as the nearest neighbour interaction~\eqref{eq:log_der_transf} of the Heisenberg chain. The remaining \textit{R}-matrices in \eqref{eq:MZ_ham_pre} arrange the transport to move either spin adjacent to the other.
This MZ$'$ chain is integrable in the sense that \eqref{eq:MZ_ham_pre} commute,
\begin{equation} \label{eq:commutativity}
	[H^\textsc{l}_{\textsc{mz}'} \,, H^\textsc{r}_{\textsc{mz}'}] = 0 \, ,
\end{equation}
and there exist additional higher hamiltonians that all commute with \eqref{eq:MZ_ham_pre} and each other. 

As we show in \textsection\ref{app:specialisations}, \ref{app:R8v} and \ref{app:MZ_chain}, our choice~\eqref{eq:def_theta2} of theta function ultimately amounts to a global spin rotation plus an additive correction, see \eqref{eq:MZ_ham_1vs2}. The two variants are thus generically physically equivalent. In \textsection\ref{sec:MZ_limits} we will show that the additive correction extends the parameter space by regularising the short-range limit.
This simple relation between the MZ$'$ and MZ chain, which is not at all clear from our derivation \cite{KL_extended} (cf.\ \textsection{G} in \cite{KL_23}), implies that the integrability is preserved and \eqref{eq:commutativity} holds \cite{MZ_23b}.

To make the structure of \eqref{eq:MZ_ham_pre} more manifest let us rewrite the factor in the middle of the spin interactions in~\eqref{eq:MZ_ham_pre}. Motivated by the undeformed level it is reasonable to seek a linear combination of the spin-spin interactions of the SZ$'$ chain~\eqref{eq:SZ_intro},
\begin{equation} \label{eq:E^alpha}
	\EE^\alpha \coloneqq 1- P \; \sigma^\alpha \otimes \sigma^\alpha \, ,
	\quad \ \
	\EE^\alpha = P-\sigma^\alpha \otimes \sigma^\alpha \ \ (\alpha \neq 0) \, .
\end{equation}
In Appendix \textsection\ref{app:RR_decomp} we show that this is indeed possible:
\begin{equation} \label{eq:RR_decomp}
	\check{R}(-u;\eta) \, \check{R}'(u;\eta) = 
	\sum_{\alpha=0}^z \mspace{2mu} \frac{1}{4} \mspace{1mu}  V\mspace{-1mu}\Bigl(\frac{u+\omega_\alpha}{2} \mspace{2mu}; \frac{\eta}{2} \Bigr) \, \EE^\alpha \, ,
\end{equation}
where we define
\begin{equation} \label{eq:V^eta}
	V\mspace{-1mu}(u;\eta) \coloneqq
	\sum_{\beta = 0}^z \mspace{-2mu} \frac{ A_\beta(\eta) \, \bigl( \wp\bigl(u + \tfrac{1}{2}\omega_\beta\bigr) -\wp\bigl(\eta + \tfrac{1}{2}\omega_\beta\bigr) \bigr) - 
	\frac{1}{2}\wp'(2\eta) }{\wp(2u) - \wp(2\eta)}
\end{equation}
with coefficients 
\begin{equation} \label{eq:A_beta}
	A_0(\eta) \coloneqq \tfrac{1}{2} \rho(\eta) \, , \quad
	A_\beta(\eta) \coloneqq \tfrac{1}{2} \rho\bigl(\eta+\tfrac{1}{2}\omega_\beta\bigr) - \tfrac{1}{2} \rho\bigl(\tfrac{1}{2}\omega_\beta\bigr) \ \ (\beta\neq 0) \, .
\end{equation}
In \eqref{eq:V^eta_to_V} we will see that it is a deformation of $V\mspace{-1mu}(u)$ defined in \eqref{eq:V}. 
Another deformation of $V\mspace{-1mu}(u)$ will appear in \textsection\ref{sec:qIno}.

Let us write e.g.\ $\tfrac{1}{4} \mspace{1mu} V\mspace{-1mu}\bigl(\tfrac{1}{2}(u+\vec{\omega});\tfrac{1}{2}\eta\bigr) \cdot \vec{\EE}$ as a short-hand for \eqref{eq:RR_decomp}. 
Then \eqref{eq:MZ_ham_pre} assumes the form
\begin{subequations} \label{eq:MZ_ham_decomp}
	\begin{gather}
	\begin{aligned} 
		H^\textsc{l}_{\textsc{mz}'} & = \sum_{i<j}^N \frac{1}{4} \mspace{1mu} V\mspace{-1mu}\biggl(\frac{i-j+\vec{\omega}}{2} \mspace{2mu};\frac{\eta}{2}\biggr) 
		\cdot \vec{\SS}_{[i,j]}^{\,\textsc{l}}(\eta) \, , \\
		H^\textsc{r}_{\textsc{mz}'} & = \sum_{i<j}^N \frac{1}{4} \mspace{1mu}  V\mspace{-1mu}\biggl(\frac{i-j+\vec{\omega}}{2}\mspace{2mu};\frac{\eta}{2}\biggr)
		\cdot \vec{\SS}_{[i,j]}^{\, \textsc{r}}(\eta) \, ,
	\end{aligned}
\intertext{with chiral long-range spin interactions in the $\alpha$-direction}
	\begin{aligned}
		\SS_{[i,j]}^{\mspace{1mu}\textsc{l},\,\alpha}(\eta) & \coloneqq 
		\! \ordprod_{(j>)k(>i)} \!\!\!\!\!\!\! \check{R}_{k,k+1}(j-k;\eta) 
		\ \, \EE^\alpha_{i,i+1} 
		\ordprodopp_{(i<)k(<j)} \!\!\!\!\!\!\! \check{R}_{k,k+1}(k-j;\eta) 
		\,, \\
		\SS_{[i,j]}^{\mspace{1mu}\textsc{r},\,\alpha}(\eta) & \coloneqq 
		\! \ordprodopp_{(i<)k(<j)} \!\!\!\!\!\!\! \check{R}_{k-1,k}(k-i;\eta) 
		\ \, \EE^\alpha_{j-1,j} 
		\ordprod_{(j>)k(>i)} \!\!\!\!\!\!\! \check{R}_{k-1,k}(i-k;\eta)
		\, .
	\end{aligned}
	\end{gather}
\end{subequations}
For example, for $N=3$ we have
\begin{equation}
	\begin{aligned}
		H_{\textsc{mz}'}^\textsc{l} = {} & \tfrac{1}{4} \mspace{1mu} V\mspace{-1mu}\bigl(\tfrac{1}{2}(-1 + \vec{\omega});\tfrac{1}{2}\eta\bigr) \cdot \bigl(\vec{\EE}_{12} + \vec{\EE}_{23}\bigr) \\
		& + \tfrac{1}{4} \mspace{1mu} V\mspace{-1mu}\bigl(\tfrac{1}{2}(-2 + \vec{\omega}\tfrac{1}{2}\eta)\bigr) \cdot \check{R}_{23}(1;\eta) \, \vec{\EE}_{12} \, \check{R}_{23}(-1;\eta) \, , \\
		H_{\textsc{mz}'}^\textsc{r} = {} & \tfrac{1}{4} \mspace{1mu} V\mspace{-1mu}\bigl(\tfrac{1}{2}(-1 + \vec{\omega});\tfrac{1}{2}\eta\bigr) \cdot \bigl(\vec{\EE}_{12} + \vec{\EE}_{23}\bigr)  \\
		& + \tfrac{1}{4} \mspace{1mu} V\mspace{-1mu}\bigl(\tfrac{1}{2}(-2 + \vec{\omega});\tfrac{1}{2}\eta\bigr) 
		\cdot \check{R}_{12}(1;\eta) \, \vec{\EE}_{23} \, \check{R}_{12}(-1;\eta) \, .
	\end{aligned}
\end{equation}

In the following we suppress the $\eta$-dependence of $\vec{\SS}_{[i,j]}^{\,\textsc{l},\textsc{r}}$.  Diagrammatically, the chiral long-range spin exchange operators in \eqref{eq:MZ_ham_decomp} have the form announced in \eqref{eq:S^LR_diagr}:
\begin{equation} \label{eq:MZ_ham_diagr}
	\SS_{[i,j]}^{\textsc{l},\,\alpha} = \!
	\tikz[baseline={([yshift=-.5*11pt*0.13+5pt]current bounding box.center)},xscale=0.35,yscale=0.15,font=\footnotesize]{
		\draw[->] (9.5,0) node[below]{$N$} -- (9.5,10); 
		\foreach \x in {-1,...,1} \draw (8.75+.3*\x,-1.4) node{$\cdot\mathstrut$};
		\draw[rounded corners=2pt,->] (8,0) node[below]{$j$} -- (8,1.5) -- (5,4.5) -- (5,5.5) -- (8,8.5) -- (8,10);
		\draw[rounded corners=2pt,->] (7,0) -- (7,1.5) -- (8,2.5) -- (8,7.5) -- (7,8.5) -- (7,10); 
		\draw[rounded corners=2pt,->] (6,0) -- (6,2.5) -- (7,3.5) -- (7,6.5) -- (6,7.5) -- (6,10);
		\foreach \x in {-1,...,1} \draw (6+.3*\x,-1.4) node{$\cdot\mathstrut$};
		\draw[rounded corners=2pt,->] (5,0) -- (5,3.5) -- (6,4.5) -- (6,5.5) -- (5,6.5) -- (5,10); 
		\draw[->] (4,0) node[below]{$i$} -- (4,10);
		\foreach \x in {-1,...,1} \draw (3.25+.3*\x,-1.4) node{$\cdot\mathstrut$};
		\draw[->] (2.5,0) node[below]{$1$} -- (2.5,10);
		\draw[style={decorate, decoration={zigzag,amplitude=.5mm,segment length=1mm}}] (4,5) -- (5,5);
		\foreach \x in {-1,...,1} \draw (3.25+.3*\x,5) node{$\cdot\mathstrut$};
		\foreach \x in {-1,...,1} \draw (8.75+.3*\x,5) node{$\cdot\mathstrut$};	
		\node[yshift=.02cm] at (4.5,6) {$\scriptsize\alpha$};
		} \! , 
	\quad
	\SS_{[i,j]}^{\textsc{r},\,\alpha} = \!
	\tikz[baseline={([yshift=-.5*11pt*0.13+5pt]current bounding box.center)},xscale=-0.35,yscale=0.15,font=\footnotesize]{
		\draw[->] (9.5,0) node[below]{$1$} -- (9.5,10); 
		\foreach \x in {-1,...,1} \draw (8.75+.3*\x,-1.4) node{$\cdot\mathstrut$};
		\draw[rounded corners=2pt,->] (8,0) node[below]{$i$} -- (8,1.5) -- (5,4.5) -- (5,5.5) -- (8,8.5) -- (8,10);
		\draw[rounded corners=2pt,->] (7,0) -- (7,1.5) -- (8,2.5) -- (8,7.5) -- (7,8.5) -- (7,10);
		\draw[rounded corners=2pt,->] (6,0) -- (6,2.5) -- (7,3.5) -- (7,6.5) -- (6,7.5) -- (6,10);
		\foreach \x in {-1,...,1} \draw (6+.3*\x,-1.4) node{$\cdot\mathstrut$};
		\draw[rounded corners=2pt,->] (5,0) -- (5,3.5) -- (6,4.5) -- (6,5.5) -- (5,6.5) -- (5,10);
		\draw[->] (4,0) node[below]{$j$} -- (4,10);
		\foreach \x in {-1,...,1} \draw (3.25+.3*\x,-1.4) node{$\cdot\mathstrut$};
		\draw[->] (2.5,0) node[below]{$N$} -- (2.5,10);
		\draw[style={decorate, decoration={zigzag,amplitude=.5mm,segment length=1mm}}] (4,5) -- (5,5);
		\foreach \x in {-1,...,1} \draw (3.25+.3*\x,5) node{$\cdot\mathstrut$};
		\foreach \x in {-1,...,1} \draw (8.75+.3*\x,5) node{$\cdot\mathstrut$};	
		\node[yshift=.02cm] at (4.5,6) {$\scriptsize\alpha$};
		} \! . 
\end{equation}
These diagrams encode the spin operators in \eqref{eq:MZ_ham_decomp} as follows. The diagrams are read from bottom to top, as the little arrows at the top of the vertical lines indicate. Each line carries an inhomogeneity parameter, indicated at the start (below) the line.\,%
\footnote{\ Unlike when one works with the \textit{R}-matrix $R(u) = P\,\check{R}(u)$, for us the spaces $\mathbb{C}^2$ do \emph{not} follow the lines: the spaces are labelled $1$ to $N$ from left to right at \emph{each} step (horizontal slice) of the diagram.} 
Each crossing (vertex) represents a deformed permutation implemented through the eight-vertex \textit{R}-matrix~\eqref{eq:Rch8v}, while the zigzag line indicates the nearest-neighbour exchange interaction in the $\alpha$-direction: 
\begin{equation} \label{eq:Rch_and_E_diagram}
	\check{R}(u-v;\eta) = \!
	\tikz[baseline={([yshift=-2*11pt*.15]current bounding box.center)},xscale=.35,yscale=.15,font=\footnotesize]{
		\draw[rounded corners=2pt,->] (1,-.5) node[below]{$v$} -- (1,.5) -- (0,1.5) -- (0,2.5) node[above,yshift=-.05cm]{\textcolor{gray!80}{$v$}};
		\draw[rounded corners=2pt,->] (0,-.5) node[below] {$u$} -- (0,.5) -- (1,1.5) -- (1,2.5) node[above,yshift=-.05cm]{\textcolor{gray!80}{$u$}};
	} , \qquad
	\EE^\alpha = \tikz[baseline={([yshift=-2*11pt*0.15]current bounding box.center)},xscale=0.35,yscale=0.15,font=\footnotesize]{
		\draw[->] (1,-.5) node[below] {$u$} -- (1,2.5) node[above,yshift=-.05cm] {\textcolor{gray!80}{$u$}};
		\draw[->] (2.4,-.5) node[below] {$v$} -- (2.4,2.5) node[above,yshift=-.05cm] {\textcolor{gray!80}{$v$}};
		\draw[style={decorate, decoration={zigzag,amplitude=.5mm,segment length=1mm}}] (1,1) -- node[above] {$\scriptsize\alpha$} (2.4,1);
		} \, .
\end{equation}
Finally, in \eqref{eq:MZ_ham_diagr} the inhomogeneity parameters are `frozen' to values $u_j = j$ for all $1\leqslant j\leqslant N$.

The deformed long-range interactions of $H_{\textsc{mz}'}^\textsc{l}$ (resp.~$H_{\textsc{mz}'}^\textsc{r}$) can thus be interpreted in the following way.
First, the spin at site~$j$ is transported to site~$i+1$ (resp.~$i$ to $j-1$) by the deformed permutations $\check{R}(u;\eta)$; this is encoded by the crossings in the lower half of the diagrams in \eqref{eq:MZ_ham_diagr}, accounting for the ordered products on the right in \eqref{eq:MZ_ham_decomp}.
Next are the components $\EE^\alpha$ of the exchange interaction between the (now adjacent) spins. Finally, the spin at site $i+1$ is transported back to $j$ (resp.~$j-1$ to $i$), corresponding to the top half of the diagram in \eqref{eq:MZ_ham_diagr} and the ordered products on the left in \eqref{eq:MZ_ham_decomp}. The potentials $V\bigl((u+\omega_\alpha)/2;\eta/2\bigr)/4$ control the range and strength of the interactions.

\subsection{Deformed translation} 
\label{sec:MZ_transl}

\noindent
The MZ$'$ spin chain is not translationally invariant because of its boundary conditions, as is evident from comparing the bulk terms $\vec{\SS}_{[i,i+1]}^{\,\textsc{l}} = \vec{\EE}_{i,i+1} = \vec{\SS}_{[i,i+1]}^{\,\textsc{r}}$ and the highly nonlocal $\vec{\SS}_{[1,N]}^{\,\textsc{l}} \neq \vec{\SS}_{[1,N]}^{\,\textsc{r}}$. However, it is invariant under a modified translation operator, which belongs to the model's hierarchy of commuting operators. This (twisted) left translation operator is given by
\begin{equation} \label{eq:MZ_trans}
	G_{\textsc{mz}'} = \!
	\tikz[baseline={([yshift=-.5*11pt*0.13+5pt]current bounding box.center)},xscale=0.35,yscale=0.15,font=\footnotesize]{
		\node at (5,7-.1) {$\tikz[baseline={([yshift=-.5*11pt*.25]current bounding box.center)},scale=.35]{\fill[black, yshift=-.2] (0,0) rectangle ++(.4,.4)}$};
		\draw[rounded corners=2pt,->] (0,0) node[below]{$1$} -- (0,1) -- (5,6) -- (5,8.5);
		\draw[rounded corners=2pt,->] (1,0) node[below]{$2$} -- (1,1) -- (0,2) -- (0,8.5);
		\foreach \x in {2,...,4} \draw[rounded corners=2pt,->] (\x,0) -- (\x,\x) -- (\x-1,\x+1) -- (\x-1,8.5);
		\foreach \x in {-1,...,1} \draw (3+.3*\x,-1.4) node{$\cdot\mathstrut$};
		\draw[rounded corners=2pt,->] (5,0) node[below]{$N$} -- (5,5) -- (4,6) -- (4,8.5);
		} 
	\!\! = \E^{-(N-1)\kappa \mspace{1mu}\eta/2} \, \sigma^x_N \!\!\! \ordprod_{N\geqslant i> 1} \!\!\!\!\! \check{R}_{i-1,i}(1-i;\eta) \, ,
\end{equation}
where $\tikz[baseline={([yshift=-.5*11pt*0.13-3pt]current bounding box.center)},xscale=0.45,yscale=0.195,font=\footnotesize]{
\node at (0,1-.05) {$\tikz[baseline={([yshift=-.5*11pt*.25]current bounding box.center)},scale=.35]{\fill[black] (0,0) rectangle ++(.4,.4)}$};
\draw[->] (0,0) -- (0,2);
} \!\! \coloneqq \E^{-(N-1)\kappa\mspace{1mu}\eta/2} \, \sigma^x$ is an antiperiodic twist, acting at site~$N$ in \eqref{eq:MZ_trans}, and we emphasise once more that below each line we indicate the (integer) value of the inhomogeneity parameter attached to that line. Like for the hamiltonians, the \textit{R}-matrix $\check{R}(u;\eta)$ plays the role of a deformed permutation taking care of the transport along the chain. The deformed translation~\eqref{eq:MZ_trans} obeys the antiperiodic boundary condition 
\begin{equation} \label{eq:MZ_quasiperiodicity}
	G_{\textsc{mz}'}^{\,N} = \prod_{i=1}^N \sigma^x_i\, .
\end{equation} 
Note that \eqref{eq:MZ_quasiperiodicity} is a central element: it is the global spin flip operator $\uparrow \,\leftrightarrow\, \downarrow$ (which is the action of the Weyl group of $\mathfrak{sl}_2$).
The inverse translation can be written as
\begin{equation} \label{eq:MZ_trans_inv}
	G^{\,-1}_{\textsc{mz}'} = \!
	\tikz[baseline={([yshift=-.5*11pt*0.13+5pt]current bounding box.center)},xscale=-0.35,yscale=0.15,font=\footnotesize]{
		\node at (5,7-.1) {$\tikz[baseline={([yshift=-.5*11pt*.25]current bounding box.center)},scale=.35]{\fill[black, yshift=-.2] (0,0) rectangle ++(.4,.4)}$};
		\draw[rounded corners=2pt,->] (0,0) node[below,xshift=.07cm]{$N$} -- (0,1) -- (5,6) -- (5,8.5);
		\draw[rounded corners=2pt,->] (1,0) node[below,xshift=-.05cm]{$N{-}1$} -- (1,1) -- (0,2) -- (0,8.5);
		\foreach \x in {2,...,4} \draw[rounded corners=2pt,->] (\x,0) -- (\x,\x) -- (\x-1,\x+1) -- (\x-1,8.5);
		\foreach \x in {-1,...,1} \draw (3+.3*\x,-1.4) node{$\cdot\mathstrut$};
		\draw[rounded corners=2pt,->] (5,0) node[below]{$1$} -- (5,5) -- (4,6) -- (4,8.5);
	}
	\!\!\! = \E^{-(N-1)\kappa\mspace{1mu}\eta/2} \, \sigma^x_1 \!\!\! \ordprodopp_{1 \leqslant i < N} \!\!\!\! \check{R}_{i,i+1}(i-N;\eta) \, .
\end{equation}
The proof of the commutativity of \eqref{eq:MZ_trans} with the hamiltonians is similar to that in \cite{lamers2022spin}, see also our upcoming \cite{KL_extended}.

\section{Limits} \label{sec:limits}

\noindent The MZ$'$ chain unifies various spin chains that reside in different corners of the parameter space.
There are three parameters: the periods $N$ and $\omega = \I\pi/\kappa$, which combine into the elliptic nome $\tau=-N/\omega = \I \mspace{1mu} N \mspace{1mu} \kappa/\pi$, together with the deformation (`crossing') parameter $\eta$. The following limits are of interest.
\begin{itemize}
	\item the \emph{long-range} limit $\kappa \to 0^+$, where the imaginary period $\omega \to \I \,\infty$ is removed to give trigonometric functions;
	\item the \emph{short-range} limit $\kappa \to \infty$, where instead $\omega \to \I \, 0^+$;
	\item the \emph{macroscopic} limit $N \to \infty$, which we will only treat formally,
	\item the \emph{undeformed} limit $\eta\to0$;
	\item its subsequent limits.
\end{itemize}

The short-range limit, where the imaginary period $\omega \to \I \, 0^+$ is sent to zero, is clearly badly behaved. It has to be treated with care. As we will see, the worst divergences are hidden by the lattice spacing, which ensures that all \textit{R}-matrices and the potential have integer arguments $\neq 0\,\mathrm{mod}\,N$. This will make it possible to evaluate the short-range limit in our conventions.
For the \textit{R}-matrix the long-range, macroscopic and undeformed limits are standard, while the short-range limit gives a well-defined and nontrivial result if we simultaneously send the crossing parameter to zero, $\eta = \omega\,\nncrossing$ with $\nncrossing$ fixed. 

Note that in the long-range limit \eqref{eq:R8v_trig} the conjugation by $U^{\otimes 2}$, switching the spin-$x$ and -$z$ directions, will extend from the six-vertex to the whole Hilbert space. We will present the results in the usual form by conjugation by $U^{\otimes N}$, and call the (thus rotated) result the trigonometric MZ (rather than MZ$'$) chain, and similarly at the undeformed level.\,%
\footnote{\ This conjugation is absent in \cite{MZ_23b}. It can be moved from the long-range to the short-range \emph{and} macroscopic limits by reordering $\vec{\omega}$ as $(0,-\omega,N-\omega,N)$. Our choice follows elliptic standards (cf.~\textsection\ref{app:ell_functions}). \label{fn:trig_lim_conj}}

A useful feature of the decomposition~\eqref{eq:RR_decomp} is that, rather than having to analyse the different entries of $\check{R}(-u;\eta) \, \check{R}'(u;\eta)$ separately, it remains to make sense of the limits of $V\mspace{-1mu}(u;\eta)$.

\subsection{Normalisation}
\noindent
In order for all limits to be finite but non-trivial we need to renormalise the the function $V\mspace{-1mu}(u;\eta)$ from~\eqref{eq:V^eta} by multiplying it by some factor $n_{\kappa,\eta}$, where we suppress the dependence on $N$. As we will see, the asymptotics of the function~\eqref{eq:V^eta} yields the following wish-list for the prefactor:
\begin{itemize}
	\item long-range limit: $n_{\kappa,\eta} \sim \kappa^0$ as $\kappa \to 0^+$;
	\item short-range limit: $n_{\kappa,\omega \nncrossing} \sim \E^\kappa \!/\kappa$ as $\kappa \to \infty$, where $\eta = \omega\mspace{1mu} \nncrossing$ for fixed $\nncrossing$;
	\item undeformed limit: $n_{\kappa,\eta} \sim \eta^{-1}$ as $\eta\to 0$;
\end{itemize}
and at the undeformed level 
\begin{itemize}
	\item long-range limit: $\text{Res}_{\eta = 0} \, n_{\kappa,\eta} \sim \kappa^0$ as $\kappa \to 0^+$;
	\item short-range limit: $\text{Res}_{\nncrossing = 0} \, n_{\kappa,\omega\mspace{1mu} \nncrossing} \sim \E^\kappa \!/\kappa^2$ as $\kappa \to \infty$;
	\item macroscopic limit: $\text{Res}_{\eta = 0} \, n_{\kappa,\eta} \sim N^0$ as $N\to\infty$.
\end{itemize}
While such a prefactor is not unique, a particularly nice choice is 
\begin{equation} \label{eq:normalisation_choice}
	n_{\kappa,\eta} =  \frac{n_\kappa}{\theta(\eta)} \, , \quad 
	n_\kappa = \frac{\sinh^2 \kappa}{\kappa^2 \cosh{\kappa}} \, . 
\end{equation}
By construction $n_\kappa = \Res_{\eta=0} \, n_{\kappa,\eta}$ is the residue at the pole at $\eta=0$. It will be the normalisation at the undeformed level. We will use bars to denote (re)normalised objects.
\medskip

\subsection{Deformed level} 
\label{sec:MZ_limits}

\noindent
In order to evaluate the limit of the renormalised hamiltonians
\begin{equation}
	\longbar{H}_{\textsc{mz}'}^\textsc{l} \coloneqq n_{\kappa,\eta} \, H_{\textsc{mz}'}^\textsc{l} \, , \quad
	\longbar{H}_{\textsc{mz}'}^\textsc{r} \coloneqq n_{\kappa,\eta} \, H_{\textsc{mz}'}^\textsc{r} \, .
\end{equation}
we will need the limits of
\begin{enumerate}
	\item[1.] the eight-vertex \textit{R}-matrix: this is standard, see \textsection\ref{app:R_matrices};
	\item[2.] crucially, the renormalised potential
\begin{equation}
	\longbar{V}\mspace{-1mu}(u/2;\eta/2) \coloneqq n_{\kappa,\eta} \, V\mspace{-1mu}(u/2;\eta/2) \, ,
\end{equation}
	\item[3.]the coefficients in the decomposition
	\eqref{eq:RR_decomp}, which readily follow from 2.
\end{enumerate}
Let us invent the shorthand (a ligature of `$V$' and `$\mspace{-2mu}\EE$')
\begin{equation} \label{eq:MZ_spin_interaction}
	\begin{aligned}
	\longbar{\VE}(u;\eta) \coloneqq {} & 
	\frac{1}{4} \mspace{1mu} \longbar{V}\mspace{-1mu}\Bigl(\frac{u+\vec{\omega}}{2}\mspace{2mu};\frac{\eta}{2}\Bigr) \cdot \vec{\EE} =
	n_{\kappa,\eta} \sum_{\alpha=0}^z \frac{1}{4} \mspace{1mu} V\mspace{-1mu}\Bigl(\frac{u+\omega_\alpha}{2}\mspace{2mu};\frac{\eta}{2}\Bigr) \, \EE^\alpha \, .
	\end{aligned}
\end{equation}
\smallskip

\textit{Long-range limit.} When $\kappa\to 0$ all elliptic functions degenerates into trigonometric ones. To make our treatments of the long- and short-range limits more symmetric we set $\eta = N\,\gamma$. As we saw in \eqref{eq:R8v_trig}, the eight-vertex \textit{R}-matrix becomes the usual (symmetric) six-vertex \textit{R}-matrix \emph{up to a spin rotation}. Therefore
\begin{equation}
	\lim_{\kappa\to0} G_{\textsc{mz}'} = G_{\text{tri}\,\textsc{mz}'} = U^{\otimes N} \, G_{\text{tri}\,\textsc{mz}} \; \bigl(U^{\otimes N} \bigr)^{\!-1} \, ,
\end{equation}
with deformed translation operator
\begin{equation} \label{eq:MZ_trans_tri}
	G_{\text{tri}\,\textsc{mz}} = \sigma^z_N \!\!\! \ordprod_{N\geqslant i> 1} \!\!\!\!\! \check{R}_{i-1,i}^\text{6v}(1-i;\gamma) \, .
\end{equation}
Note that the twist is rotated into the spin-$z$ direction.

The limit of $\longbar{V}\mspace{-1mu}(u;N\,\gamma)$ is straightforward. For a nice result admitting an undeformed limit we need $n_{\kappa,N\gamma} \sim \pi/(N \sin \pi \gamma)$. The result is a `point-splitting' of the trigonometric potential \eqref{eq:V_tri_intro}:
\begin{equation} \label{eq:V_MZ_trig}
	\begin{aligned}
	\longbar{V}_\text{\!tri}(u;\gamma) \coloneqq {} &
	\lim_{\kappa \to 0} \longbar{V}\mspace{-1mu}(u;N\mspace{1mu}\gamma) = \frac{(\pi/N)^2}{\sin\bigl(\tfrac{\pi}{N} \mspace{1mu} u + \pi\mspace{1mu}\gamma\bigr) \sin \bigl(\tfrac{\pi}{N} \mspace{1mu} u - \pi\mspace{1mu}\gamma\bigr) }\, . 
	\end{aligned}
\end{equation}
For $\alpha=x$ we shift $u\to u+N$, which changes  $\sin(u+\pi/2) = \cos(u)$ in \eqref{eq:V_MZ_trig}.
Since both $\omega_y = N-\omega$ and $\omega_z = -\omega$ diverge, the terms with $\alpha=y,z$ are suppressed in the limit $\omega \to \I\,\infty$. 
Hence
\begin{equation} \label{eq:trig_spin_int}
	\begin{aligned}
	\lim_{\kappa \to 0} 
	\longbar{\VE}(u;N\mspace{1mu}\gamma) & = \frac{1}{4} \mspace{1mu} \longbar{V}_{\!\text{tri}}\Bigl(\frac{u}{2}\mspace{2mu};\frac{\gamma}{2}\Bigr) \EE^0 + \frac{1}{4} \mspace{1mu} \longbar{V}_{\!\text{tri}} \Bigl(\frac{u+N}{2}\mspace{2mu};\frac{\gamma}{2}\Bigr) \EE^x \\ 
	& = U^{\otimes 2} \; \longbar{V}_\text{\!tri}(u;\gamma) \, E^\text{tri}(u;\gamma) \; \bigl(U^{\otimes 2} \bigr)^{\!-1} \, ,
	\end{aligned}
\end{equation}
where we use $\EE^x = U^{\otimes 2} \, \EE^z \, \bigl(U^{\otimes 2} \bigr)^{-1}$ to arrive at
\begin{equation} \label{eq:trig_spin_nn}
	E^\text{tri}(u;\gamma) \coloneqq \cos\bigl(\tfrac{\pi}{N}\mspace{1mu}u\bigr) \, \EE^{\mspace{1mu}\textsc{xx}} + \cos(\pi\,\gamma) \, \EE^\diag
\end{equation}
where the \textsc{xx} (hopping, i.e.\ kinetic) and diagonal (potential) spin-spin interactions are
\begin{equation} \label{eq:E^xx}
	\begin{aligned}
		\EE^{\mspace{1mu}\textsc{xx}} \coloneqq \frac{1}{2} \bigl(\EE^0 - \EE^z\bigr) & = - \frac{1}{2} \bigl( \sigma^x \otimes \sigma^x + \sigma^y \otimes \sigma^y \bigr) \, , \\
		\EE^\diag \coloneqq \frac{1}{2} \bigl(\EE^0 + \EE^z\bigr) & = - \frac{1}{2} \bigl( \sigma^z \otimes \sigma^z - 1 \bigr) \, .
	\end{aligned}
\end{equation}
The sum over $\alpha$ has disappeared in \eqref{eq:trig_spin_int}, and $\sum_\alpha \EE^\alpha$ ($= 2 \; \id\otimes\id$) is replaced by \eqref{eq:trig_spin_nn}, which deforms the isotropic interaction $1-P$. 

Since Pauli matrices always appear as $\sigma^\alpha \otimes \sigma^\alpha$
the minus signs in the spin rotation \eqref{eq:U_mat} cancel. We thus arrive, up to a conjugation
\begin{equation} \label{eq:MZ_trig_lim}
	\lim_{\kappa \to 0}  \longbar{H}_{\mspace{-2mu}\textsc{mz}'}^{\textsc{l},\textsc{r}} \big|_{\eta = N \gamma} = \longbar{H}_{\mspace{-2mu}\text{tri}\,\textsc{mz}'}^{\textsc{l},\textsc{r}}  = \, U^{\otimes N} \, \longbar{H}_{\mspace{-2mu}\text{tri}\,\textsc{mz}}^{\textsc{l},\textsc{r}} \; \bigl( U^{\otimes N} \bigr)^{-1} \, ,
\end{equation}
at the trigonometric MZ chain \cite{MZ_23b} in the form
\begin{subequations} \label{eq:MZ_trig}
	\begin{gather}
	\begin{aligned}
	\longbar{H}^\textsc{l}_{\mspace{-1mu}\text{tri}\,\textsc{mz}}& = \sum_{i<j}^N  \longbar{V}_{\!\text{tri}}(i-j;\gamma) \, S_{[i,j]}^{\textsc{l},\text{tri}}(\gamma) \, , \qquad
	\longbar{H}^\textsc{r}_{\mspace{-1mu}\text{tri}\,\textsc{mz}} &= \sum_{i<j}^N \longbar{V}_{\!\text{tri}}(i-j;\gamma) \, S_{[i,j]}^{\textsc{r},\text{tri}}(\gamma) \, .
	\end{aligned}
\intertext{Here the potential (without shifts!) is \eqref{eq:V_MZ_trig}, and the interactions}
	\begin{aligned}
	S_{[i,j]}^{\textsc{l},\,\text{tri}}(\gamma) \coloneqq {}& \Bigg( \ordprod_{(j>)k(>i)} \!\!\!\!\!\!\! \check{R}_{k,k+1}^\text{6v}(j-k;\gamma) \Bigg) E^\text{tri}_{i,i+1}(i-j;\gamma) \,\Bigg( \ordprodopp_{(i<)k(<j)} \!\!\!\!\!\!\! \check{R}_{k,k+1}^\text{6v}(k-j;\gamma) \Bigg) \,, \\
	S_{[i,j]}^{\textsc{r},\,\text{tri}}(\gamma) \coloneqq {}& \Bigg( \ordprodopp_{(i<)k(<j)} \!\!\!\!\!\!\! \check{R}_{k-1,k}^\text{6v}(k-i;\gamma) \Bigg) E^\text{tri}_{j-1,j}(i-j;\gamma) \Bigg( \ordprod_{(j>)k(>i)} \!\!\!\!\!\!\! \check{R}_{k-1,k}^\text{6v}(i-k;\gamma) \Bigg) \, ,
	\end{aligned}
	\end{gather}
\end{subequations}
are built from \eqref{eq:trig_spin_nn} and the six-vertex \textit{R}-matrix \eqref{eq:6v_wts}--\eqref{eq:R8v_trig}. Unlike $\SS_{[i,j]}^{\textsc{l},\alpha}$ and $\SS_{[i,j]}^{\textsc{r},\alpha}$, both $S_{[i,j]}^{\textsc{l}}$ and $S_{[i,j]}^{\textsc{r}}$ deform $1-P_{ij}$. Graphically, they look as in \eqref{eq:MZ_ham_diagr} without $\alpha$, where
\begin{equation} \label{eq:E_diagr}
	E^\text{tri}(u-v;\gamma) = \tikz[baseline={([yshift=-2*11pt*0.15]current bounding box.center)},xscale=0.35,yscale=0.15,font=\footnotesize]{
		\draw[->] (1,-.5) node[below] {$u$} -- (1,2.5) node[above,yshift=-.05cm] {\textcolor{gray!80}{$u$}};
		\draw[->] (2.4,-.5) node[below] {$v$} -- (2.4,2.5) node[above,yshift=-.05cm] {\textcolor{gray!80}{$v$}};
		\draw[style={decorate, decoration={zigzag,amplitude=.5mm,segment length=1mm}}] (1,1) -- (2.4,1);
	} \, .
\end{equation}

To compare with \cite{MZ_23b} we note that, cf.\ \eqref{eq:V_MZ_trig},
\begin{equation} \label{eq:RR_decomp_tri}
	\begin{aligned} 
	\check{R}^{\text{6v}}(-u;\gamma) \, \partial_u \check{R}^{\text{6v}}(u;\gamma) & = V_{\text{tri}}(u;\gamma) \, E^\text{tri}(u;\gamma) \\
	& = \tfrac{N}{\pi} \sin(\pi \gamma) \, \longbar{V}_{\!\text{tri}}(u;\gamma) \, E^\text{tri}(u;\gamma) \, .
	\end{aligned}
\end{equation}
Hence \eqref{eq:MZ_trig} differs by a factor of $\lim_{\kappa\to0} n_{N\gamma,\kappa} = \pi/(N \sin \pi \gamma)$ from the trigonometric MZ chain in the form \eqref{eq:MZ_ham_pre} with six-vertex \textit{R}-matrix, as in \cite{MZ_23b}.
We further note that the potential~\eqref{eq:V_MZ_trig} and the structure of \eqref{eq:MZ_trig} are identical to that of the $q$-deformed Haldane--Shastry chain \cite{Lam_18,lamers2022spin}. 
We will get back to this comparison in \textsection\ref{sec:practical_comparison}.
\medskip

\textit{Short-range limit.} 
The limit $\kappa \to \infty$ is subtle and hinges on a delicate balance between the different factors in the hamiltonians. This time we set $\eta = \omega\,\nncrossing$. For the \textit{R}-matrix we use $\check{R}(\omega\,u;\omega\,\nncrossing) \to \check{R}^{\text{6v}}(N \, u;\nncrossing)$ 
as computed in \textsection\ref{app:R_matrices}. However, we cannot simply rescale the spectral (inhomogeneity) parameters, as their values are fixed to integers in the definitions \eqref{eq:MZ_ham_pre} and \eqref{eq:MZ_ham_decomp}. Nevertheless, from this limit we find for real $u$
\begin{equation} \label{eq:R_nn_lim}
	\lim_{\kappa \to \infty} \! \check{R}(u;\omega\mspace{2mu}\nncrossing) = 
	\lim_{\kappa \to \infty} \! \check{R}^{\text{6v}}\bigl(\tfrac{N}{\omega}\mspace{2mu} u;\nncrossing\bigr) =  P\bigl(-\mathrm{sgn}(u)\mspace{1mu}\nncrossing\bigr) \, , 
\end{equation}
where we define the braid operator
\begin{equation} \label{eq:R_XX}
	P(\nncrossing) \coloneqq 
	\begin{pmatrix}
		1 & \color{gray!80}{0} & \color{gray!80}{0} & 0 \\
		\color{gray!80}{0} & \color{gray!80}{0} & \!\!\E^{\I \pi \nncrossing}\!\! & \color{gray!80}{0} \\[2pt]
		\color{gray!80}{0} & \!\!\E^{\I \pi \nncrossing}\!\! & \color{gray!80}{0} & \color{gray!80}{0} \\[3pt]
		0 & \color{gray!80}{0} & \color{gray!80}{0} & 1
	\end{pmatrix} \, . 
\end{equation}
Thus the short-range, or `nearest neighbour' (`nn'), limit of the deformed translation operator \eqref{eq:MZ_trans} is 
\begin{equation} \label{eq:MZ_trans_nn}
	G_{\text{nn}\,\textsc{mz}'} \coloneq \E^{-\I (N-1)
	\pi \mspace{1mu} \nncrossing \!/2} \, \sigma^x_N \!\!\! \ordprod_{N\geqslant i> 1} \!\!\!\!\! P_{i-1,i}(\nncrossing) \, , \ \,
\end{equation}
obeying 
\begin{equation}
	G_{\text{nn}\,\textsc{mz}'}^{-1} = \sigma^x_1 \!\!\! \ordprodopp_{1 \leqslant i < N} \!\!\!\!\! P_{i,i+1}(\nncrossing) \, , \quad
	G_{\text{nn}\,\textsc{mz}'}^N = 
	\prod_{i=1}^N \sigma^x_i \, . \!\!
\end{equation}

For the hamiltonian we also need the limits of the function~$V\mspace{-1mu}(u;\eta)$ from \eqref{eq:V^eta} and its versions with shifted arguments. 
Rewriting \eqref{eq:V^eta} in terms of $V\mspace{-1mu}(u) = \wp(u) + \text{cst}$ from \eqref{eq:V} allows us to use the expansions
\begin{equation} \label{eq:wp_expansions_first_order}
	\begin{aligned}
		V\mspace{-1mu}\Bigl(\frac{u}{2}\Bigr) & = \hphantom{-}4\,\kappa^2  \, t^{|u|_{2N}^{\vphantom{n}}} + O\bigl(t^{2 \mspace{2mu} |u|_{2N}^{\vphantom{n}}}\bigr)  \, , \\ 
		V\mspace{-1mu}\Bigl(\frac{u-\omega}{2}\Bigr) & = -4\,\kappa^2  \, t^{|u|_{2N}^{\vphantom{n}}} + O\bigl(t^{2\mspace{2mu}|u|_{2N}^{\vphantom{n}}}\bigr) \, ,
	\end{aligned}
	\qquad t \coloneqq \E^{-\kappa} \, ,
\end{equation}
where $|u|_{2N} \coloneqq | u \text{ mod } 2N|$ is the distance on a circle of circumference $2N$. 
The complete series in \eqref{eq:wp_expansions_first_order} will be given in \eqref{eq:wp/2_expansion}.
From the representation~\eqref{eq:wp_as_periodicised_sinh} of $\wp$ as a periodised $\sinh$ we find
\begin{equation}
	\lim_{\kappa \to \infty} \biggl( \frac{1}{\kappa^2} \, V\mspace{-1mu}\Bigl(\frac{\omega\,\nncrossing + \vec{\omega}}{2}\Bigr) \biggr)
	= \biggl( \frac{-1}{\sin^2 \pi \nncrossing},0,0,\frac{1}{\cos^2 \pi \nncrossing} \biggr) \, , 
\end{equation}
and 
\begin{equation}
	\lim_{\kappa \to \infty} \Bigl( \frac{1}{\kappa^3} \, \wp'( \omega\mspace{2mu} \nncrossing) \Bigr) = - 2\mspace{2mu}\I \, \frac{\cot \pi \nncrossing}{\sin \pi \nncrossing}\, . 
\end{equation}
Similarly we derive 
\begin{equation}
	\lim_{\kappa \to \infty} \Bigl( \frac{1}{\kappa} \, \vec{A}(\omega\mspace{2mu}\nncrossing )\, \Bigr) = \biggl(\frac{\cot(\pi \nncrossing/2)}{2\mspace{2mu}\I},0,0, \frac{-\!\tan(\pi \nncrossing/2)}{2\mspace{2mu}\I} \biggr) \, .
\end{equation}
Putting these pieces together we find that all contributions independent of $u$ cancel, leaving us with a simple lowest order
\begin{equation} \label{eq:sr_limit}
	\frac{1}{4} \mspace{1mu} V\mspace{-1mu} \Bigl(\frac{u}{2}\mspace{2mu};\frac{\omega\mspace{2mu}\nncrossing}{2}\Bigr) = \I \mspace{2mu} \kappa\sin(\pi \nncrossing) \, t^{|u|_{2N}^{\vphantom{n}}} + O\bigl(t^{2 \mspace{2mu} |u|_{2N}^{\vphantom{n}}}\bigr) \, .
\end{equation}
We divide by $\kappa \, t$ to get a non-zero limit for $u \equiv \pm 1 \ \text{mod}\ 2N$ sufficiently far from the origin, $|u|_{2N} \geqslant 1$.
Using $\lim_{\kappa \to \infty} \theta(\omega \nncrossing)/\omega = \sin(\pi \nncrossing)/\pi$ we obtain the nearest-neighbour potential 
\begin{equation}
	\lim_{\kappa \to \infty} \frac{1}{4} \mspace{1mu} \longbar{V}\mspace{-1mu} \Bigl(\frac{u}{2};\frac{\omega\mspace{2mu}\nncrossing}{2}\Bigr) =  \frac{1}{2} \, \delta_{|u|_{2N}\mspace{-1mu},\mspace{1mu}1} \, .
\end{equation}
The limit for $\alpha=z$ is analogous, except that the roles of $\omega_0$ and $\omega_z$ are interchanged, yielding an overall minus sign. The limits for $\alpha=x,y$ follow by replacing $u \to u+N$ in these results. Hence
\begin{equation} \label{eq:spin_interaction_xx}
	\lim_{\kappa \to \infty} \longbar{\VE}(u;\omega\mspace{2mu}\nncrossing) 
	= \delta_{|u|_{2N}\mspace{-1mu},\mspace{1mu}1} \, \EE^{\textsc{xx}} + \delta_{|u-N|_{2N}\mspace{-1mu},\mspace{1mu}1} \, \frac{1}{2} \bigl(\EE^x - \EE^y\bigr)
\end{equation}
with $\EE^{\textsc{xx}}$ from \eqref{eq:E^xx}.

Finally using $\EE^x - \EE^y = (\id \otimes \sigma^x) \, \EE^{\textsc{xx}} \, (\id \otimes \sigma^x)$ we arrive at the nearest-neighbour hamiltonian
\begin{equation} \label{eq:HXXZ}
	\begin{aligned}
	\longbar{H}_{\text{nn}\,\textsc{mz}'} \coloneqq {} & \lim_{\kappa \to \infty} \longbar{H}^\textsc{l}_{\textsc{mz}'} \big|_{\eta = \omega \nncrossing} \\
	= {} & \sum_{i=1}^{N-1} \! \EE_{i,i+1}^{\textsc{xx}} \; + G_{\text{nn}\,\textsc{mz}'}^{-1} \, \EE^{\mspace{1mu}\textsc{xx}}_{12} \, G_{\text{nn}\,\textsc{mz}'} \\
	= {} & \sum_{i=1}^{N-1} \! \EE_{i,i+1}^{\textsc{xx}} \; + G_{\text{nn}\,\textsc{mz}'} \, \EE^{\mspace{1mu}\textsc{xx}}_{N-1,N} \, G_{\text{nn}\,\textsc{mz}'}^{-1} \\
	= {} & \lim_{\kappa \to \infty} \longbar{H}^\textsc{r}_{\text{nn}\,\textsc{mz}'} \big|_{\eta = \omega \nncrossing} \, . 
	\end{aligned}
\end{equation}
Here the (nonlocal) boundary terms are conveniently described in terms of the limit of the deformed translation~\eqref{eq:MZ_trans_nn}.
The result is no longer chiral. It is a special quasiperiodic version of the \textsc{xx} model. Rather than playing its usual role as parametrising the anisotropy~$\Delta$, the deformation parameter~$\nncrossing$ here appears in the boundary term, which resembles a braid translation \cite{martin1993algebraic}. 
\medskip

\textit{Macroscopic limit.}
In the macroscopic (thermodynamic) limit $N\to\infty$ the elliptic functions degenerate into hyperbolic ones, $\theta(u) \to \sinh(\kappa u)/\kappa$, etc. 
This directly gives the resulting building blocks of the MZ$'$ chain, such as the potential
\begin{equation} \label{eq:V_hyp}
	\longbar{V}_{\!\text{hyp}}(u;\eta) \coloneqq \lim_{N\to\infty} \!\! \longbar{V}\mspace{-1mu}(u;\eta) = n_{\kappa} \, \frac{\kappa^2}{\sinh\bigl(\kappa( x+\eta)\bigr) \sinh\bigl(\kappa (x-\eta)\bigr)} \, ,  
\end{equation}
and the entries of the \textit{R}-matrix \eqref{eq:R8v},
\begin{equation}
	\begin{aligned}
		a^\text{hyp}(u;\eta) & = 1 \, , && c^{\text{hyp}}(u;\eta) = \frac{\sinh(\kappa \, \eta)}{\sinh(\kappa (x +\eta))} \, ,\! \\
		b^{\text{hyp}}(u;\eta) & =  \frac{\sinh(\kappa\, x)}{\sinh(\kappa (x +\eta))}  \, , \quad && d^{\mspace{1mu}\text{hyp}}(u;\eta) = 0 \, . 
	\end{aligned}
\end{equation}
These expressions are also obtained by replacing (`Wick rotating') $N \rightarrow \omega = \I\pi/\kappa$ in the long-range limits, setting $\gamma =\eta/N$ to undo the rescaling from \eqref{eq:R8v_trig} and \eqref{eq:V_MZ_trig}.
This gives the hyperbolic counterpart of \eqref{eq:trig_spin_int},
\begin{equation}
	\begin{aligned}
	\lim_{N\to \infty} \!\! \longbar{\VE}(u;\eta) & = \frac{1}{4} \mspace{2mu} \longbar{V}_{\!\text{hyp}}\Bigl(\frac{u}{2};\frac{\eta}{2}\Bigr) \, F^0 - \frac{1}{4} \mspace{2mu} \longbar{V}_{\!\text{hyp}}\Bigl(\frac{u-\omega}{2};\frac{\eta}{2}\Bigr) \, F^z \\
	& = \longbar{V}_{\!\text{hyp}}(u) \, E^\text{hyp}(u;\eta) \, ,
	\end{aligned} 
\end{equation}
except that, unlike in the long-range limit, no spin rotation is needed to get
\begin{equation} \label{eq:hyp_spin_nn}
	E^\text{hyp}(u;\eta) \coloneqq \cosh(\kappa\,u) \, \EE^{\mspace{1mu}\textsc{xx}} + \cosh(\kappa\,\eta) \, \EE^\diag \, .
\end{equation}
Thus our chiral hamiltonians \eqref{eq:MZ_ham_decomp} have a clear (formal) limit, namely the hyperbolic counterpart of \eqref{eq:MZ_trig} with sum over all pairs of sites $i<j$ in $\mathbb{Z}$. To make sense of these hamiltonians as operators on a suitable infinite-dimensional Hilbert space one must show that these interactions fall off sufficiently quickly. The intricate non-local and multi-spin nature of the long-range interactions make this a nontrivial task, since any $\vec{S}_{[i,j]}^{\,\textsc{l},\textsc{r}}$ contributes to all $\sigma^{\alpha_1}_{i_1} \cdots \sigma^{\alpha_l}_{i_l}$ with $\{i_1,\dots,i_l\} \subseteq \{i,i+1,\dots,j-1,j\}$. While numerics suggests that the matrix elements converge, a proper analysis is beyond our scope.
\medskip

\textit{Limit to undeformed level.} 
Next we compute the  limit $\eta \to 0$. From \eqref{eq:def_theta2} it is easy to see that the \textit{R}-matrix \eqref{eq:Rch8v} becomes the permutation operator $\check{R}(u;\eta) \to P$   as $\eta \to 0$. The remaining $\eta$-dependence of the MZ$'$ chain resides in the function~\eqref{eq:V^eta}. The denominator behaves as 
\begin{equation}
	\frac{1}{\wp(u) - \wp(\eta)} = \frac{1}{\phi(u,\eta)\,\phi(u,-\eta)} = -\frac{\theta(u)^2 \, \theta(\eta)^2}{\theta(u+\eta)\,\theta(u-\eta)} \sim \eta^2 \, ,
\end{equation}
the coefficients $A_\beta(\eta)$ with $\beta\neq 0$ are regular, whilst $A_0(\eta)$ has a simple pole with residue 1 at $\eta=0$. 
This means that for a normalisation constant that behaves as $n_{\kappa,\eta} \sim 1/\eta$ we have
\begin{equation} \label{eq:V^eta_to_V}
	\lim_{\eta \to 0} \longbar{V}\mspace{-1mu}(u;\eta) = \longbar{V}(u) \coloneqq n_\kappa \, V\mspace{-1mu}(u) \, .
\end{equation} 
At last we see that \eqref{eq:V^eta} indeed deforms \eqref{eq:V}. 
Together this implies that the chiral hamiltonians have the same non-chiral limit
\begin{equation}
\label{eq:undeformed_limit}
	\lim_{\eta \to 0} \longbar{H}_{\textsc{mz}'}^{\textsc{l}} = \longbar{H}_{\textsc{sz}'} = \lim_{\eta \to 0}  \longbar{H}_{\textsc{mz}'}^{\textsc{r}} \, , \quad
	\longbar{H}_{\textsc{sz}'} \coloneqq n_\kappa \, H_{\textsc{sz}'} \, .
\end{equation}
The resulting spin chain is our next topic.

\subsection{Undeformed level} 
\label{sec:SZ_limits}

\noindent
Consider the (elliptic) SZ$'$ chain from \eqref{eq:SZ_intro}, i.e.
\begin{equation} \label{eq:SZ}
	\longbar{H}_{\mspace{-2mu}\textsc{sz}'} = n_\kappa \, H_{\textsc{sz}'} \, , \quad
	H_{\textsc{sz}'} = \sum_{i<j}^N \frac{1}{4} \mspace{1mu} V\mspace{-1mu}\biggl(\frac{i-j + \vec{\omega}}{2}\biggr) \cdot \vec{\EE}  \, ,
\end{equation}
where we recall that $V\mspace{-1mu}(u)$ was defined in \eqref{eq:V}, the shifts $\omega_\alpha$ in \eqref{eq:omega_alpha}, and $\EE^\alpha$ in \eqref{eq:E^alpha}. The prefactor $1/4$, which arises naturally from the prepotential as $-\partial_u^2 \log\theta(u/2) = V\mspace{-1mu}(u/2)/4$, sets the residue of the double pole equal to one.

In analogy with \eqref{eq:MZ_spin_interaction} we abbreviate
\begin{equation} \label{eq:SZ_spin_interaction}
	\longbar{\VE}(u) \coloneqq n_\kappa \, \VE(u) =
	\frac{1}{4} \mspace{1mu} \longbar{V}\mspace{-1mu}\Bigl(\frac{u+\vec{\omega}}{2}\Bigr) \cdot \vec{\EE} \, .
\end{equation}

The shift operator obtained from \eqref{eq:MZ_trans} is
\begin{equation} \label{eq:SZ trans op}
	G_{\textsc{sz}'} \coloneqq \lim_{\eta \to 0} G_{\textsc{mz}'} = \, \sigma^x_N \!\!\ordprod_{N\geqslant i> 1} \!\!\!\!\! P_{i-1,i}\, , \quad 
	G_{\textsc{sz}'}^{\,N} = \prod_{i=1}^N \! \sigma^x_i \, ,
\end{equation}
which is the \emph{anti}periodic version of the usual translation operator~\eqref{eq:homogeneous_translation}, so that any spin is flipped when carried once around the chain. For a way to understand this antiperiodicity see \textsection\ref{sec:wrapping}.
\medskip

\textit{Long-range limit.}
When $\kappa \to 0$ the function
$V\mspace{-1mu}(u)$ becomes $V_\text{tri}(u)$ from \eqref{eq:V_tri_intro}, which is also the $\gamma\to0$ limit of \eqref{eq:V_MZ_trig}.
The limit of the coefficient in \eqref{eq:SZ} for $\alpha=x$ follows using $V_\text{tri}(u+N/2) = (\pi/N)^2 / \cos^2(\pi \mspace{1mu} u/N)$. For the two remaining spin directions the $\kappa$-dependent periods $\omega_y, \omega_z \to -\I\mspace{2mu}\infty$ suppress the limit. Since $n_\kappa \sim 1$ we find
\begin{equation} \label{eq:FK_spin_interaction}
	\begin{aligned}
	\lim_{\kappa\to0} \longbar{\VE}(u) & = \frac{1}{4} \mspace{1mu} V_\text{tri}\Bigl(\frac{u}{2}\Bigr) \EE^0 + \frac{1}{4} \mspace{1mu} V_\text{tri}\Bigl(\frac{u+N}{2}\Bigr) \EE^x \\
	& = U^{\otimes 2} \, V_\text{tri}(u) \, E^\text{tri}(u) \, \bigl( U^{\otimes 2} \bigr)^{-1} \, , 
	\end{aligned}
\end{equation}
where
\begin{equation} \label{eq:trig_spin}
	E^\text{tri}(u) \coloneqq \cos\bigl(\tfrac{\pi}{N}\mspace{1mu}u\bigr) \, \EE^{\mspace{1mu}\textsc{xx}} + \EE^\diag \, ,
\end{equation}
is the limit of \eqref{eq:trig_spin_nn} as $\gamma \to 0$. Therefore we find
\begin{equation} \label{eq:SZ_trig_lim}
	\lim_{\kappa \to 0}
	\longbar{H}_{\mspace{-2mu}\textsc{sz}'} = H_{\textsc{fk}'} 
	 = U^{\otimes N} \, H_{\textsc{fk}} \, \bigl( U^{\otimes N} \bigr)^{-1} \, ,
\end{equation}
where $H_\textsc{fk}$ is the (antiperiodic) Fukui--Kawakami chain \eqref{eq:FK}.
Its translation operator is as in \eqref{eq:SZ trans op} but with $\sigma^z$ instead of $\sigma^x$.
Note that we also obtain \eqref{eq:SZ_trig_lim} from the trigonometric MZ$'$ chain when $\gamma \to 0$, since \eqref{eq:V^eta} deforms \eqref{eq:V_tri_intro}, \eqref{eq:trig_spin_int} becomes \eqref{eq:FK_spin_interaction}, and $\check{R}^{\text{6v}}(u) \to P$. We will get back to $H_{\textsc{fk}}$ in \textsection\ref{sec:wrapping}.
\medskip

\textit{Short-range limit.}
To take the limit $\kappa \to \infty$ of the SZ$'$ chain we recall \eqref{eq:wp_expansions_first_order} and work out the four cases. 
One readily sees that \eqref{eq:SZ_spin_interaction} has the same limit as \eqref{eq:spin_interaction_xx}, meaning that the only contributions are those coming from the cases $i-j=-1$ and $i-j=-N+1$. In complete analogy with the discussion following \eqref{eq:spin_interaction_xx} we find the antiperiodic \textsc{xx} chain~\eqref{eq:antiperiodic_xx}:
\begin{equation} \label{eq:ham_XX}
	\lim_{\kappa \to \infty} \longbar{H}_{\mspace{-2mu}\textsc{sz}'} = 
	H_{\textsc{xx}'} \coloneqq \sum_{i=1}^{N-1} \! \EE^{\mspace{1mu}\textsc{xx}}_{i,i+1} \; + \sigma^x_1 \, \EE^{\mspace{1mu}\textsc{xx}}_{N,1} \, \sigma^x_1 \, ,
\end{equation}
This coincides with the short-range limit \eqref{eq:HXXZ} of the MZ$'$ chain, since sending $\nncrossing\to0$ just simplifies the translation operators in the boundary terms to \eqref{eq:SZ trans op}.
This model is also the free-fermion point of the antiperiodic Heisenberg \textsc{xxz} chain studied in e.g.\ \cite{batchelor_exact_1995,niccoli_antiperiodic_2015}. Note that the antiperiodic boundary conditions spoil weight-preservation.
\medskip

\textit{Macroscopic limit.}
The simple spin-spin interactions of the SZ$'$ chain make it possible to evaluate the limit $N\to \infty$ at the undeformed level. The potentials have direct hyperbolic limits, 
\begin{equation}
	V_\text{hyp}(u) \coloneqq \lim_{N\to\infty} \! V\mspace{-1mu}(u) = \frac{\kappa^2}{\sinh^2(\kappa\mspace{1mu}u)} \, .
\end{equation}
This time the coefficients for $\alpha=x,y$ with diverging real shifts vanish. In complete analogy with \eqref{eq:FK_spin_interaction} we find
\begin{equation}
	\begin{aligned}
	\lim_{N\to\infty} \VE(u) & = \frac{1}{4} \mspace{1mu} V_\text{hyp}\biggl(\!\frac{u}{2}\biggr) \EE^0 - \frac{1}{4} \mspace{1mu} V_\text{hyp}\biggl(\!\frac{u -\omega}{2}\biggr) \EE^z \\
	& = V_\text{hyp}(u) \,
	\EE^\text{hyp}(u) \, ,
	\end{aligned}
\end{equation}
where this time no spin rotation is needed, and
\begin{equation}
	E^\text{hyp}(u) \coloneqq
	\cosh(\kappa \mspace{2mu} u) \, \EE^\textsc{xx} + \EE^\diag \, .
\end{equation}
This is consistent with the limit $\eta\to0$ of the (formal) macroscopic limit of the MZ$'$ chain. Thus $H_{\textsc{sz}'}$ formally limits to 
\begin{equation} \label{eq:SZ_hyp}
	H_{\text{hyp}\,\textsc{sz}'} =  \sum_{i<j}^\infty V_\text{hyp}(i-j) \, E^\text{hyp}(i-j) \, . 
\end{equation}
One can make sense of such hamiltonians as operators on an infinite-dimensional Hilbert space that is a suitable closure of the span of vectors differing from some reference state at finitely many sites. Note that the antiperiodic boundary conditions are `pushed to infinity' so that \eqref{eq:SZ_hyp} is $S^z$-symmetric. This makes it, at least in principle, amenable to (e.g.\ coordinate) Bethe-ansatz techniques, cf.\ e.g.\ \cite{klabbers2015inozemtsev}.
\medskip

\textit{Rational limit.} Finally we consider the limit in which we remove both periods. Taking either the formal limit $N\to \infty$ in \eqref{eq:FK_spin_interaction} or $\kappa \to 0$ in \eqref{eq:SZ_hyp} yields
\begin{equation} \label{eq:ratHS}
	H_{\text{rat}\,\textsc{hs}} = -\frac{1}{2} \sum_{i<j}^\infty \frac{1}{(i-j)^2} \, \bigl( \sigma^x_i  \sigma^x_j + \sigma^y_i  \sigma^y_j + \sigma^z_i  \sigma^z_j - 1 \bigr) \, .
\end{equation}
As in \eqref{eq:SZ_hyp} the antiperiodic boundary conditions have disappeared. The result is identical to the macroscopic limit $N \to \infty$ of the Haldane--Shastry chain!
In \textsection\ref{sec:wrapping} we will show how, conversely, the periods $N,\omega$ can be (re)introduced in \eqref{eq:ratHS}, so as to retrieve the Sechin--Zotov chain.

\subsection{Summary: vertex-type landscape} \label{sec:vx-landscape}

\begin{figure}[h]
	\!\!\begin{tikzpicture}[x={(-0.866cm,-0.5cm)}, y={(1cm,0cm)}, z={(0cm,1cm)}, scale=.6, font=\footnotesize]
		\fill [gray!10] (0,0,0) -- (4,0,0) -- (4,6,0) -- (0,6,0) -- cycle;
		\draw [dotted] (0,0,4-.35) -- (0,0,0) -- (0,6,0) -- (0,6,4-.35);
		\fill [gray!05,fill opacity=.9] (0,0,2.5) -- (4,0,2.5) -- (4,6,2.5) -- (0,6,2.5) -- cycle;
		\draw (.2,0,0) -- (4,0,0) -- (4,6,0) -- (.2,6,0);
		\draw (4,0,0) -- (4,0,5) (4,6,0) -- (4,6,5);
		\node at (0,6,2) [shift={(.2cm,-.2cm)}] {\rotatebox{90}{\textcolor{gray!60}{unphysical}}};
		\node at (2,3-.5,0) {\textcolor{gray!60}{(ell)} SZ$'$};
		\node at (2-.6,0,0) [shift={(-.4cm,-.05cm)}] {\rotatebox{30}{antiper.\ \textsc{xx}}};
		\node at (2,6,0) [shift={(.3cm,-.05cm)}] {\rotatebox{30}{\textcolor{gray!60}{(tri)} FK, rot'd}};
		\node at (4,3,0) [yshift=-.2cm]{hyp\,SZ};
		\node at (2-.05,0,2.5) [shift={(-.15cm,.125cm)}] {\rotatebox{30}{deformed-}};
		\node at (2-.05,0,2.5) [yshift=-.125cm] {\rotatebox{30}{antiper.\ \textsc{xx}}};
		\node at (2,3-.5,2.5) {\textcolor{gray!60}{(ell)} MZ$'$};
		\node [yshift=.03cm] at (2,6,2.5) {\rotatebox{30}{tri\,MZ, rot'd}};
		\node [xshift=.18cm] at (4,3,2.5) {\textcolor{gray!60}{hyp\,MZ}};
		\fill[black] (4,6,0) circle (.06cm); 
		\node [shift={(-.1cm,-.2cm)}] at (4,6,0) {rat\,HS};
		\draw [->] (0,7+.6,0) -- (4.5,7+.6,0);
		\node at (4.9,7.6+.6+.1,0) {$N$};
		\draw (0,7+.6,0) -- (0,7+.2+.6,0);
		\node [xshift = -.1cm] at (0,7.6+.6,0) {$0$};
		\draw (4,7+.6,0) -- (4,7+.2+.6,0);
		\node at (4,7.6+.6,0) {$\infty$};
		\draw [->] (5+.55,0,0) -- (5+.55,6.4,0);
		\node at (5.7+.55,6.9+.1,0) {$\omega$};
		\draw (5+.55,0,0) -- (5.2+.55,0,0);
		\node [shift={(.05cm,.05cm)}] at (5.55+.55,0,0) {$0$};
		\draw (5+.55,6,0) -- (5.2+.55,6,0);
		\node [xshift = .1cm] at (5.55+.55,6,0) {$\I\mspace{1mu}\infty$};
		\draw [->] (4,-1,0) -- (4,-1,5.3);
		\node [shift={(-.15cm,.1cm)}] at (4,-1,5.3) {$\eta$};
		\draw (4,-1,0) -- (4,-1-.2,0);
		\node [xshift = -.3cm] at (4,-1,0) {$0$};
		\node at (4,0,2.5) [xshift=-.2cm] {\rotatebox{90}{\textsc{xx} model}};
		\draw [rounded corners=3pt] (0+.25+.1,8.8-.2,0) -- (0+.25+.1,8.8,0) -- (2,8.8,0) -- (2,8.8+.2,0) (2,8.8+.2,0) -- (2,8.8,0) -- (4-.25,8.8,0) -- (4-.25,8.8-.2,0);
		\node at (2,8.8+.2,0) [right, shift={(.07cm,.05cm)}] {system};
		\node at (2,8.8+.2,0) [right, shift={(-.38cm,-.2cm)}] {on a circle};
		\draw [->, >=stealth] (4,8.8+.1,0) -- (4,8.8-.3,0);
		\node at (4,8.8+.1,0) [right, shift={(-.05cm,-.05cm)}] {system on a line};
		\draw [->, >=stealth] (6.6+.2,0+.05,0) -- (6.6-.25,0+.05,0);
		\node at (6.6+1,0+.05,0) [shift={(.05cm,.1cm)}] {nearest\vphantom{l}};
		\node at (6.6+1,0+.05,0) [shift={(.23cm,-.15cm)}] {neighbour};
		\draw [rounded corners=3pt] (6.6-.2,0+.3,0) -- (6.6,0+.25,0) -- (6.6,3,0) -- (6.6+.1,3,0) (6.6+.1,3,0) -- (6.6,3,0) -- (6.6,6-.25,0) -- (6.6-.2,6-.25,0);
		\node at (6.6+1,3,0) [shift={(.5cm,.1cm)}] {intermediate};
		\node at (6.6+1,3,0) [shift={(.5cm,-.15cm)}] {\vphantom{l}interaction range};
		\draw [->, >=stealth] (6.6+.2,6,0) -- (6.6-.25,6,0);
		\node at (6.6+1,6,0) [shift={(.5cm,.1cm)}] {long};
		\node at (6.6+1,6,0) [shift={(.42cm,-.15cm)}] {\vphantom{l}range};
	\end{tikzpicture}
	\caption{Landscape of the elliptic Matushko--Zotov chain, with the Sechin--Zotov and (antiperiodic) Fukui--Kawakami chains at the undeformed level. In the long-range limits we use a spin rotation. The short-range limit is the \textsc{xx} chain with antiperiodic boundary conditions, deformed for $\eta\neq 0$.}
	\label{fg:landscape_vertex}
\end{figure}
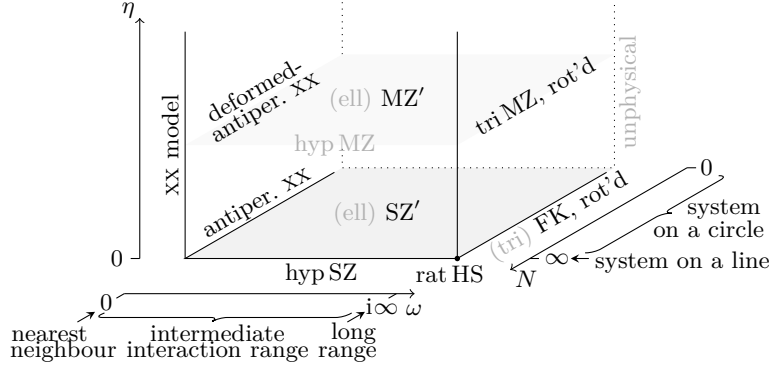

\noindent
We thus conclude our tour through the rich landscape of the MZ$'$ chain, which is built from the \emph{eight-vertex} \textit{R}-matrix. The highlights included corners of the parameter space, powered by our choice of theta function~\eqref{eq:def_theta2}. See Fig.~\ref{fg:landscape_vertex} for an overview.

The limits of the imaginary period $\omega = \I\pi/\kappa$, with $\kappa >0$, are of special interest. The long-range limit $\kappa \to 0^+$ gives the (symmetric: `principal') six-vertex \textit{R}-matrix, yielding the trigonometric MZ chain~\eqref{eq:MZ_trig} up to a conjugation, see \eqref{eq:MZ_trig_lim}. 
Our setup unlocks the short-range limit $\kappa\to\infty$. The result~\eqref{eq:HXXZ} is a version of the \textsc{xx} model, i.e.\ the \emph{free-fermion} point $\Delta=0$ of the Heisenberg \textsc{xxz} chain, with very specific quasiperiodic boundary conditions that  deform \emph{antiperiodic} boundaries and resemble braid translations.

Regarding the real period, the macroscopic (thermodynamic) limit $N\to\infty$ formally yields a hyperbolic counterpart of the trigonometric MZ chain. We did not perform the functional analysis required to make sense of the result as a well defined operator on an infinite-dimensional Hilbert space. We ignored the nonphysical limit $N\to0$.

Concerning the deformation (crossing) parameter we focussed on the undeformed limit $\eta\to0$, which yields the SZ$'$ chain~\eqref{eq:SZ}. Its long-range limit~\eqref{eq:SZ_trig_lim} gives, up to a conjugation, the (antiperiodic) Fukui--Kawakami chain~\eqref{eq:FK}. The short-range limit is the antiperiodic \textsc{xx} model~\eqref{eq:antiperiodic_xx}. Its (formal) macroscopic limit is the hyperbolic version of the SZ chain~\eqref{eq:SZ_hyp}. Combined, the long-range and macroscopic limit yield the rational Haldane--Shastry chain~\eqref{eq:ratHS}. We will return to this observation in \textsection\ref{sec:wrapping}.

In order to write the trigonometric MZ$'$ and SZ$'$ chains in the standard convention where the spin-$z$ axis is singled out we applied a global spin rotation in the long-range limit. In which limit(s) such a spin rotation appears depends on conventions; in ours it is absent in the macroscopic and short-range limits. 
\medskip

\textit{Global symmetries.} Let us turn to the spatial and spin symmetry of the MZ$'$ chain and its various limits. See Table~\ref{tab:MZ!=qIno} (p.\,\pageref{tab:MZ!=qIno}) for an overview.
The MZ chain is not parity invariant, having two chiral hamiltonians whose spin interactions are each other's mirror image. Parity symmetry is restored in the short-range limit as well as at the undeformed level.
None of the chains with finite $N$ are invariant under ordinary translations. Instead, they commute with a (deformed or ordinary) \emph{antiperiodic} version of the translation operator, see \eqref{eq:MZ_trans} and \eqref{eq:SZ trans op}, such that a full spatial rotation over $N$ sites is the same as flipping all spins.

A particularly notable property of the chains in the vertex landscape is their amount of spin symmetry, which is lower than one would expect for \textit{q}-deformed and undeformed spin chains based on experience with nearest-neighbour integrability. This is why we  simply call $\eta\to0$ the `undeformed limit'.

The elliptic MZ$'$ and SZ$'$ chains are both \emph{anisotropic}: they only have some discrete spin symmetry, viz.\ invariance under simultaneously flipping all spins $\uparrow \,\longleftrightarrow\, \downarrow$. There are multiple reasons for the anisotropy. First, the distinct coefficients $V\mspace{-1mu}\bigl((u+\omega_\alpha)/2,\dots\bigr)$ for each of the spin operators $F^\alpha = 1 - P \, \sigma^\alpha \otimes \sigma^\alpha$, see \eqref{eq:RR_decomp} and \eqref{eq:SZ}. Second, at the deformed level, the eight-vertex \textit{R}-matrices arranging the transport. Third, the (possibly deformed) antiperiodic boundary conditions, visible in the corresponding translation operator, which in particular spoil the $S^z$-symmetry of the ordinary \textsc{xx} model in the short-range limit.

In the long-range and macroscopic limits one period is sent to infinity. Since $\vec{\omega} = (0,N,N-\omega,\omega)$ and the potentials are inverse squares this suppresses two of the coefficients. The limiting trigonometric and hyperbolic chains thus become partially isotropic, viz.\ invariant under $\mathit{U}\mspace{-1mu}(1) \subset \mathit{SU}\mspace{-1mu}(2)$ generated by $S^x$ in the long-range limit and $S^z$ in the macroscopic limit. The distinct \emph{partial} (an)isotropy in the two limits underlines the \emph{full} anisotropy in the general (elliptic) case. At the undeformed level, the FK chain's partial isotropy is enhanced to the Yangian \cite{fukui_exact_1996}.

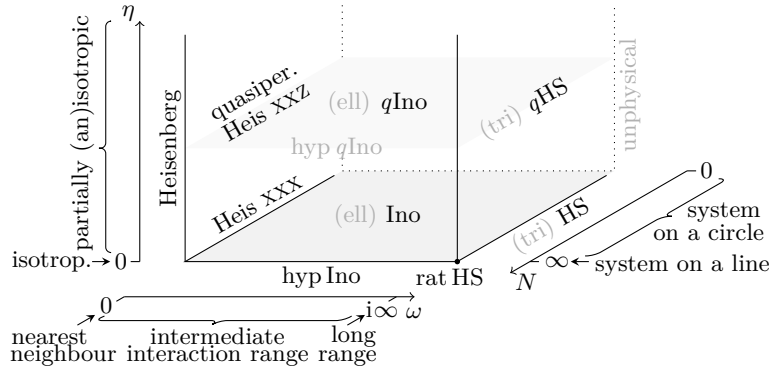
\begin{figure}[h]
	\!\!\begin{tikzpicture}[x={(-0.866cm,-0.5cm)}, y={(1cm,0cm)}, z={(0cm,1cm)}, scale=.6, font=\footnotesize]
		\fill [gray!10] (0,0,0) -- (4,0,0) -- (4,6,0) -- (0,6,0) -- cycle;
		\draw [dotted] (0,0,4-.35) -- (0,0,0) -- (0,6,0) -- (0,6,4-.35);
		\fill [gray!05,fill opacity=.9] (0,0,2.5) -- (4,0,2.5) -- (4,6,2.5) -- (0,6,2.5) -- cycle;
		\draw (.2,0,0) -- (4,0,0) -- (4,6,0) -- (.2,6,0);
		\draw (4,0,0) -- (4,0,5) (4,6,0) -- (4,6,5);
		\node at (0,6,2) [shift={(.2cm,-.2cm)}] {\rotatebox{90}{\textcolor{gray!60}{unphysical}}};
		\node at (2,3-.5,0) {\textcolor{gray!60}{(ell)} Ino};
		\node at (2-.6,0,0) [xshift=-.4cm] {\rotatebox{30}{Heis \textsc{xxx}}};
		\node at (2+.1,6,0) [shift={(.25cm,-.1cm)}] {\rotatebox{30}{\textcolor{gray!60}{(tri)} HS}};
		\node at (4,3,0) [yshift=-.2cm]{hyp\,Ino};
		\node at (2-.05,0,2.5) [shift={(-.15cm,.125cm)}] {\rotatebox{30}{quasiper.}};
		\node at (2-.05,0,2.5) [yshift=-.125cm] {\rotatebox{30}{Heis \textsc{xxz}}};
		\node at (2,3-.5,2.5) {\textcolor{gray!60}{(ell)} \textit{q}Ino};
		\node at (2+.3,6,2.5) {\rotatebox{30}{\textcolor{gray!60}{(tri)} \textit{q}HS}};
		\node [xshift=.2cm] at (4,3,2.5) {\textcolor{gray!60}{hyp\,\textit{q}Ino}};
		\fill[black] (4,6,0) circle (.06cm); 
		\node [shift={(-.1cm,-.2cm)}] at (4,6,0) {rat\,HS};
		\draw [->] (0,7+.6,0) -- (4.5,7+.6,0);
		\node at (4.9,7.6+.6+.1,0) {$N$};
		\draw (0,7+.6,0) -- (0,7+.2+.6,0);
		\node [xshift = -.1cm] at (0,7.6+.6,0) {$0$};
		\draw (4,7+.6,0) -- (4,7+.2+.6,0);
		\node at (4,7.6+.6,0) {$\infty$};
		\draw [->] (5+.55,0,0) -- (5+.55,6.4,0);
		\node at (5.7+.55,6.9+.1,0) {$\omega$};
		\draw (5+.55,0,0) -- (5.2+.55,0,0);
		\node [shift={(.05cm,.05cm)}] at (5.55+.55,0,0) {$0$};
		\draw (5+.55,6,0) -- (5.2+.55,6,0);
		\node [xshift = .1cm] at (5.55+.55,6,0) {$\I\mspace{1mu}\infty$};
		\draw [->] (4,-1,0) -- (4,-1,5.3);
		\node [shift={(-.15cm,.1cm)}] at (4,-1,5.3) {$\eta$};
		\draw (4,-1,0) -- (4,-1-.2,0);
		\node [xshift = -.25cm] at (4,-1,0) {$0$};
		\node at (4,0,2.5) [xshift=-.2cm] {\rotatebox{90}{Heisenberg}};
		\draw [rounded corners=3pt] (0+.25+.1,8.8-.2,0) -- (0+.25+.1,8.8,0) -- (2,8.8,0) -- (2,8.8+.2,0) (2,8.8+.2,0) -- (2,8.8,0) -- (4-.25,8.8,0) -- (4-.25,8.8-.2,0);
		\node at (2,8.8+.2,0) [right, shift={(.07cm,.05cm)}] {system};
		\node at (2,8.8+.2,0) [right, shift={(-.38cm,-.2cm)}] {on a circle};
		\draw [->, >=stealth] (4,8.8+.1,0) -- (4,8.8-.3,0);
		\node at (4,8.8+.1,0) [right, shift={(-.05cm,-.05cm)}] {system on a line};
		\draw [->, >=stealth] (6.6+.2,0+.05,0) -- (6.6-.25,0+.05,0);
		\node at (6.6+1,0+.05,0) [shift={(.05cm,.1cm)}] {nearest\vphantom{l}};
		\node at (6.6+1,0+.05,0) [shift={(.23cm,-.15cm)}] {neighbour};
		\draw [rounded corners=3pt] (6.6-.2,0+.3,0) -- (6.6,0+.25,0) -- (6.6,3,0) -- (6.6+.1,3,0) (6.6+.1,3,0) -- (6.6,3,0) -- (6.6,6-.25,0) -- (6.6-.2,6-.25,0);
		\node at (6.6+1,3,0) [shift={(.5cm,.1cm)}] {intermediate};
		\node at (6.6+1,3,0) [shift={(.5cm,-.15cm)}] {\vphantom{l}interaction range};
		\draw [->, >=stealth] (6.6+.2,6,0) -- (6.6-.25,6,0);
		\node at (6.6+1,6,0) [shift={(.5cm,.1cm)}] {long};
		\node at (6.6+1,6,0) [shift={(.42cm,-.15cm)}] {\vphantom{l}range};
		\draw [->, >=stealth] (4,-1.75-.3,0) -- (4,-1.75+.1,0);
		\node at (4,-1.75-.35,0) [xshift=-.5cm] {isotrop.};
		\draw [rounded corners=2pt] (4,-1.75+.1,0+.2) -- (4,-1.75,0+.2) -- (4,-1.75,2.5) -- (4,-1.75-.15,2.5) (4,-1.75-.15,2.5) -- (4,-1.75,2.5) -- (4,-1.75,5.3) -- (4,-1.75+.1,5.3);
		\node at (4,-1.75-.1,2.5) [shift={(-.25cm,.2cm)}] {\rotatebox{90}{partially (an)isotropic}};
	\end{tikzpicture}
	\caption{Landscape of the \textit{q}-deformed Inozemtsev chain. The undeformed level contains the Heisenberg, Inozemtsev and Haldane--Shastry chains.} 
	\label{fg:landscape_face}
\end{figure}

\section{Comparison with \textit{q}-deformed Inozemtsev chain}
\label{sec:comparison}

\noindent
Now that we understand the landscape provided by the MZ$'$ chain, built from the eight-vertex \textit{R}-matrix (with `principal' six-vertex limit), we turn to the comparison with another long-range spin chain: the \textit{q}-deformed Inozemtsev chain \cite{KL_23}.

\subsection{Summary of face-type landscape} \label{sec:qIno}

\noindent
We start with a brief overview of the \textit{q}-deformed Inozemtsev chain \cite{KL_23}. Its structure is very similar to that of the MZ$'$ chain, with a family of commuting operators including two chiral hamiltonians and a deformed translation operator.

One description of the chiral hamiltonians is as in \eqref{eq:MZ_ham_pre}, except that the eight-vertex \textit{R}-matrix~\eqref{eq:R8v} is replaced by Felder's (face-type) \emph{dynamical} \textit{R}-matrix
\begin{equation} \label{eq:Rdyn}
	\check{R}(u,a;\eta) \coloneqq 
	\begin{pmatrix}
		\, 1 & \color{gray!80}{0} & \color{gray!80}{0} & 0 \, \\
		\, \color{gray!80}{0} & \displaystyle \!\frac{\theta(\eta) \, \theta(\eta\,a-u)}{\theta(u+\eta)\,\theta(\eta\,a)}\! & \displaystyle \!\frac{\theta(u)\,\theta(\eta\mspace{1mu}(a+1))}{\theta(u+\eta)\,\theta(\eta\,a)}\! & \color{gray!80}{0} \, \\[1em]
		\, \color{gray!80}{0} & \displaystyle  \!\frac{\theta(u)\,\theta(\eta\mspace{1mu}(a-1))}{\theta(u+\eta)\,\theta(\eta\,a)}\! & \displaystyle  \!\frac{\theta(\eta)\,\theta(\eta\,a+u)}{\theta(u+\eta)\,\theta(\eta\,a)}\! & \color{gray!80}{0} \, \\[.8em]
		\, 0 & \color{gray!80}{0} & \color{gray!80}{0} & 1 \,
	\end{pmatrix} \, .
\end{equation}
It preserves $S^z$, unlike the eight-vertex \textit{R}-matrix~\eqref{eq:R8v}. The modest price one pays is that the `dynamical parameter'~$a$ of any operator is shifted by the weight (spin-$z$) of all spins to its left:
$\check{R}_{i,i+1}\bigl(u,a-(\sigma^z_1 + \cdots + \sigma^z_{i-1});\eta\bigr)$ sends $\ket{s_1,\dots,s_N} \in (\mathbb{C}^2)^{\otimes N}$ to $\check{R}_{i,i+1}\bigl(u,a-(s_1+\dots+s_{i-1});\eta\bigr)\,\ket{s_1,\dots,s_N}$ if $\sigma^z_j \, \ket{s_1,\dots,s_N} = s_j \, \ket{s_1,\dots,s_N}$.
For example, the Yang--Baxter equation~\eqref{eq:YBE} is replaced by the dynamical Yang--Baxter equation
\begin{align} \label{eq:DYBE}
	\check{R}_{12}(u,a;\eta) \, & \check{R}_{23}(u+v,a-\sigma^z_1;\eta)\, \check{R}_{12}(v;\eta) \nonumber \\
	= {} & \check{R}_{23}(v,a+\sigma^z_1;\eta) \, \check{R}_{12}(v+u,a;\eta) \, \check{R}_{23}(u,a-\sigma^z_1;\eta) \, .
\end{align}
Having thus accounted for the shifts, everything works as usual.

The chiral hamiltonians can be rewritten using 
\begin{equation} \label{eq:RRdyn_decomp}
	\check{R}(-u,a;\eta)\,\check{R}'(u,a;\eta) = \theta(\eta) \,  V\mspace{-1mu}(u,a;\eta) \, E(u,a;\eta) \, .
\end{equation}
This decomposition, of which \eqref{eq:RR_decomp} is a direct analogue, yields hamiltonians of the form \eqref{eq:MZ_trig}, as in \cite{KL_23}.
The potential is a finite-difference version of \eqref{eq:V} and actually independent of $a$:
\begin{equation} \label{eq:VqIno}
	V\mspace{-1mu}(u,a;\eta) \coloneqq 
	-\frac{\rho(u+\eta) - \rho(u-\eta)}{\theta(2\eta)} \sim \frac{1}{\mathrm{sn}(u+\eta) \, \mathrm{sn}(u-\eta)} \, . \!\!
\end{equation} 
The operator $E(u,a;\eta)$ is \emph{defined} by \eqref{eq:RRdyn_decomp}, see \textsection{}C in \cite{KL_23} for an explicit expression. Its definition is justified by its limits, which we will give shortly; in particular, it \textit{q}-deforms $1-P$. 

The deformed translation operator of the \textit{q}-deformed Inozemtsev chain is as in \eqref{eq:MZ_trans}, of course with dynamical \textit{R}-matrices, and now a \emph{diagonal} twist $\tikz[baseline={([yshift=-.5*11pt*0.13-3pt]current bounding box.center)},xscale=0.45,yscale=0.195,font=\footnotesize]{
	\node at (0,1-.05) {$\tikz[baseline={([yshift=-.5*11pt*.25]current bounding box.center)},scale=.35]{\fill[black] (0,0) rectangle ++(.4,.4)}$};
	\draw[->] (0,0) -- (0,2);
} \!\! = \exp(-\kappa\,\eta\,a\,\sigma^z)$ that acts at site~$N\mspace{-2mu}$, with dynamical parameter appropriately shifted. To wit,
\begin{equation} \label{eq:qIno_trans}
	G_{\mathit{q}\text{Ino}} \coloneqq {} \exp\bigl(-\kappa\,\eta\,(a-(\sigma^z_1 + \cdots + \sigma^z_{N-1})\,\sigma^z\bigr) \!\! \ordprod_{N\geqslant i> 1} \!\!\!\! \check{R}_{i-1,i}\bigl(1-i,a-(\sigma^z_1 + \cdots + \sigma^z_{i-2});\eta\bigr) \, .
	\end{equation}

The \textit{q}-deformed Inozemtsev chain is integrable in the sense that these chiral hamiltonians and deformed translation operator  belong to a hierarchy of commuting hamiltonians. The proof of commutativity sketched in \cite{KL_23} will be given in detail in a forthcoming publication.
\medskip

\textit{Long-range limit.} The long-range limit consists of sending $\kappa \to 0$ as before. Setting $\eta=N\,\gamma$, the potential \eqref{eq:VqIno} limits to \eqref{eq:V_MZ_trig}. In the dynamical \textit{R}-matrix and $E(x,a;\eta)$ we further remove $a\to -\I\mspace{2mu}\infty$. 
Then \eqref{eq:Rdyn} becomes the asymmetric  (`homogeneous') six-vertex \textit{R}-matrix
\begin{equation} \label{eq:R6v_homog}
	\begin{aligned} 
	\lim_{a\to-\I\infty} \, \lim_{\kappa\to 0} \check{R}(u,a;N\gamma)   ={}& \check{R}^{\mspace{1mu}\text{a6v}}(u;\gamma)  \coloneqq{} \id\otimes\id - b^\text{6v}(u;\gamma) \, e(\gamma) \, ,
	\end{aligned}
\end{equation}
where
\begin{equation}\label{eq:TL}
	\lim_{a\to-\I\infty} \, \lim_{\kappa\to 0} E(x,a;N\gamma) = e(\gamma) \coloneqq 
	\begin{pmatrix}
		\, 0 & \color{gray!80}{0} & \!\color{gray!80}{0} & 0 \, \\
		\, \color{gray!80}{0} & \!\!\hphantom{-}\E^{-\I\mspace{1mu} \pi\mspace{1mu}\gamma} & \!-\E^{\I\mspace{1mu} \pi\mspace{1mu}\gamma}\!\! & \color{gray!80}{0} \, \\[.8ex]
		\, \color{gray!80}{0} & \!\!-\E^{-\I\mspace{1mu} \pi\mspace{1mu}\gamma} & \!\hphantom{-}\E^{\I\mspace{1mu} \pi\mspace{1mu}\gamma}\!\! & \color{gray!80}{0} \, \\
		\, 0 & \color{gray!80}{0} & \!\color{gray!80}{0} & 0 \,
	\end{pmatrix}
	\!\!
\end{equation}
doubles as the limit of $E(x,a;\eta)$. We get the \textit{q}-deformed Haldane--Shastry (HS) chain \cite{Ugl_95u,Lam_18,lamers2022spin} with $q=\E^{\I\mspace{1mu} \pi\mspace{1mu}\gamma}$. Its chiral hamiltonians are as in \eqref{eq:MZ_trig} with the symmetric six-vertex \textit{R}-matrix replaced by \eqref{eq:R6v_homog}, and \eqref{eq:E_diagr} by \eqref{eq:TL}. The twist in \eqref{eq:qIno_trans} disappears when $\kappa\to0$, unlike for \eqref{eq:MZ_trans_tri}.

The \textit{q}-deformed HS chain has a lot of algebraic structure. The matrix~\eqref{eq:TL} is the Temperley--Lieb generator. From it, we can define the Hecke generator $T(\gamma) \coloneqq \E^{\I\mspace{1mu} \pi\mspace{1mu}\gamma} \, \id\otimes\id - e(\gamma)$, in terms of which \eqref{eq:R6v_homog} becomes
\begin{equation}
	\check{R}^{\mspace{1mu}\text{a6v}}(u;\gamma) = b^\text{6v}(u;\gamma) \, T(\gamma) + c^\text{6v}(u;\gamma)\,\id\otimes\id \, .
\end{equation}
This link with Hecke algebras (`Baxterisation'), enabled by the asymmetry of \eqref{eq:R6v_homog}--\eqref{eq:TL}, is a \emph{crucial} feature of the \textit{q}-deformed HS chain. It paves the way for a connection to affine Hecke algebras, which in turn underpins the model's enhanced (quantum-loop) spin symmetry \cite{bernard1993yang,drinfel1986degenerate} and its explicit eigenvectors featuring Macdonald polynomials \cite{lamers2022spin}.
\medskip

\textit{Short-range limit.}
Set $\eta = \I\mspace{2mu}\pi\,\nncrossing\!/\kappa$ and let $\kappa\to\infty$. This requires a factor of e.g.\ $\sinh^2(\kappa)/\kappa^2$ for \eqref{eq:VqIno}, yielding a nearest-neighbour potential.
In the bulk only the terms with $j=i+1$ remain. These are determined by the limit of $E(x,a;\eta)$, which is
\begin{equation} \label{eq:E a gamma}
	E(a;\nncrossing) = 
	\begin{pmatrix}
		\, 0 & \!\!\color{gray!80}{0} & \color{gray!80}{0} & 0 \, \\
		\, \color{gray!80}{0} & \displaystyle \!\!\!\hphantom{-}\frac{\sin\bigl(\pi \mspace{2mu} \nncrossing (a-1)\bigr)}{\sin(\pi \mspace{2mu} \nncrossing a)}\!\! & \displaystyle \!-\frac{\sin\bigl(\pi \mspace{2mu} \nncrossing (a+1)\bigr)}{\sin(\pi \mspace{2mu} \nncrossing a)}\!\! & \color{gray!80}{0} \, \\[2ex]
		\, \color{gray!80}{0} & \displaystyle \!\!\!-\frac{\sin\bigl(\pi \mspace{2mu} \nncrossing (a-1)\bigr)}{\sin(\pi \mspace{2mu} \nncrossing a)}\!\! & \displaystyle \!\hphantom{-}\frac{\sin\bigl(\pi \mspace{2mu} \nncrossing (a+1)\bigr)}{\sin(\pi \mspace{2mu} \nncrossing a)}\!\! & \color{gray!80}{0} \, \\[1.5ex]
		\, 0 & \!\!\color{gray!80}{0} & \color{gray!80}{0} & 0 \,
	\end{pmatrix} . \!\!
\end{equation}
When acting at sites $i,i+1$ its dynamical parameter is shifted to $a-(\sigma^z_1 + \dots + \sigma^z_{i-1})$.
The dynamical \textit{R}-matrix becomes
\begin{equation}
	\check{R}(a;\nncrossing) = \id \otimes \id - \E^{-\I \mspace{1mu}\pi\mspace{1mu} \nncrossing} E(a;\nncrossing) \, .
\end{equation}
This determines the limit $G_{\text{nn}}$ of the deformed translation~\eqref{eq:qIno_trans}, and of the boundary terms ($i=1,j=N$) in the chiral hamiltonians, which coincide in the short-range limit:
\begin{equation} \label{eq:bdry_term}
	G_{{\text{nn}}}^{-1} \, E_{12}(a;\nncrossing) \, G_{\text{nn}} = G_{\text{nn}} \, E_{N-1,N}\bigl(a-(\sigma^z_1 + \cdots + \sigma^z_{N-1});\nncrossing\bigr) \, G_{\text{nn}}^{-1} .
\end{equation}
The result is a dynamical variant of the Heisenberg \textsc{xxz} chain that looks similar to \eqref{eq:HXXZ} \cite{KL_23}.

This limit has a rich algebraic structure as well. If $a\to-\I\mspace{2mu}\infty$ then \eqref{eq:E a gamma} would look like the Temperley--Lieb generator \eqref{eq:TL}. However, here we cannot remove the dynamical parameter because the twist still depends on $a$ and would diverge. Nonetheless, \eqref{eq:E a gamma} generates a representation of the Temperley--Lieb algebra for any value of $a$. 
Moreover, $G_{\text{nn}}$ extends this to a representation of the \emph{affine} Temperley--Lieb algebra \cite{KL_23}. 
\medskip

\textit{Undeformed limit.} If we let $\eta\to 0$ the potential~\eqref{eq:VqIno} becomes \eqref{eq:V}. Once more sending $a\to-\I\mspace{2mu}\infty$ the dynamical \textit{R}-matrix becomes $\check{R}(x,a;\eta) \to P$ and $E(x,a;\eta) \to 1-P$. Both chiral hamiltonians thus reduce to the Inozemtsev chain~\eqref{eq:Ino}, justifying the name `\textit{q}-deformed Inozemtsev chain' when $\eta\neq 0$. The subsequent long- and (upon renormalisation) short-range limits finally yield the HS and Heisenberg \textsc{xxx} chains, respectively.

\subsection[\emph{No} global face-vertex transformation]{\emph{N$\mspace{-1mu}$o$\mspace{-1mu}$} global face-vertex transformation} \label{sec:FV}

\noindent
The eight-vertex and dynamical \textit{R}-matrices are connected by the face-vertex transformation
\begin{equation} \label{eq:FV}
	\begin{aligned}
	\check{R}(u - v;\eta) \, & \Phi(u,v,a;\eta) = \Phi(v,u,a;\eta) \, \check{R}(u-v,a;\eta) \, .
	\end{aligned}
\end{equation}
The explicit form of $\Phi$ is given in \textsection\ref{app:face_vertex}. It is generically invertible.

One might thus expect the \textit{q}-deformed Inozemtsev and MZ chains to be face-vertex transforms as well. This is, however, \emph{not} the case.
The reason is that the face-vertex transformation depends on the spectral parameter. As a result, the nearest-neighbour spin interactions \eqref{eq:RR_decomp} and \eqref{eq:RRdyn_decomp} are not conjugate. Indeed, suppressing $\eta$ for brevity, we have
\begin{equation} \label{eq:FV_does_not_extend}
	\begin{aligned}
	\mathllap{\check{R}(v-u)} \, \check{R}'(u-v) ={}& \Phi(u,v,a) \, \check{R}(v-u,a) \, \Phi(v,u,a)^{-1} \\[-.2ex]
	& \times \partial_u \bigl(  \Phi(v,u,a) \, \check{R}(u-v,a) \, \Phi(u,v,a)^{-1} \bigr) \\
	= {} & \Phi(u,v,a) \, \check{R}(v-u,a) \, \check{R}'(u-v,a) \, \Phi(u,v,a)^{-1} \\
	& + \Phi(u,v,a) \, \check{R}(v-u,a) \,  \Phi(v,u,a)^{-1} \\[-.3ex]
	& \hphantom{+ \Phi(u,v,a) } \mspace{-11mu} \times \partial_u \Phi(v,u,a) \, \check{R}(u-v,a) \, \Phi(u,v,a)^{-1} \!\!\!\!\!\! \\
	& + \Phi(u,v,a) \, \partial_u \Phi(u,v,a)^{-1} \, .
	\end{aligned}
\end{equation}
The first line of the result is naive face-vertex transformation, which is supplemented by two terms containing the derivative of the face-vertex transformation. There is no reason why these additional terms should commute with the naive transform.

The upshot is that, despite the fact that the face- and vertex-type \textit{R}-matrices are closely related, this does not extend to the corresponding long-range spin chains: the face-vertex transformation does not like derivatives.

\subsection{Practical comparison} \label{sec:practical_comparison}

\noindent
The MZ and \textit{q}-deformed Inozemtsev chain have a lot in common. Their structure is very similar. Both have chiral hamiltonians whose long-range interactions feature transport by \textit{R}-matrices and a nearest-neighbour interaction. The resemblance is clear in the form \eqref{eq:MZ_ham_pre}. While the two look less similar in the form \eqref{eq:MZ_ham_decomp}, the similarity reappears in the long-range (trigonometric) and macroscopic (hyperbolic) limits. There the sum over $\alpha$ of the MZ chain disappears, see~\eqref{eq:trig_spin_int}, yielding long-range spin interactions of the form \eqref{eq:S^LR_diagr}. In addition, both chains have deformed twisted translation operators. 

Despite these striking similarities, the MZ and \textit{q}-deformed Inozemtsev chains truly differ. As we have just seen, this is so despite the face-vertex relation between the \textit{R}-matrices from which the chains are built. The two spin chains have distinct physical properties and spectra. The differences are particularly pronounced in the symmetries and limits of the two long-range spin chains. By comparing the vertex-type landscape of the MZ$'$ chain, summarised in \textsection\ref{sec:vx-landscape} and Figure~\ref{fg:landscape_vertex}, with the face-type landscape of the \textit{q}-deformed Inozemtsev chain, see \textsection\ref{sec:qIno} and Figure~\ref{fg:landscape_face}, we come to the following key differences. Table~\ref{tab:MZ!=qIno} contains an overview.
\medskip

\textit{Celebrities.} 
The famous HS and Heisenberg \textsc{xxx} chains live on the face side. The vertex-type landscape contains interesting \emph{relatives} of these spin chains, but no \emph{generalisation} of either. The only common point is the rational HS chain, in the long-range plus macroscopic limit; we will return to this in \textsection\ref{sec:wrapping}.
\medskip

\textit{Spin symmetry.} The entire face-type landscape of Figure~\ref{fg:landscape_face} is partially isotropic, i.e.\ preserves spin-$z$. The undeformed limit is the \emph{isotropic} limit, where the chains become $\mathfrak{sl}_2$-invariant. In the long-range limit the spin symmetry is drastically enhanced: to the Yangian (resp.\ loop algebra) of $\mathfrak{gl}_2$ for (\textit{q}-deformed) HS chain. For generic $q$, the eigenspaces of the \textit{q}-deformed HS chain are as large as for the HS chain, except that the latter has additional `accidental' degeneracies due to its parity invariance and (half)integral eigenvalues. At special (root-of-unity) values of $q$, however, the \textit{q}-deformed HS chain has \emph{even larger} degeneracies than the HS chain~\cite{BMLST_24}.

In contrast, the vertex-type landscape of Figure~\ref{fg:landscape_vertex} is (fully) anisotropic, except in the  long-range or macroscopic limit, where the transverse twists disappear and the chains become partially isotropic. The amount of spin symmetry is the same in the undeformed limit, except in the long-range case, where it is enhanced to the Yangian of $\mathfrak{gl}_1$.

Thus, the spin symmetry is always higher on the face side than the vertex side, except at the single common point. The significance is that partial isotropy allows one to diagonalise the spin chain at fixed $2\,S^z = N - 2\,M$ (weight). This immediately yields some simple eigenvectors at low $M$, such as $\ket{\uparrow\dots\uparrow}$. More importantly, it paves the way for a description of the spectrum at any $M$ via a connection to a Calogero--Sutherland or Ruij-senaars system. On the face side, this underpins the appearance of Jack or Macdonald polynomials in the long-range limit, see \cite{Hal_91a} (cf.\ \cite{lamers2022fermionic}), \cite{bernard1993yang} and \cite{lamers2022spin}, and of elliptic Calogero--Sutherland wave functions for the Inozemtsev chain \cite{Inozemtsev:2002vb,KL_23}. We expect this connection to extend to the \textit{q}-deformed Inozemtsev chain, thanks to its partial isotropy. On the vertex side, it is less clear how such a connection might work for the anistropic (elliptic) MZ and SZ chains. Here the spin-chain Hilbert space only has an even and an odd sector, and it seems very hard to find any eigenvectors, cf.~\cite{matushko2022matrix}.
\medskip

\textit{Boundary conditions.}
The face-type spin chains are formally periodic, up to a diagonal twist at the deformed level away from the long-range limit. In contrast, the vertex side is formally \emph{anti}periodic, with a particular transverse twist.\,%
\footnote{\ By `formal' we mean that 
for $q\neq \pm1$ integrability requires the translation operator to be modified\,---\,by deforming permutations to \textit{R}-matrices\,---\,but $G^N$ remains the identity, or at least a central element.}
\medskip

\textit{Algebraic structure.} While both sides feature \textit{R}-matrices and are thus related to quantum groups, the face side seems to have a more transparent algebraic structure. The long-range limit is intimately related to (degenerate or ordinary) affine Hecke algebras, which explains its enhanced spin symmetry. The short-range limit instead features a representation of the affine Temperley--Lieb algebra.
We have not been able to identify such extra algebraic structures on the vertex side.

\begin{table}[h]
	\centering
	\begin{tabular}{lllll}
		\toprule
		& parity$\!$ & \rlap{spin symmetry} & & boundary cond.$\!\!$ \\ \midrule
		$\!\!$MZ$'$ & chiral* & none$\!\!$ & hyp: $S^z$, tri: $S^x$ & $q$-def.\ antiper.$\!\!$ \\ 
		$\!\!$SZ$'$ & \checkmark & none$\!\!$ & hyp: $S^z$, tri: Yangian & antiperiodic \\ 
		\midrule
		$\!\!q$Ino & chiral* & $S^z$ & tri: quantum-loop alg. & $q$-def.\ periodic \\
		$\!\!$Ino & \checkmark & $\mathit{SU}\mspace{-1mu}(2)\!\!\!\!$ & tri: Yangian & periodic \\
		\bottomrule
	\end{tabular} 
	\bigskip
	\caption{Global symmetries of the vertex- and face-type long-range spin chains. At the deformed level the chiral hamiltonian break parity, *$\,$except in the short-range limit. The spin symmetry is enhanced for the FK chain to the Yangian of $\mathfrak{gl}_1$, for ($q$)HS the (quantum-loop algebra, resp.)\ Yangian of $\mathfrak{gl}_2$.} 
	\label{tab:MZ!=qIno}
\end{table}

\subsection{SZ is antiperiodic Inozemtsev}
\label{sec:wrapping}

\noindent 
The vertex and face landscapes of Figures \ref{fg:landscape_vertex} and~\ref{fg:landscape_face} intersect in a single point: the rational HS chain \eqref{eq:ratHS}, which has no (periodicity or deformation) parameters left. Let us use this to provide yet another perspective for comparing the two landscapes. We will stay at the undeformed level.

The HS and Inozemtsev chains can be (re)constructed from their macroscopic limits by a procedure known as \emph{wrapping}. The idea is to wrap the infinite chains on a circle with $N$ sites to create a `comb of particles' (method of images), cf.\ \cite{sutherland1971exact}.
Namely, starting from the rational HS chain~\eqref{eq:ratHS} impose periodic boundary conditions by identifying $\sigma^\alpha_{i+N} \equiv \sigma^\alpha_i$ for all $i$. This turns the rational potential $V_\text{rat}(u) \coloneqq 1/u^2$ into
\begin{equation} \label{eq:wrapping_rat}
	\sum_{k \in \mathbb{Z}} 
	V_\text{rat}(u+k\,N) = V_\text{tri}(u) \quad \text{from\ } \eqref{eq:V_tri_intro} \, , 
\end{equation}
to yield the (trigonometric) HS chain. We can go further, as follows. The `Wick rotation' $\omega \leftrightarrow N$ (formally) turns the HS chain into the infinite Inozemtsev chain, with the hyperbolic potential $V_\text{hyp}(u) = \kappa^2/\sinh^2(\kappa\,u)$ defined in \eqref{eq:V_hyp}. We can again $N$-periodise, which yields
\begin{equation}
	\sum_{k \in \Z} V_\text{hyp}(u + k\,N) = V\mspace{-1mu}(u) \quad \text{from\ } \eqref{eq:V} \, ,
\end{equation}
to arrive at the (elliptic) Inozemtsev chain, cf.\ e.g.\ \cite{klabbers2022coordinate}. Note that this procedure does \emph{not} guarantee that the resulting chains are integrable.

As observed in \cite{fukui_exact_1996}, one can also use wrapping to construct \emph{twisted} Haldane--Shastry chains. Given a parameter twist $\varphi \in \pi \,\mathbb{Q}$ consider the diagonally twisted boundary conditions 
\begin{subequations} \label{eq:twisted_bc}
	\begin{gather}
	\sigma^\alpha_{i + k\mspace{1mu}N} = \exp(\I \mspace{2mu}\varphi \, k \, S^z) \; \sigma^\alpha_i \, \exp(-\I \mspace{2mu} \varphi \, k \, S^z) \, ,
\intertext{which twists the spins in the $x,y$-directions,}
	\sigma^\pm_{i +k\mspace{1mu}N} = \bigl(\E^{\pm \I \mspace{2mu} \varphi} \bigr)^{\!k} \, \sigma^\pm_i \, ,\quad \sigma_{i+k\mspace{1mu} N}^z = \sigma_i^z \, .		
	\end{gather}
\end{subequations}
For $\varphi=0$ we recover periodic boundaries as above. The simplest nontrivial case is $\varphi=\pi$. Since $\exp(\pm\I\, \pi \, \sigma^z\!/2) = \pm\I\,\sigma^z$ this corresponds to the \emph{anti}\-periodic boundary conditions
\begin{subequations} \label{eq:antiperiodic_bc}
	\begin{gather}
	\sigma^\alpha_{i + k\mspace{1mu}N} = \biggl( \prod_{j=1}^N \! \sigma^z_j \biggr)^{\!\!k} \, \sigma^\alpha_i \, \biggl( \prod_{j=1}^N \! \sigma^z_j \biggr)^{\!\!k} \, ,
\shortintertext{i.e.}
	\sigma^\pm_{i +k\mspace{1mu}N} = (-1)^k \, \sigma^\pm_i \, , \quad \sigma_{i+k\mspace{1mu} N}^z = \sigma_i^z \, .
	\end{gather}
\end{subequations}
Again starting from the rational HS chain we still get the potential~\eqref{eq:wrapping_rat} for the terms proportional to the identity ($\alpha=0$) and in the spin-$z$ direction, but in the $x,y$-directions the potentials are modified to 
\begin{equation} \label{eq:FK_1}
\begin{aligned}
	\sum_{k \in \mathbb{Z}} (-1)^k \, V_\text{rat}(u+k\,N) = \cos\bigl(\tfrac{\pi}{N} \mspace{1mu} u \bigr) \, V_\text{tri}(u) \, .
\end{aligned}
\end{equation}
We thus get the FK chain~\eqref{eq:FK} as the antiperiodic counterpart of the HS chain.

As before we can continue by `Wick rotating' $\omega \leftrightarrow N$, to (formally) obtain the hyperbolic SZ$'$ chain~\eqref{eq:SZ_hyp}. In order to arrive at the elliptic SZ$'$ chain we need to impose the antiboundary conditions
\begin{equation} \label{eq:yz_bounds}
	\sigma^\alpha_{i+k\mspace{1mu}N} = \biggl( \prod_{j=1}^N \! \sigma^x_j \biggr)^{\!\!k} \, \sigma^\alpha_i \, \biggl( \prod_{j=1}^N \! \sigma^x_j \biggr)^{\!\!k} \, ,
\end{equation}
which are obtained by conjugating \eqref{eq:antiperiodic_bc} by $U^{\otimes N}$. Note that the products in \eqref{eq:yz_bounds} naturally arise from the SZ$'$ translation operator~\eqref{eq:SZ trans op}. To see that wrapping \eqref{eq:yz_bounds} indeed yields the SZ$'$ chain~\eqref{eq:SZ} we rewrite the latter in terms of the spin operators 
\begin{equation}
	\vec{\sigma}^{\,\otimes 2} \coloneqq \bigl( \id\otimes\id, \sigma^x \otimes \sigma^x, \sigma^y \otimes \sigma^y, \sigma^z \otimes \sigma^z\bigr)\, .
\end{equation}
That is, we seek coefficients $C_\alpha(i-j)$ such that 
\begin{equation} \label{eq:SZ_via_Is}
	H_{\textsc{sz}'} = \sum_{i<j}^N \frac{1}{4} V\mspace{-1mu}\Big(\frac{i-j + \vec{\omega}}{2}\Big) \cdot \vec{\EE}_{ij}
	= \sum_{i<j}^{N} \vec{C}(i-j) \cdot \vec{\sigma}^{\,\otimes 2} _{ij} \, . 
\end{equation}
As $P_{ij} = \vec{\sigma}_i \cdot \vec{\sigma}_j /2$ and $V\mspace{-1mu}\big((-u +\omega_\alpha)/2\big) = V\mspace{-1mu}\big((u +\omega_\alpha)/2\big)$ we get
\begin{equation} \label{eq:coeff_fns}
	 \vec{C}(u) = -\frac{1}{4} 
	\begin{pmatrix}
		\hphantom{-}1 & \hphantom{-}1 & \hphantom{-}1 & \hphantom{-}1 \\
		-1 & -1 & \hphantom{-}1 & \hphantom{-}1 \\
		-1 & \hphantom{-}1 & -1 & \hphantom{-}1 \\
		-1 & \hphantom{-}1 & \hphantom{-}1 & -1
	\end{pmatrix} \cdot V\mspace{-1mu}\Bigl(\frac{u+\vec{\omega}}{2}\Bigr) \, ,  
\end{equation}
The expression for $C_0$ is precisely (minus) the $n$-tuple relation for the $\wp$-function with $n=2$ \cite{DLMF}, so that
\begin{equation}
	C_0(u) = -V\mspace{-1mu}(u) = -\sum_{k \in \mathbb{Z}} V_\text{hyp}(u+k\,N) 
	\, . 
\end{equation}
Similarly expressing the other $C_\alpha$ in terms of infinite sums over shifts of $V_\text{hyp}$ and manipulating summands we arrive at
\begin{equation} \label{eq:SZ_C_coeffs_hyps}
\begin{aligned}
	C_x(u) 	& = \sum_{k \in \mathbb{Z}} \mspace{2mu} \cosh\bigl(\kappa (u+ k\,N)\bigr) \, V_\text{hyp}(u + k\,N) \,, \\
	C_y(u) 	& = \sum_{k \in \mathbb{Z}} \mspace{2mu} (-1)^k \cosh\bigl(\kappa (u+ k\,N)\bigr) \, V_\text{hyp}(u + k\,N) \,, \\
	C_z(u) & =  \sum_{k \in \mathbb{Z}} \mspace{2mu} (-1)^k \, V_\text{hyp}(u + k\,N)\, .
\end{aligned}
\end{equation}
This confirms that wrapping with the antiperiodic boundaries~\eqref{eq:yz_bounds} yields the (elliptic) SZ$'$ chain. 
\medskip

Summarising, we find that, starting from the rational HS chain we can either
\begin{enumerate}
	\item[1.] $N$-periodise to obtain the (trig) HS chain, 
	\item[2.] `Wick rotate' $\omega \leftrightarrow N$ to (formally) obtain the hyperbolic Inozemtsev chain,
	\item[3.] $N$-periodise again, to find the (elliptic) Inozemtsev chain,
\end{enumerate}
or alternatively
\begin{enumerate}
	\item[1$'\!\!$.] $N$-antiperiodise in $\sigma^x, \sigma^y$ to get the FK chain,
	\item[2$'\!\!$.] `Wick rotate' to (formally) get the hyperbolic SZ$'$ chain,
	\item[3$'\!\!$.] $N$-antiperiodise in $\sigma^y, \sigma^z$, to find the SZ$'$ chain.
\end{enumerate}
We stress again that wrapping does not guarantee the integrability if the resulting chain, but in the preceding cases the results are known to be integrable.
It would be interesting to investigate whether any other combination of (anti)periodic, or more generally twisted, boundary conditions likewise yield elliptic spin chains that are integrable too.

\subsection{A view at long-range deformations} \label{sec:long-range_deformations}

\noindent
As we have just seen, starting from the long-range limit, the Ino\-zemtsev (respectively SZ$'$) chain can be reconstructed by doubly (anti)periodising the rational HS chain. Here we investigate how they relate to their short-range limits, by computing perturbative series around the nearest-neighbour limit. This is interesting for three reasons. First, this is the regime of particular interest to neutron-scattering experiments. Second, this will facilitate comparison in future work with methods that seek to construct integrable long-range spin chains in a systematic way starting from nearest-neighbour spin chains~\cite{gombor2021integrable,gombor2022wrapping,de_leeuw_lifting_2023}.
\medskip

\textit{Inozemtsev chain.} We start with the elliptic Inozemtsev spin chain, and group together all terms in the hamiltonian~\eqref{eq:Ino} at a fixed distance $n=|i-j|$,
\begin{equation} \label{eq:Ino_fixed_distance}
	H_\text{Ino} = \frac{1}{2}\sum_{n=1}^{N-1} \! V\mspace{-1mu}(n) \, I(n) \, , \ \ 
	I(n) \coloneqq -\frac{1}{2} \sum_{i=1}^N \vec{\sigma}_i \cdot \vec{\sigma}_{i+n \text{ mod } N} \, , \!\!
\end{equation}
where we note that $I(1)$ is precisely the \textsc{xxx} chain in \eqref{eq:Heis}. Recall that, in terms of $t\coloneqq \E^{-\kappa}$, for $u\in (-N/2,N/2)$ we have
\begin{equation} \label{eq:wp_expansion}
	\frac{1}{\kappa^2} \, V\mspace{-1mu}(u) = 4 \sum_{k\in \mathbb{Z}} \frac{t^{2(u+k N)}}{(t^{2(u+kN)}-1)^2} 
	 = 4 \sum_{k\in \mathbb{Z}} \sum_{l=1}^\infty l \, t^{2\,l\,|u+k N|}\, . 
\end{equation}
Collecting in powers of $t$, we can immediately write
\begin{equation} \label{eq:Ino_expansion_around_xxx}
	\begin{aligned}
	\mathllap{\frac{1}{4\,\kappa^2}} \, H_\text{Ino} = {} & \frac{1}{2}\sum_{n =1}^{N-1} \sum_{k\in \mathbb{Z}} \sum_{l=1}^\infty l \, t^{2\,l\,|n+k N|} \, I(n) =  \! \sum_{m=1}^\infty t^{2m} \!\!\! \sum_{\substack{d=1 \\ m/d \, \in \mathbb{N}}}^m \!\!\! d \; I(m/d) \, ,
	\end{aligned}
\end{equation}
where the final sum runs over all (proper and improper) divisors of the power of $t^2$, and in the second equality we get an extra factor of $2$ because for every $k>0$ there is a $k'<0$ such that $l \, |n+k \, N| = l' \, | n+k' \, N|$. 
The first few orders read
\begin{align}
	\frac{1}{4\,\kappa^2} \, H_\text{Ino} = {} &  t^2 I(1) + t^4 \bigl( 2 \, I(1) + I(2) \bigr) + t^6 \bigl( 3 \, I(1) + I(3) \bigr) \!\!\!\!\!\!\! \nonumber \\
	& + t^8 \bigl(4\, I(1) + 2 \, I(2) + I(4) \bigr) + O\bigl(t^{10}\bigr) \, .
\end{align}

Note that the power of $t^2$ records the maximal interaction range at that order. These series show that $H_\text{Ino}$ times a factor behaving as $t^{-2}/(4\kappa^2)$ for $\kappa\to\infty$, e.g.\ $n_\kappa =  \sinh^2(\kappa)/\kappa^2$ as in \cite{inozemtsev1996solution,klabbers2022coordinate}, becomes the Heisenberg \textsc{xxx} chain from~\eqref{eq:Heis} in the short-range limit. Next, the higher-order corrections take a very regular shape, where the $I(n)$ appearing with $t^{2m}$ have range $n\leqslant m$ and integer coefficients, which count how many times $n$ fits in $m$.
Finally, \eqref{eq:Ino_expansion_around_xxx} may naturally appear in periodic wrapping. To see this, let us briefly return to the hyperbolic model by sending $N\to \infty$, which is equivalent to dropping the `mod $N$' in the definition~\eqref{eq:Ino_fixed_distance} of $I(n)$. Without this mod~$N$, \eqref{eq:Ino_expansion_around_xxx} takes the form of an \emph{all-order} long-range deformation of the infinite Heisenberg \textsc{xxx} chain that is expected to be `order-by-order integrable' in the sense of \cite{de_leeuw_lifting_2023}. To wrap such a model around a chain of finite length $N$, as in \cite{gombor2022wrapping}, we impose periodic boundary conditions which simply reintroduces the `mod $N$' in the definition \eqref{eq:Ino_fixed_distance}. 
\medskip

\textit{SZ\/$'$ chain.} In order to perform the same analysis for the SZ$'$ chain we collect all terms in the hamiltonian~\eqref{eq:SZ} at fixed $|i-j|$. We will need to replace \eqref{eq:Ino_fixed_distance} by
\begin{equation} \label{eq:I^alpha(n)}
	I^\alpha(n) \coloneqq -\frac{1}{2} \sum_{i=1}^N \sigma^\alpha_i \, \sigma^\alpha_{i+n \text{ mod } N} \, , \quad \alpha \in \{0,x,y,z\} \, .
\end{equation}
According to \eqref{eq:SZ_via_Is} we then have 
\begin{equation}
	H_{\textsc{sz}'} = \sum_{i<j}^N \frac{1}{4} V\mspace{-1mu}\Big(\frac{i-j + \vec{\omega}}{2}\Big) \cdot \vec{\EE}
	= \frac{1}{2} \! \sum_{n=1}^{N-1} \vec{C}(n) \cdot \vec{I}(n) \, ,
\end{equation}
with coefficients $C_\alpha(n)$ given in \eqref{eq:coeff_fns}. Now use \eqref{eq:wp_expansion} to write 
\begin{equation} \label{eq:wp/2_expansion}
	\begin{aligned}
	\frac{1}{\kappa^2} \, \frac{1}{4} \mspace{1mu} V\mspace{-1mu}\Bigl(\frac{u}{2}\Bigr) & = \sum_{k\in \mathbb{Z}} \sum_{l=1}^\infty l \, t^{l\,|u+2k N|} \, , \\
	\frac{1}{\kappa^2} \, \frac{1}{4} \mspace{1mu} V\mspace{-1mu}\Bigl(\frac{u - \omega}{2}\Bigr) & = \sum_{k\in \mathbb{Z}} \sum_{l=1}^\infty l \, (-1)^l \, t^{l\,|u+2k N|}\, , 
	\end{aligned}
\end{equation}
which facilitates the rewriting 
\begin{align} \label{eq:SZ_C0_coeff}
	C_0(n) & = \kappa^2 \sum_{k\in \mathbb{Z}} \sum_{l=1}^\infty l \, \Bigl( t^{l |n+2 k N|} + t^{l \, |n+N+2 k N|} \nonumber  + (-1)^l \, t^{l \, |n+N+2 k N|} + (-1)^l \, t^{l \, |n+2 k N|} \Bigr) \nonumber \\
	& = -2 \,\kappa^2 \sum_{k\in \mathbb{Z}} \sum_{l=1}^\infty 2 \, l \, t^{2l \, |n+k N|}\, . 
\end{align}
In a similar fashion the other coefficients are
\begin{equation} \label{eq:SZ_C_coeffs}
\begin{aligned}
	C_x(n) & = 2 \, \kappa^2 \sum_{k\in\mathbb{Z}} \sum_{l=1}^\infty \, (2l-1) \, t^{(2l-1) \, |n+k N|}  \, , \\
		C_y(n) & = 2 \, \kappa^2 \sum_{k\in \mathbb{Z}} (-1)^k \sum_{l=1}^\infty \, (2l-1) \, t^{(2l-1) \, |n+k N|} \, , \\
	C_z(n) & = 2 \, \kappa^2 \sum_{k\in \mathbb{Z}} (-1)^k \sum_{l=1}^\infty 2 \, l \, t^{2l \, |n+k N|}\, .
	\end{aligned}
\end{equation}
To write $H_{\textsc{sz}'}$ as a series in $t$ analogous to \eqref{eq:Ino_expansion_around_xxx} we impose antiperiodic boundaries \eqref{eq:yz_bounds}, extend the range of~\eqref{eq:I^alpha(n)} to arbitrary $n$, and define the combinations 
\begin{equation} \label{eq:SZ_interaction_hams}
	\begin{aligned} 
	I^\textsc{xx}(n) & \coloneqq I^x(n) + I^y(n) \, , \\
	I^\diag\!(n) & \coloneqq I^z(n) - I^0(n) \, , 
	\end{aligned} \qquad 
	n>0 \, . 
\end{equation}
We recognise $I^{\textsc{xx}}(1)$ as the antiperiodic \textsc{xx} chain defined in \eqref{eq:ham_XX}. The definition~\eqref{eq:SZ_interaction_hams} using \eqref{eq:yz_bounds} neatly accounts for the signs in \eqref{eq:SZ_C_coeffs}: the SZ$'$ hamiltonian becomes
\begin{align} \label{eq:SZ_order_by_order}
\frac{1}{2 \, \kappa^2}	H_{\textsc{sz}'} 
	& =  \sum_{m=1}^\infty t^m \Biggl(\ \,
	\!\!\! \sum_{\substack{d=1 \\ m/d \, \in \mathbb{N} \\ d \ \text{odd}}}^m \!\!\! 
	d \, I^\textsc{xx}(m/d) + 
	\!\!\! \sum_{\substack{d=1 \\ m/d \, \in \mathbb{N} \\ d \ \text{even}}}^m \!\!\! d \, I^\diag\!(m/d) \Biggr) \, . \!
\end{align}
This is the analogue of our expression \eqref{eq:Ino_expansion_around_xxx} from the Inozemtsev chain. 
The structure is exactly the same, except that here we have a series in $t$ rather than $t^2$, and the coefficients feature $I^\textsc{xx}$ or $I^\diag$ depending on the parity of $d$.
The first few orders are
\begin{align}
	\frac{1}{2\kappa^{2}} \, H_{\textsc{sz}'} = {} & t \, I^\textsc{xx}(1) + t^2 \, \bigl( 2 \, I^\diag\!(1) + I^\textsc{xx}(2) \bigl)  + t^3 \, \bigl( 3 \, I^\textsc{xx}(1) + I^\textsc{xx}(3) \bigl) \nonumber \\
	& \! + t^4 \, \bigl( 4 \, I^\diag\!(1) + 2  \, I^\diag\!(2) + I^\textsc{xx}(4) \bigl) \,+\, O\bigl(t^5\bigr) \, ,
\end{align}
containing the antiperiodic \textsc{xx} chain~\eqref{eq:ham_XX} at lowest order. 

As a check note that the macroscopic limit of \eqref{eq:SZ_order_by_order} reproduces the hyperbolic SZ chain. Indeed, rewriting \eqref{eq:SZ_order_by_order} as a sum over fixed interaction distance $n=m/d$ and computing the sum over divisors we recover 
\begin{align}
	H_{\text{hyp}\,\textsc{sz}} & = \kappa^2 \sum_{n=1}^\infty \sum_{l \in \mathbb{Z}} \, \bigl( |2\,l|\, t^{n \mspace{1mu} |2l|} I^\textsc{xx}(n) \: + \: |2\,l+1|\, t^{n \mspace{1mu} |2l+1|} I^\diag\!(n) \bigr) \nonumber \\
	& = \sum_{n=1}^\infty V_\text{hyp}(n) \, \bigl( I^\textsc{xx}(n) + \cosh(\kappa\, n) \, I^\diag\!(n) \bigr)  \, ,
\end{align}
which indeed equals \eqref{eq:SZ_hyp}. 

Let us finally return to wrapping. The macroscopic limit $N\to\infty$ amounts to forgetting the (antiperiodic) boundary conditions of the $I^\alpha(n)$ in \eqref{eq:SZ_order_by_order}. Viewed in this way, \eqref{eq:SZ_order_by_order} looks like an all-order long-range deformation of the antiperiodic \textsc{xx} chain, whose integrability follows from that of the elliptic SZ$'$ chain. It would be interesting to investigate whether it can also arise from an order-by-order construction as in \cite{gombor2021integrable,de_leeuw_lifting_2023}.

\section{Conclusion} 
\label{sec:concl}

\subsection{Summary} 

\noindent
We considered the elliptic long-range spin chain of Matushko and Zotov (MZ) \cite{MZ_23b}, which is built from Baxter's eight-vertex \textit{R}-matrix. By decomposing the nearest-neighbour spin interaction, see~\eqref{eq:RR_decomp}, we rewrote its chiral hamiltonians as~\eqref{eq:MZ_ham_decomp}. Its building blocks have a clear physical interpretation, like for other known long-range spin chains \cite{Lam_18,lamers2022spin,KL_23,sechin2018r}. In addition we presented suitable translation operators, see \eqref{eq:MZ_trans}.

More precisely, we presented a small modification of the MZ chain, which we dubbed the MZ$'$ chain, by using the Jacobi theta function~\eqref{eq:def_theta2}, which is a Jacobi imaginary transform of the one used in \cite{MZ_23b}. Concretely, this choice amounts to a global spin rotation along with an additive constant that regularises the short-range limit. By evaluating various limits of the MZ$'$ chain (Figure~\ref{fg:landscape_vertex} on p.\,\pageref{fg:landscape_vertex}) we obtained the following results.
\begin{itemize}
	\item In the long-range (trigonometric) and macroscopic (hyperbolic) limits the MZ hamiltonians can be rewritten in the form \eqref{eq:MZ_trig} like the \textit{q}-deformed HS chain; 
	\item The (new) short-range limit is an \textsc{xx} model with deformed antiperiodic boundary conditions, see \eqref{eq:HXXZ};
	\item The undeformed limit is a variant of the spin chain~\eqref{eq:SZ_intro} of Sechin and Zotov (SZ)~\cite{sechin2018r}.
\end{itemize}
We called the latter the SZ$'$ chain.
\begin{itemize}
	\item Its long-range limit is, up to a global spin rotation, the (antiperiodic) Fukui--Kawakami chain~\eqref{eq:FK};
	\item Its short-range limit is the antiperiodic \textsc{xx} model~\eqref{eq:antiperiodic_xx}.
\end{itemize}
A more detailed summary of the limits was given in  \textsection\ref{sec:vx-landscape}. As we saw in \textsection\ref{sec:wrapping}, the antiperiodicity is not merely a recurring theme, but can be used to construct the SZ$'$ chain by `wrapping'. For more about the original MZ chain see \textsection\ref{app:MZ_chain}. 
\medskip

In \textsection\ref{sec:comparison} we compared the landscape of the MZ$'$ chain with that of the \textit{q}-deformed Inozemtsev chain, which we recently introduced in~\cite{KL_23}. It unifies the Inozemtsev chain~\eqref{eq:Ino} with the \textit{q}-deformed HS chain, and thus contains the Heisenberg \textsc{xxx} and HS chains as special cases. Its limits are summarised in Figure~\ref{fg:landscape_face}. The \textit{q}-deformed Inozemtsev chain is based on the (face-type)  dynamical \textit{R}-matrix. Since the latter is related to the eight-vertex \textit{R}-matrix by the face-vertex transformation, one might naively expect the \textit{q}-deformed Inozemtsev and MZ$'$ chains to be related by a conjugation too, and thus equivalent. In \textsection\ref{sec:FV} we showed that this is not the case. We supplemented this with a practical comparison between the two long-range spin chains, in terms of basic properties and symmetries, in \textsection\ref{sec:practical_comparison}. At the undeformed level we provided yet another perspective by constructing both landscapes from a common starting point by periodic or antiperiodic wrapping in \textsection\ref{sec:wrapping}, and determined how they might look in the framework of long-range deformations in \textsection\ref{sec:long-range_deformations}.

\subsection{Redux}

\noindent
In \cite{MZ_23b}, the MZ chain was presented as an elliptic, anisotropic generalisation of the \textit{q}-deformed Haldane--Shastry (HS) chain. So how are the two spin chains precisely related? The MZ chain does indeed look like an elliptic version of the \textit{q}-deformed HS chain, as it has precisely the same kind of long-range interactions. Moreover, being built from the eight-vertex \textit{R}-matrix, the MZ chain is certainly both elliptic and anisotropic. 

The \textit{q}-deformed HS chain is expected to have two different elliptic generalisations (see Figure~1 in \cite{KL_23}): one in terms of `coordinates' (potential), and one in terms of `momenta' (expected to be related to spin symmetry). Being anisotropic, the MZ chain looks more like the latter. However, the eight-vertex \textit{R}-matrix limits to the symmetric (`principal') six-vertex \textit{R}-matrix (cf.\ endnote [49] in \cite{KL_23}), rather than its asymmetric (`homogeneous') counterpart. The latter appears in the \textit{q}-deformed HS chain, enabling a connection to Hecke algebras that underlies the model's enhanced spin symmetry, whence high degeneracies, and the appearance of Macdonald polynomials. Could the MZ chain be the other elliptic generalisation of the \textit{q}-deformed HS chain, and instead deform the Inozemtsev chain? Not quite: its undeformed limit is the SZ chain, see \textsection7 of \cite{MZ_23b} and \textsection\ref{sec:SZ_limits}. A \textit{q}-deformed chain that does have the Inozemtsev chain as its undeformed limit is, well, the \textit{q}-deformed Inozemtsev chain \cite{KL_23}.
The MZ chain, instead, has a landscape of its own. It does not contain the (trigonometric) HS chain, let alone its \textit{q}-deformation, but rather comprises other interesting spin chains.
\medskip

For long-range spin chains, at the deformed level integrability requires spin permutations to become \textit{R}-matrices leading to chiral interactions as in \eqref{eq:S^LR_diagr} \cite{Lam_18,lamers2022spin,MZ_23b,KL_23}.
Moreover, an elliptic potential requires these \textit{R}-matrices to be elliptic in view of the decompositions \eqref{eq:RR_decomp}, \eqref{eq:RR_decomp_tri} and \eqref{eq:RRdyn_decomp}. The two standard elliptic \textit{R}-matrices yield two long-range spin chains. At the deformed elliptic level, it thus seems natural to use the terms `face' and `vertex' to distinguish the two sides. In the long-range (trigonometric) limit, they correspond to the `homogeneous' and `principal' six-vertex \textit{R}-matrices. At the undeformed level the spin interactions simplify drastically. As we saw in \textsection\ref{sec:wrapping}, here the terms `periodic' and `antiperiodic' seem more natural. While these boundary conditions persist, suitably \textit{q}-deformed and twisted, at the deformed level, it is not clear if the two sides can similarly be obtained by wrapping, and the antiperiodicity was not at all clear from \cite{MZ_23b}.

\subsection{Outlook}

\noindent 
Many interesting venues remain. We highlight three of them.

First, to give an exact characterisation of the spectrum of the \textit{q}-deformed Inozemtsev and MZ$'$ chains. We hope to tackle the former soon. For the latter initial steps were taken in \cite{matushko2022matrix}, but it remains a challenge due to the lack of partial isotropy. 

Second, to determine and leverage the pertinent elliptic quantum groups to construct the spectrum of the long-range spin chains by Bethe-ansatz methods. This is well understood in the special cases of the Heisenberg \textsc{xxx} and antiperiodic \textsc{xx} chains. Recently it was shown that it can be adapted to the HS chain \cite{ferrando2023bethe}, which should readily \textit{q}-deform. 
Can this be extended to the elliptic case?

Third, to construct the doubly deformed, fully anisotropic \emph{generalisation} of the \textit{q}-deformed HS chain.

\subsection*{Acknowledgements} 

\noindent
We are grateful to the organisers of the 16th MSJ-SI school and conference \textit{Elliptic Integrable Systems, Representation Theory and Hypergeometric Functions}, held in Tokyo on 26--28 July and 31 July--4 August 2023, for the stimulating environment and hospitality.
JL was funded by LabEx Mathématique Hadamard (LMH), and in the later stages of this work by ERC-2021-CoG\,--\,BrokenSymmetries 101044226.

\appendix

\section{Elliptic functions and their limits}
\label{app:ell_functions}

\noindent 
In this work we make use of several elliptic functions. Here we summarise these functions and their most important properties. Standard references are ancient \cite{DLMF,abramowitz1948handbook,whittaker1904course}.

We will start with general $\tau \in \mathbb{C}$ with $\mathrm{Im}\,\tau>0$, associated to the period lattice $\mathbb{Z} + \tau \, \mathbb{Z}$. Then we will specialise in two ways, related by a Jacobi imaginary transformation, that are associated to the rectangular lattice $N \, \mathbb{Z} + \omega \, \mathbb{Z}$ with real period $N\in \mathbb{Z}_{\geqslant 2}$ and imaginary period $\omega \coloneqq \I \pi/\kappa$ with $\kappa>0$. These provide the building blocks for the spin chains that we discuss. One specialisation gives the functions $\theta(x), V(x), \dots$ that we use in the main text. The other specialisation gives the conventions of Matushko and Zotov \cite{MZ_23a}, which we will denote in \textsf{sans serif}, e.g.\ $\stheta(x), \mathsf{V}(x),\dots$, and we will use in \textsection\ref{app:MZ_chain}.

\subsection{General setup}  
\noindent 

For $\tau \in \mathbb{C}$ in the upper half-plane, $\mathrm{Im}\,\tau>0$, consider the lattice $\mathbb{Z} + \tau \, \mathbb{Z}$ in $\mathbb{C}$. At this stage we will explicitly indicate the dependence on $\tau$ to ensure an unambiguous notation. The corners of the fundamental domain are 
\begin{equation} \label{eq:periods}
	\bigl(\omega_0(\tau),\omega_x(\tau),\omega_y(\tau),\omega_z(\tau)\bigr) = (0,\tau,1+\tau,1) \, .
\end{equation} 
The odd Jacobi theta function with nome $p(\tau) \coloneqq \E^{\I \pi \tau}$ is
\begin{align} \label{eq:theta_def}
	\vartheta(u \,|\, \tau ) \coloneqq {} & 2 \sum_{n=0}^{\infty} (-1)^n \, p(\tau)^{(n+1/2)^2} \sin\bigl(\pi \mspace{2mu} (2n+1) \mspace{2mu} u\bigr) \nonumber \\
	= {}& 2 \,\sin(\pi u) \, p(\tau)^{1/4} \prod_{n=1}^\infty \! \bigl(1-p(\tau)^{2n}\bigr) \! \bigl(1 - \E^{2\pi\I u} \mspace{2mu} p(\tau)^{2n}\bigr) \! \bigl(1 - \E^{-2\pi\I u} \mspace{2mu} p(\tau)^{2n}\bigr) \, .
\end{align}
We prefer to work with its rescaled version
\begin{equation} \label{eq:theta_normalised}
	\theta(u \,|\, \tau) \coloneqq 
	\frac{\vartheta(u \,|\, \tau)}{\vartheta'(0 \,|\, \tau )}  
	= \tfrac{1}{\pi} \sin(\pi\mspace{2mu} u) + O\bigl(p(\tau)^2\bigr) \, ,
\end{equation}
normalised such that $\theta'(0 \,|\, \tau)=1$. This function is odd and entire, has a simple zero at the origin, and is doubly quasiperiodic
\begin{equation} \label{eq:th_quasiper}
	\begin{aligned}
	\theta(u + 1 \,|\, \tau) &= -\theta(u \,|\, \tau) \, , \qquad \theta(u + \,\tau \,|\, \tau)  = - p(\tau)^{-1} \, \E^{-2\pi\I u} \, \theta(u \,|\, \tau) \, .
	\end{aligned}
\end{equation} 
Together, these properties uniquely characterise $\theta(u \,|\, \tau)$.

From the odd Jacobi theta function we can construct any other function that we will need. 
We define the `prepotential' as the logarithmic derivative of \eqref{eq:theta_def}, i.e.\ the Weierstraß zeta function up to a linear term, 
\begin{equation}
	\rho(u \,|\, \tau) \coloneqq \frac{\theta'(u \,|\, \tau)}{\theta(u \,|\, \tau )} = \zeta(u \,|\, \tau) - \eta_1(\tau) \, u \, ,
\end{equation}
with constant $\eta_1(\tau) \coloneqq 2 \, \zeta(1/2 \,|\, \tau)$.
This function is odd and has a simple pole at $u=0$ with residue $1$. It is doubly (quasi)periodic, $\rho(u+1 \,|\, \tau) = \rho(u \,|\, \tau)$ and $\rho(u+\tau \,|\, \tau) = \rho(u \,|\, \tau)-2\pi \I$. It is odd around the zero $u=1/2$, i.e.\ $\rho(1/2-u \,|\, \tau) = -\rho(1/2+u \,|\, \tau)$.

Taking another derivative gives the Weierstraß $\wp$ function,
\begin{equation} \label{eq:-rho'}
	V(u\,|\,\tau) \coloneqq  {-}\rho'(u \,|\, \tau) = \wp(u \,|\, \tau) + \eta_1(\tau) \, ,
\end{equation}
which has periods $1,\tau$ and double pole at $u=0$.

The Kronecker elliptic function is 
\begin{align} \label{eq:Kronecker_def}
	\phi(u,v \,|\, \tau) \coloneqq \frac{\theta(u+v \,|\, \tau)}{\theta(u \,|\, \tau) \, \theta(v \,|\, \tau)} & = \frac{\pi \sin \bigl(\pi (u+v)\bigr)}{\sin(\pi u) \, \sin(\pi v)} + O\bigl(p(\tau)\bigr) \nonumber \\
	& = \pi \bigl(\cot (\pi u) + \cot(\pi v)\bigr) + O\big(p(\tau)\bigr) \, .
\end{align} 
This symmetric function is doubly (quasi)periodic, $\phi(u+1,v \,|\, \tau) = \phi(u,v \,|\, \tau)$ and $\phi(u+\tau,v \,|\, \tau) = \E^{-2\pi \I v} \, \phi(u,v \,|\, \tau)$. It is related to \eqref{eq:-rho'} via $\phi(u,v \,|\, \tau) \, \phi(u,-v \,|\, \tau) = \wp(u \,|\, \tau) - \wp(v \,|\, \tau)$.

Finally, we will occasionally need the three other Jacobi theta functions, defined in terms of \eqref{eq:theta_def} as
\begin{equation} \label{eq:theta_234}
	\begin{aligned}
	&\vartheta_2(z \,|\, \tau) \coloneqq \vartheta\bigl(z+ \tfrac{1}{2} \,\big|\, \tau\bigr)\, ,\\
	& \vartheta_3(z \,|\, \tau) \coloneqq 
	p(\tau)^{1/4} \, \E^{\I \pi x} \, \vartheta\bigl(z + \tfrac{1}{2}(1+\tau) \,\big|\, \tau\bigr) \, , \\
	& \vartheta_4(z \,|\,\tau) \coloneqq -\I \, 
	p(\tau)^{1/4} \, \E^{\I \pi x} \, \vartheta\bigl(z+\tfrac{1}{2}\tau \,\big|\, \tau\bigr) \, .
	\end{aligned}
\end{equation}
Hence one can think of these as versions of the odd Jacobi theta function \eqref{eq:theta_def} shifted over half-periods, generalising the identity $\cos(x) = \sin(x+\pi/2)$ between the elementary trigonometric functions. Expanding in the nome we find 
\begin{equation}
	\begin{aligned}
	p(\tau)^{-1/4} \, \vartheta_2(z \,|\, \tau) & = 2 \cos(\pi x) + O\bigl(p(\tau)^2\bigr) \, , \\
	\vartheta_3(z \,|\, \tau) &= 1 + 2\,p(\tau) \, \cos(\pi x) + O\bigl(p(\tau)^2\bigr) \, ,\\
	\vartheta_4(z \,|\, \tau) & = 1 - 2\, p(\tau) \, \cos(\pi x) + O\bigl(p(\tau)^2\bigr) \, . 
	\end{aligned} 
\end{equation}

\subsection{Two specialisations} \label{app:specialisations}

\noindent
Now we specialise the above general setting to the period lattice $N \, \mathbb{Z} + \omega \, \mathbb{Z}$ by fixing $\tau$ and rescaling $u$. We will use two choices out of an infinite family generated by an action of the modular group $\mathit{SL}(2,\mathbb{Z})$, whose significance in the present context will be discussed in \cite{KL_extended}. We choose
\begin{equation}
\label{eq:taus_and_scalings}
	\begin{aligned}
	& \text{main text, \cite{KL_23}} \qquad && \tau \coloneqq -N/\omega \, , \quad u\to -u/\omega \, , \\
	& \text{\textsection\ref{app:MZ_chain}, \cite{MZ_23b}} && \mathrlap{\stau}\hphantom{\tau} \coloneqq \hphantom{+}\omega/N \, , \quad u\to \hphantom{+}u/N \, ,
	\end{aligned}
\end{equation}
so that the periods \eqref{eq:periods} become
\begin{equation}
	\begin{aligned}
	(\omega_0, \omega_x, \omega_y, \omega_z) & \coloneqq (0,N, N-\omega, -\omega) \, , \\
	(\somega_0, \somega_x, \somega_y, \somega_z) & \coloneqq (0,\omega, N+\omega,  N)\, .
	\end{aligned}
\end{equation}
Note the minus signs! They originate from the minus sign in $\tau$ in \eqref{eq:taus_and_scalings}, which is necessary to ensure that Im$(\tau)>0$. 

The theta function \eqref{eq:theta_def} is thus specialised to\,%
\footnote{\ In \cite{klabbers2022coordinate} we used the more compact and uniform notation $\theta_1(u) \coloneqq \theta(u)$ and $\theta_2(u) \coloneqq \stheta(u)$, and called $N\,\mathbb{Z} + \omega\,\mathbb{Z}$ the `coordinate lattice'. 
Here we opt to avoid a proliferation of subscripts. Our $\theta(u)$ is as in \cite{KL_23}.}
\begin{align} \label{eq:theta_def_typed}
	\theta(u) \coloneqq {} & {-}\omega \, \theta\bigl(\tfrac{u}{-\omega } \big| \tfrac{N}{-\omega} \bigr) = \omega \,\theta\bigl(\tfrac{u}{\omega } \big| \tfrac{N}{-\omega} \bigr) \nonumber \\
	= {} & \tfrac{\omega}{\pi}\sin\bigl(\tfrac{\pi}{\omega} \mspace{2mu} u\bigr) \prod_{n=1}^{\infty} \! \frac{ \sin\bigl(\tfrac{\pi}{\omega}(n \mspace{2mu} N + u)\bigr) \sin\bigl(\frac{\pi}{\omega}(n \mspace{2mu} N - u)\bigr)}{\sin^2 \bigl(\tfrac{\pi}{\omega} \mspace{2mu} n \mspace{2mu} N\bigr)} \; , \nonumber \\
	\stheta(u) \coloneqq {}& N \, \theta\bigl(\tfrac{u}{N} \big| \tfrac{\omega}{N}\bigr) \nonumber \\
	= {}& \tfrac{N}{\pi} \sin\bigl(\tfrac{\pi}{N} \mspace{2mu} u \bigr) \prod_{n=1}^{\infty} \! \frac{ \sin\bigl(\tfrac{\pi}{N} (n\, \omega + u) \bigr) \sin\bigl(\tfrac{\pi}{N} (n \,\omega - u) \bigr)}{\sin^2 \bigl(\tfrac{\pi}{N} \mspace{2mu} n \, \omega \bigr)} \; .
\end{align} 
Rewriting the former using $\omega = \I\pi/\kappa$, with $\kappa>0$, gives \eqref{eq:def_theta2}. Note that \eqref{eq:theta_def_typed} are quasiperiodisations of $\sinh$ and $\sin$, respectively. 
Indeed, define the corresponding nomes  
\begin{equation} \label{eq:nomes}
	p \coloneqq \E^{-\I \pi N/\omega} = \E^{-N\mspace{1mu}\kappa} \, , \qquad \mathsf{p} \coloneqq \E^{\I \pi \omega/N} = \E^{-\pi^2/(N\mspace{1mu}\kappa)} \, .
\end{equation}
Then \eqref{eq:th_quasiper} implies that 
\begin{equation}
	\begin{aligned}
	\theta(u + \omega) & = -\theta(u) \, , \quad \theta(u + N) = - p^{-1} \, \E^{- 2\pi\I u/N} \, \theta(u) \, , \\
	\stheta(u + N) & = -\stheta(u) \, , \quad \, \stheta(u + \omega) = - {\text{\sffamily p}}^{-1} \, 
	\E^{- 2\kappa\mspace{1mu} u} \, \stheta(u) \, .  
	\end{aligned}
\end{equation}
The trigonometric and rational degenerations are
\begin{equation}
	\begin{aligned}
	\theta(u) & = \mspace{2mu} \tfrac{1}{\kappa} \sinh(\kappa \mspace{2mu} u) \mspace{1mu} + O(p^{2}) = u + O\bigl(\kappa^2\bigr) \, , \\
	\stheta(u) &= \tfrac{N}{\pi} \sin\bigl(\tfrac{\pi}{N} \mspace{2mu} u\bigr) + O(\mathsf{p}^{2}) = u + O\bigl(N^{-2}\bigr) \, .
	\end{aligned}
\end{equation}
These correspond to removing the periods, as $N \to \infty$ and then $\kappa \to 0$ for $\theta(u)$, and the other way around for $\stheta(u)$. By \eqref{eq:nomes}, the first equalities furthermore describe $\kappa\to\infty$ (small $p$) or $\kappa\to 0$ (small $\mathsf{p}$).
These two regimes are related by a Jacobi imaginary transformation,
\begin{equation} \label{eq:Jacobi_imaginary_b=12}
	\theta(u) = \E^{\kappa \mspace{1mu} u^2/N} \, \stheta(u) \, .
\end{equation}

Now consider the Weierstraß functions $\sigma$, $\zeta(u)$ and $\wp(u) = -\zeta'(u)$ associated to the period lattice $N\,\mathbb{Z} + \omega\,\mathbb{Z}$.
We define the lattice constants $\eta_z \coloneqq 2\zeta(-\omega/2) = {-}2 \,\zeta (\omega/2)$ and $\seta_{\mspace{1mu}z} \coloneqq 2 \,\zeta (N/2)$, slightly deviating from \cite{Inozemtsev_1995} so that our two specialisations are on equal footing. Indeed,
\begin{equation}
	\sigma(u) = \E^{\eta_z \mspace{1mu} u^2 \mspace{-2mu}/(2\mspace{1mu} \omega)} \, \theta(u) = \E^{\seta_{\mspace{1mu}z}  \mspace{1mu} u^2 \mspace{-2mu}/(2 N)} \, \stheta(u) \, .
\end{equation}
Compatibility with \eqref{eq:Jacobi_imaginary_b=12} is ensured by the Legendre relation $\omega\,\seta_{\mspace{1mu}z} - N\, \eta_z = 2\pi\I$. 
The corresponding prepotentials are
\begin{equation} \label{eq:rho}
	\begin{aligned}
	\rho (u) & \coloneqq \frac{\theta'(u)}{\theta(u)} = \zeta(u) -  \frac{\eta_z}{\omega}\, u \, ,\\ 
	\srho (u) & \coloneqq \frac{\stheta'(u)}{\stheta(u)} = \zeta(u) - \frac{\seta_{\mspace{1mu}z}}{N}
	\, u = \rho(u) + \frac{2\kappa}{N} \, u \, ,
	\end{aligned}
\end{equation}
with degenerations
\begin{equation}
	\begin{aligned}
	\rho (u) & = \kappa \coth(\kappa \mspace{2mu} u) \mspace{3mu} + O\bigl(p^2\bigr) = \frac{1}{u} + O\bigl(\kappa^2\bigr) \, , \\
	\srho (u) & = \tfrac{\pi}{N} \cot\bigl(\tfrac{\pi}{N} \mspace{2mu} u\bigr) + O\bigl(\mathsf{p}^2\bigr) = \frac{1}{u} + O\bigl(N^{-2}\bigr) \, .
	\end{aligned}
\end{equation}
The associated potentials only differ by an additive constant,
\begin{equation} \label{eq:pot_typed}
	\begin{aligned}
	V\mspace{-1mu}(u) & \coloneqq -\rho'(u) = \wp(u) + \frac{\eta_z}{\omega} \, , \\
	\mathsf{V}(u) & \coloneqq -\srho'(u) = \wp(u) + \frac{\seta_{\mspace{1mu} z}}{N} = V\mspace{-1mu}(u) - \frac{2\kappa}{N} \, ,
	\end{aligned}
\end{equation}
and degenerate as 
\begin{equation}
	\begin{aligned}
	V(u) &= \frac{\kappa^2}{\sinh^2(\kappa u)} + O\bigl(p^2\bigr) = \frac{1}{u^2}  + O\big(\kappa^2\bigr)\, , \\
	\mathsf{V}(u) &= \frac{(\pi/N)^2}{\sin^2\bigl(\tfrac{\pi}{N} \mspace{2mu}u\bigr)} + O\bigl(\mathsf{p}^2\bigr) = \frac{1}{u^2}  + O\big(\kappa^2\bigr)\, . 
	\end{aligned}
\end{equation}
It has a nice series representations in terms of $t^2 \coloneqq \E^{-2\kappa}$,
\begin{equation} \label{eq:wp_as_periodicised_sinh}
	V\mspace{-1mu}(u) = \sum_{n\in \mathbb{Z}} \frac{\kappa^2}{\sinh^2 \bigl( \kappa (u+n\mspace{1mu} N)\bigr)} = 4\, \kappa^2 \sum_{n \in \mathbb{Z}} \frac{t^{2(u+n N)}}{(t^{2u}-t^{2nN})^2} \; ,
\end{equation}
obtained from the representation \eqref{eq:theta_def_typed} as a periodicised $\sinh$.

Finally, the associated Kronecker elliptic functions are
\begin{equation}
\label{eq:Kronecker_typed}
	\begin{aligned}
	\phi(u,v) & \coloneqq \frac{\theta(u+v)}{\theta(u) \, \theta(v)} \, , \\
	\sphi(u,v) & \coloneqq \frac{\stheta(u+v)}{\stheta(u) \, \stheta(v)} = \E^{-2\kappa \mspace{1mu} u \mspace{1mu} v/N} \phi(u,v) \, .
	\end{aligned}
\end{equation} 
To illustrate how the final equality relates the regimes of small and large $\kappa$, note that, whereas the trigonometric expansion
\begin{equation} \label{eq:phi1_trig}
	\begin{aligned}
	\sphi(u,v) & = \frac{\pi}{N} \frac{ \sin\bigl(\tfrac{\pi}{N} (u+v)\bigr)}{ \sin\bigl(\tfrac{\pi}{N} u\bigr) \sin\bigl( \tfrac{\pi}{N} v \bigr)} + O(\mathsf{p}^{2}) \\
	& = \frac{\pi}{N} \Bigl( \cot\bigl(\tfrac{\pi}{N} u\bigr) + \cot\bigl( \tfrac{\pi}{N} v \bigr) \Bigr) + O(\mathsf{p}^{2})
	\end{aligned}
\end{equation}
describes the long-range limit $\kappa \to 0$ (or $N\to 0$), in order to take the short-range limit $\kappa \to \infty$ (or $N\to \infty$) we need
\begin{equation} \label{eq:phi2_short}
	\begin{aligned}
	\phi(u,v) & = \kappa \, \frac{ \sinh\bigl(\kappa(u+v)\bigr)}{ \sinh(\kappa\mspace{2mu} u) \sinh(\kappa\mspace{2mu} v)} + O(p^{2}) \\
	& = \kappa \, \bigl( \coth(\kappa\mspace{2mu}u) + \coth(\kappa\mspace{2mu}v) \bigr)+ O(p^{2}) \, . 
	\end{aligned}
\end{equation}
When interpreted as removing both periods ($\omega\to\I\mspace{2mu} \infty$, $N\to\infty$) their common limit is $(u+v)/(uv) = 1/u + 1/v$.

\section{Eight-vertex \textit{R}-matrix and its limits}
\label{app:R_matrices}

\noindent
In this appendix we define all \textit{R}-matrices appearing in our discussion and compute their limits.

\subsection{Baxter's eight-vertex \textit{R}-matrix} \label{app:R8v}

\noindent 
We define $\vec{g}(u |\tau)$ to be the four-tuple with components
\begin{equation}
	\label{eq:g_tau}
	g^\alpha(u;\eta \mspace{1mu}|\mspace{1mu} \tau) \coloneqq
	\begin{cases} 
		\displaystyle \hphantom{\E^{\I \pi u} \,} \frac{\phi\bigl(u,(\eta+\omega_\alpha)/2 \mspace{1mu}|\mspace{1mu} \tau\bigr)}{\phi(u,\eta \mspace{1mu}|\mspace{1mu} \tau)} \quad & \alpha = 0,z \, ,  \\[1.6ex]
		\displaystyle \E^{\I \pi u} \, \frac{\phi\bigl(u,(\eta+\omega_\alpha)/2 \mspace{1mu}|\mspace{1mu} \tau\bigr)}{\phi(u,\eta \mspace{1mu}|\mspace{1mu} \tau)}\quad & \alpha=x,y \, .
	\end{cases} 
\end{equation}
Then Baxter's eight-vertex \textit{R}-matrix reads
\begin{equation} \label{eq:R8v_tau}
	R(u;\eta \mspace{1mu}|\mspace{1mu} \tau) \coloneqq \frac{1}{2} \, \vec{g}(u \mspace{1mu}|\mspace{1mu} \tau) \cdot \vec{\sigma}^{\,\otimes 2} \, , \quad 
	(\vec{\sigma}^{\,\otimes 2})^\alpha \coloneqq (\sigma^{\alpha})^{\otimes 2} \, .
\end{equation}
This yields $\check{R}(u \mspace{1mu}|\mspace{1mu} \tau) = P \, R(u \mspace{1mu}|\mspace{1mu} \tau)$, which satisfies the quantum Yang--Baxter equation, given in braid-like form in \eqref{eq:YBE}. 

The two specialisations of \textsection\ref{app:specialisations} lead us to
\begin{equation}
	\label{eq:g_typed1}
	g^\alpha(u;\eta) \coloneqq
	\begin{cases} 
		\displaystyle \hphantom{\E^{-\kappa u} \,} \frac{\phi\bigl(u,(\eta+\omega^\alpha)/2\bigr)}{\phi(u,\eta)} \quad & \alpha = 0,z \, ,  \\[1.6ex]
		\displaystyle \E^{-\kappa u} \, \frac{\phi\bigl(u,(\eta+\omega^\alpha)/2\bigr)}{\phi(u,\eta)}\quad & \alpha = x,y
		\, ,
	\end{cases} 
\end{equation}
and 
\begin{equation}
	\label{eq:g_typed2}
	\mathsf{g}^\alpha(u;\eta) \coloneqq
	\begin{cases} 
		\displaystyle \hphantom{\E^{\I \pi u/N} \,} \frac{\sphi\bigl(u,(\eta+\somega^\alpha)/2\bigr)}{\sphi(u,\eta)} \quad & \alpha = 0,z \, ,  \\[1.6ex]
		\displaystyle \E^{\I \pi u/N} \, \frac{\sphi\bigl(u,(\eta+\somega^\alpha)/2\bigr)}{\sphi(u,\eta)}\quad & \alpha = x,y
		\, ,
	\end{cases} 
\end{equation}
with corresponding eight-vertex \textit{R}-matrices
\begin{equation} \label{eq:R8v_typed}
\begin{aligned}
	R(u;\eta) &\coloneqq \frac{1}{2} \, \vec{g}(u;\eta) \cdot \vec{\sigma}^{\,\otimes 2} = R\bigl(\tfrac{u}{-\omega};\tfrac{\eta}{-\omega} \big| \tfrac{N}{-\omega}\bigr) \, ,\\
	\mathsf{R}(u;\eta) &\coloneqq \frac{1}{2} \, \vec{\mathsf{g}}(u;\eta) \cdot \vec{\sigma}^{\,\otimes 2} = R\bigl(\tfrac{u}{N};\tfrac{\eta}{N} \mspace{1mu}|\mspace{1mu} \tfrac{\omega}{N} \bigr) \, ,
\end{aligned}
\end{equation}
along with $\check{R}(u;\eta) = P \, R(u;\eta)$, given in \eqref{eq:Rch8v}, and $\check{\mathsf{R}}(u;\eta) = P \, \mathsf{R}(u;\eta)$. By \eqref{eq:Kronecker_typed} we have 
\begin{equation} 
	R(u;\eta) = \frac12 \, \E^{-\kappa \eta u/N} \, \vec{g}(u;\eta) \cdot M \cdot \vec{\sigma}^{\,\otimes 2} \, , 
\end{equation}
where
\begin{equation}
	M =
	\begin{pmatrix}
		1 & \color{gray!80}{0} & \color{gray!80}{0} & \color{gray!80}{0} \\
		\color{gray!80}{0} & \color{gray!80}{0} & \color{gray!80}{0} & 	1 \\[2pt]
		\color{gray!80}{0} & \color{gray!80}{0}& 1 & \color{gray!80}{0} \\[3pt]
		\color{gray!80}{0} & 1 & \color{gray!80}{0} & \color{gray!80}{0}
	\end{pmatrix} \, , 
\end{equation}
swaps the spin-$x$ and -$z$ directions.

It will be convenient to realise the action of the swap $M$ on $\vec{\sigma}$ via a change of basis for the spin-space $\mathbb{C}^2$ on which the $\sigma^\alpha$ act. Although there does not exist a $U \in \text{Mat}(2,\mathbb{C})$ such that conjugation by $U$ swaps $\sigma^x$ and $\sigma^z$ while leaving $\sigma^y$ (and of course $\sigma^0 = \id$) untouched, the (Hadamard-type) matrix 
\begin{equation} 
	U \coloneqq \frac{1}{\sqrt{2}} 
	\begin{pmatrix}
	1 & -1 \\ 1 & 1 
	\end{pmatrix}
\end{equation}
obeys
\begin{equation} \label{eq:sigma_conj_U}
	\begin{aligned}
	& U \, \sigma^0 \, U^{-1} = \sigma^0 \, , \qquad 
	&& U \, \sigma^x \, U^{-1} = -\sigma^z \, , \\ 
	& U \, \sigma^y \, U^{-1} = \sigma^y \, , \qquad 
	&& U \, \sigma^z \, U^{-1} = \sigma^x \, .
	\end{aligned}
\end{equation}
On linear combinations of $(\sigma^\alpha)^{\otimes 2}$, such as the eight-vertex \textit{R}-matrix \eqref{eq:R8v}, the sign disappears when conjugating by $U^{\otimes 2}$. This conjugation extends to the spin-chain space $(\mathbb{C}^2)^{\otimes N}$ via $U^{\otimes N}$. 
Hence we find 
\begin{equation} \label{eq:8v1_as-2}
	\mathsf{R}(u;\eta) = \E^{\kappa \eta u /N} \left( U^{\otimes 2} \right)^{-1} \! R(u;\eta) \, U^{\otimes 2} \, , 
\end{equation}
implying that, up to a unitary transformation, the two \textit{R}-matrices only differ by an overall exponent. 
The braid-like versions $\check{\mathsf{R}}(u;\eta)$ and $\check{R}(u;\eta)$ are related in exactly the same way.

\subsection{Limits}

\noindent
Rather than analysing the two versions \eqref{eq:R8v_typed} of the eight-vertex \textit{R}-matrix separately we will use \eqref{eq:8v1_as-2} to relate the results. 
\medskip

\noindent
\textit{Trigonometric limit.} 
To let $\kappa \to 0$ it is convenient to start from $\mathsf{R}(u;\eta)$. Using \eqref{eq:phi1_trig} we find
\begin{equation}
	\mathsf{R}(u; N\gamma) = R^{\text{6v}}(u;\gamma) + O(\mathsf{p})\, , 
\end{equation}	
where
\begin{gather} \label{eq:R6v_princ}
	R^{\text{6v}}(u;\gamma) \coloneqq 
	\begin{pmatrix}
		1 & \color{gray!80}{0} & \color{gray!80}{0} & 0 \\
		\color{gray!80}{0} & b^{\text{6v}}(u;\gamma) & c^{\text{6v}}(u;\gamma) & \color{gray!80}{0} \\
		\color{gray!80}{0} & c^{\text{6v}}(u;\gamma) & b^{\text{6v}}(u;\gamma)  & \color{gray!80}{0} \\
		0 & \color{gray!80}{0} & \color{gray!80}{0} & 1
	\end{pmatrix} \, , \qquad\quad \\
	b^{\text{6v}}(u;\gamma) \coloneqq \frac{\sin \tfrac{\pi}{N} u}{\sin( \tfrac{\pi}{N} u + \pi\gamma)} \, , \quad
	c^{\text{6v}}(u;\gamma) \coloneqq \frac{\sin \pi \gamma}{\sin( \tfrac{\pi}{N} u +\pi \gamma)} \, , \nonumber
\end{gather}
is the \emph{principal grading} of the six-vertex \textit{R}-matrix. Note that it is symmetric. 

Then \eqref{eq:8v1_as-2} gives
\begin{equation}
	\begin{aligned}
	\lim_{\kappa \to 0} R(u; N \gamma ) & = U^{\otimes 2} \, \lim_{\kappa \to 0} 
	\E^{\kappa \gamma u } \,  \mathsf{R}(u; N \gamma) \; \bigl( U^{\otimes 2} \bigr)^{-1} \\
	& = U^{\otimes 2} \, R^{\text{6v}}(u;\gamma) \, \bigl( U^{\otimes 2} \bigr)^{-1} \\
	& = R^{\text{tri\,8v}}(u;\gamma) \, , 
	\end{aligned}
\end{equation}
where the explicit result is of eight-vertex form:
\begin{equation} \label{eq:app_R8v_trig}
	R^{\text{tri\,8v}}(u;\gamma) \coloneqq  	
	\begin{pmatrix}
	\frac{\cos\tfrac{\pi u}{2 N} \cos\tfrac{\pi \gamma}{2}}{\cos\bigl(\tfrac{\pi}{2}\big(\tfrac{u}{N} +\gamma\big)\bigr)} & \color{gray!80}{0} & \color{gray!80}{0} & -\!\frac{\sin\tfrac{\pi u}{2 N} \sin\tfrac{\pi \gamma}{2}}{\cos\bigl(\tfrac{\pi}{2}\big(\tfrac{u}{N} +\gamma\big)\bigr)} \\
	\color{gray!80}{0} & \frac{\cos\tfrac{\pi u}{2 N} \sin\tfrac{\pi \gamma}{2}}{\sin\bigl(\tfrac{\pi}{2}\big(\tfrac{u}{N} +\gamma\big)\bigr)} & \frac{\sin\tfrac{\pi u}{2 N} \cos\tfrac{\pi \gamma}{2}}{\sin\bigl(\tfrac{\pi}{2}\big(\tfrac{u}{N} +\gamma\big)\bigr)}  & \color{gray!80}{0} \\[1ex]
	\color{gray!80}{0} & \frac{\sin\tfrac{\pi u}{2 N} \cos\tfrac{\pi \gamma}{2}}{\sin\bigl(\tfrac{\pi}{2}\big(\tfrac{u}{N} +\gamma\big)\bigr)} & \frac{\cos\tfrac{\pi u}{2 N} \sin\tfrac{\pi \gamma}{2}}{\sin\bigl(\tfrac{\pi}{2}\big(\tfrac{u}{N} +\gamma\big)\bigr)} & \color{gray!80}{0} \\
	-\frac{\sin\tfrac{\pi u}{2 N} \sin\tfrac{\pi \gamma}{2}}{\cos\bigl(\tfrac{\pi}{2}\big(\tfrac{u}{N} +\gamma\big)\bigr)} & \color{gray!80}{0} & \color{gray!80}{0} & \frac{\cos\tfrac{\pi u}{2 N} \cos\tfrac{\pi \gamma}{2}}{\cos\bigl(\tfrac{\pi}{2}\big(\tfrac{u}{N} +\gamma\big)\bigr)}
	\end{pmatrix} 
	\, . 
\end{equation}
\smallskip

\textit{Short-range limit.} 
To investigate what happens as $\kappa \to \infty$, we start from $R(u;\eta)$ and use \eqref{eq:phi2_short} to deduce that
\begin{equation}
	R(\omega \, u ; \omega \, \nncrossing) = R^{\text{6v}}(N u ; \nncrossing) + O(p) \, .
\end{equation}
This time we can use \eqref{eq:8v1_as-2} to see that  
\begin{align}
	\lim_{\kappa \to \infty} \mathsf{R}(\omega \, u; \omega \, \nncrossing) & = \left( U^{\otimes 2} \right)^{-1} \!\! \lim_{\kappa \to \infty}  \E^{-\pi^2 \nncrossing u /(N\kappa)}   R(\omega \, u ; \omega \, \nncrossing) \; U^{\otimes 2} \nonumber \\
	&= \bigl( U^{\otimes 2} \bigr)^{-1}  R^{\text{6v}}(N \, u ; \nncrossing) \; U^{\otimes 2} \nonumber \\
	& = R^{\text{tri\,8v}}(N\, u;\nncrossing) \, . 
\end{align}

\textit{Summary of $\kappa$-limits.}
The limits of $R(u;\eta)$ and $\mathsf{R}(u;\eta)$ are swapped in the short- and long-range limits:
\begin{align}
	R^{\text{6v}}(N u;\nncrossing) \leftarrow & \, R(u;\eta) \rightarrow  
	U^{\otimes 2} \, R^{\text{6v}}(u;\gamma) \, (U^{-1})^{\otimes 2} \, , \\
	\bigl( U^{\otimes 2} \bigr)^{-1}  R^{\text{6v}}(N \, u ; \nncrossing) \; U^{\otimes 2} 
	\leftarrow & \, \mathsf{R}(u; \eta) \rightarrow R^{\text{6v}}(u; \gamma)\, . \nonumber
\end{align}
On the left we have the short-range limit $\kappa \to \infty$, with $u \to \omega \, u$ rescaled and $\eta = \omega \, \nncrossing$, and on the right the long-range limit $\kappa \to 0$ with $\eta = N \gamma$. The explicit result of the conjugated six-vertex \textit{R}-matrices is of eight-vertex form, see \eqref{eq:app_R8v_trig}. 
\medskip

\textit{Undeformed limit.}
To take the (`classical') limit $\eta \to 0$ we note that the pole of $\phi(u,v)$ at $u=0$ (or $v=0$) means that at small $\eta$ we can expand $g^\alpha(u) = 2 \, \delta_{\alpha,0} + O(\eta)$ as only the numerator of $g^0$ has a pole at $\eta=0$ to compensate that in its denominator. The same holds for its variant $\mathsf{g}^\alpha(u)$. Hence
\begin{equation}
	\lim_{\eta \to 0 } \mathsf{R}(u;\eta) =  \lim_{\eta \to 0 } R(u;\eta)= \frac{1}{2} \lim_{\eta \to 0 } g^0(u;\eta) \, \id\otimes \id = \id\otimes \id \, .
\end{equation}
\bigskip

\section{Face-vertex transformation}
\label{app:face_vertex}

\noindent In \textsection\ref{sec:comparison} we compare the landscapes of long-range spin chains associated to the eight-vertex and dynamical \textit{R}-matrices. These two types of elliptic \textit{R}-matrices are connected by Baxter's \emph{face-vertex transformation}~\cite{baxter1973eight123}, which can be written as
\begin{equation} \label{eq:FV_app}
	\begin{aligned}
	\check{R}(u - v;\eta) \, & \Phi(u,v,a;\eta) = \Phi(v,u,a;\eta) \, \check{R}(u-v,a;\eta) \, .
	\end{aligned}
\end{equation}
It removes the entries of the eight-vertex \textit{R}-matrix $\check{R}(u;\eta)$ at the price of modifying the Yang--Baxter equation~\eqref{eq:YBE} to its dynamical version~\eqref{eq:DYBE}.

We like the exposition in \cite{FELDER1996485}, but adapt the expressions to our conventions: we use braid-like \textit{R}-matrices, and functions adapted to the specialisations from \textsection\ref{app:specialisations}--\ref{app:R_matrices}. We use the theta functions \eqref{eq:theta_234} to define
\begin{equation} \label{eq:theta_23_spec}
	\psi_b(u) \coloneqq \E^{-\kappa u/2} \, \vartheta_{4-b}\biggl( \frac{u - N/2}{\omega} \mspace{2mu}\bigg|\mspace{2mu} \frac{-2 N}{\omega} \biggr) \, , 
	\qquad b\in \{1,2\} \, . 
\end{equation}
By arranging these functions in $2\times 2$ matrices as 
\begin{equation} \label{eq:fv_per_vs}
	\Psi(u,a;\eta) \coloneqq 
	\begin{pmatrix} 
		\psi_1(u +\eta\, a) & \psi_2(u+\eta \, a) \\
		\psi_1(u - \eta \,a) & \psi_2(u - \eta \, a) 
	\end{pmatrix} \, ,
\end{equation} 
the face-vertex transformation on $\mathbb{C}^2 \otimes \mathbb{C}^2$ can be factorised as
\begin{equation} \label{eq:fv_two_vs}
	\Phi(u,v,a;\eta) \coloneqq \Psi_1(u,a;\eta) \, \Psi_2(v,a-\sigma_1^z;\eta)\, , 
\end{equation}
with the subscripts labelling the tensor space on which $\Phi$ acts non-trivially. This is just a tensor product of $2 \times 2$ matrices, with the dynamical parameter shifted as appropriate in the dynamical context, cf.~\textsection\ref{sec:qIno}. 
Since \eqref{eq:fv_per_vs} does \emph{not} preserve spin-$z$, the two factors in \eqref{eq:fv_two_vs}, one with shifted dynamical parameter, do not commute.
This is the face-vertex transformation \eqref{eq:FV} used in the main text.

If we instead take 
\begin{equation}
\label{eq:fv_functions_type1}
	\spsi_b(u) \coloneqq \E^{\I \pi  u/(2N)} \, \vartheta_{4-b}\biggl( \frac{u +\omega/2}{N} \mspace{2mu}\bigg|\mspace{2mu} \frac{2 \mspace{2mu} \omega}{N} \biggr) \, ,
	\qquad b\in \{1,2\} \, . 
\end{equation} 
to form $\mathsf{\Psi}(u,a;\eta)$ and $\mathsf{\Psi}(u,v,a;\eta)$ exactly as in \eqref{eq:fv_per_vs}--\eqref{eq:fv_two_vs}
we obtain the face-vertex transformation
\begin{equation} \label{eq:FV_type1_app}
	\begin{aligned}
	\check{\mathsf{R}}(u - v;\eta) \, & \mathsf{\Phi}(u,v,a;\eta) = \mathsf{\Phi}(v,u,a;\eta) \, \check{\mathsf{R}}(u-v,a;\eta) \, .
	\end{aligned}
\end{equation}
It relates the second specialisation of the eight-vertex \textit{R}-matrix, see \eqref{eq:R8v_typed}, to the corresponding dynamical \textit{R}-matrix, which is obtained from \eqref{eq:Rdyn} by replacing $\theta(x)$ by $\stheta(x)$ from  \eqref{eq:theta_def_typed},
\begin{equation} \label{eq:Rdyn_type2}
	\check{\mathsf{R}}(u,a;\eta) \coloneqq 
	\begin{pmatrix}
		\, 1 & \color{gray!80}{0} & \color{gray!80}{0} & 0 \, \\
		\, \color{gray!80}{0} & \displaystyle \!\frac{\stheta(\eta) \, \stheta(\eta\,a-u)}{\stheta(u+\eta)\,\stheta(\eta\,a)}\! & \displaystyle \!\frac{\stheta(u)\,\stheta(\eta\mspace{1mu}(a+1))}{\stheta(u+\eta)\,\stheta(\eta\,a)}\! & \color{gray!80}{0} \, \\[1em]
		\, \color{gray!80}{0} & \displaystyle  \!\frac{\stheta(u)\,\stheta(\eta\mspace{1mu}(a-1))}{\stheta(u+\eta)\,\stheta(\eta\,a)}\! & \displaystyle  \!\frac{\stheta(\eta)\,\stheta(\eta\,a+u)}{\stheta(u+\eta)\,\stheta(\eta\,a)}\! & \color{gray!80}{0} \, \\[.8em]
		\, 0 & \color{gray!80}{0} & \color{gray!80}{0} & 1 \,
	\end{pmatrix} \, .
\end{equation}

\section{The nearest-neighbour exchange}
\label{app:RR_decomp}

\noindent
The deformed long-range spin chains discussed in this paper all feature interactions built from \textit{R}-matrices playing the role of spin permutations along with a nearest-neighbour exchange interaction $\check{R}(-u;\eta) \, \check{R}'(u;\eta)$. The latter also appears as the two-site hamiltonian of the nearest-neighbour Heisenberg spin chains, where it is specialised to a simple value of the spectral parameter, e.g.\ $u=0$. Since for the long-range case $u=i-j$ takes various values it is worth it to work a little and massage $\check{R}(-u;\eta) \, \check{R}'(u;\eta)$ into a simpler and physically meaningful form. In this appendix we show how to derive the result~\eqref{eq:RR_decomp}. 

Because it is technically simpler we work with general elliptic nome~$\tau$, with functions associated to the (quasi)period lattice $\Lambda \coloneqq \mathbb{Z} \oplus \tau \, \mathbb{Z}$, see \textsection\ref{app:ell_functions}. The results can readily be specialised to either \textit{R}-matrix of \eqref{eq:R8v_typed}. In this appendix (and only here) we suppress the $\tau$-dependence.

\subsection{Decomposition}

\noindent
Motivated by the fact that in the short-range and undeformed limit the spin interaction takes the form 
$1-P \, \sigma^\alpha \otimes \sigma^\alpha$ let us seek a similar form at the deformed level. By \eqref{eq:Rch8v} we have
\begin{align} \label{eq:RR'_pre}
	\check{R}(-u;\eta) \, \check{R}'(u;\eta) & = \frac{1}{4} \sum_{\alpha,\beta=0}^z \!\! g_\alpha(-u;\eta) \, g_\beta'(u;\eta) \bigl(\sigma^\alpha \otimes \sigma^\alpha\bigr)  \bigl(\sigma^\beta \otimes \sigma^\beta\bigr) \nonumber \\
	& = \frac{1}{\phi(-u,\eta) \, \phi(u,\eta)} \, \sum_{\alpha=0}^z h_\alpha(u) \,\frac{1-P \, \sigma^\alpha \otimes \sigma^\alpha}{2} \, , 
\end{align}
for some unknown functions $h_\alpha(u;\eta)$, and where we extracted an overall factor. Using some Pauli-matrix algebra we find
\begin{align} \label{eq:h_expression}
	h_\alpha(u;\eta) = {} & \phi(-u,\eta) \, \phi(u,\eta)  \!\! \sum_{\beta,\gamma=0}^z \!\! \bigl(1-2 \, \delta_{\alpha \beta}\bigr) \, g_\beta(-u;\eta) \, \bigl(1-2 \, \delta_{\alpha \gamma}\bigr) \, g_\gamma'(u;\eta) \, , 
\end{align}
with $\delta_{\alpha \beta}$ the Kronecker symbol.
To find a more compact expression for \eqref{eq:h_expression} we analyse its periodicity properties. Using the (quasi)periodicity of the Kronecker elliptic function it is not difficult to derive
\begin{equation}
	h_\alpha(u+2;\eta) = h_\alpha(u;\eta)\, , \quad
	h_\alpha(u+2\tau;\eta) = h_\alpha(u;\eta) \, , 
\end{equation}
valid for all $\alpha \in \{0,\ldots,z\}$.
The fact that the combinations \eqref{eq:h_expression} conspire to be \emph{elliptic} with periods $(2,2\,\tau)$ is rather special and seems to result from our choice to peel off an extra permutation~$P$ in~\eqref{eq:RR'_pre}. From the definition~\eqref{eq:g} of the $g_\alpha$ we moreover have
\begin{equation} \label{eq:hgamma_vs_h0}
	h_\alpha(u;\eta) = h_0(u+\omega_\alpha;\eta) \, . 
\end{equation}
We thus focus on $\alpha=0$.
Since elliptic functions are rare we can simplify $h_0$ by studying its pole structure. The possible poles are $u = -\eta + \Lambda$ and the eight points 
$u = \omega_\alpha + \Lambda$ and $u =  -(\eta+\omega_\alpha)/2 + \Lambda$.
Expanding the Kronecker and Weierstrass-$\zeta$ functions in $h_0$ we find that the residues at $u=-\eta$  and $u= -(\eta+\omega_\alpha)/2$ all vanish, while at $u=0$ we can expand $h_0$ as 
\begin{align}
	h_0(u;\eta) = & \sum_{\alpha=0}^z \Bigl(\delta_{\alpha\gamma}-\frac{1}{2}\Bigr) \, \frac{ \rho( (\eta+\omega_\alpha)/2) - \rho( \eta) -\rho(\omega_\alpha/2)}{u^2} + O\bigl(u^0\bigr) \, , 
\end{align}
where we abuse notation setting $\rho(\omega_0/2) \coloneqq 0$  for brevity. All simple poles of the individual summands cancel when added together. For the other poles we compute the residues using the quasiperiodicity of the summands in \eqref{eq:h_expression}, finding 
\begin{equation}
\begin{aligned}
	& g_\alpha(-u-\omega_\gamma;\eta) \, g_\beta'(u+\omega_\gamma;\eta) = \E^{(\rho\mspace{-1mu}(\omega_\alpha/2)-\rho\mspace{-1mu}(\omega_\beta/2)) \, \omega_\gamma \, - \, \rho\mspace{-1mu}(\omega_\gamma/2) \, ( \omega_\alpha - \omega_\beta)} \, g_\alpha(-u;\eta) \, g_\beta'(u;\eta)\, . 
	\end{aligned}
\end{equation}
The Legendre relation allows us to simplify the exponent to
\begin{equation}
	\begin{aligned}
	& \bigl(\rho\mspace{-1mu}(\tfrac{\omega_\alpha}{2})-\rho\mspace{-1mu}(\tfrac{\omega_\beta}{2})\bigr) \, \omega_\gamma \, - \, \rho\mspace{-1mu}(\tfrac{\omega_\gamma}{2}) \, ( \omega_\alpha - \omega_\beta) = \I\mspace{1mu}\pi  \, \bigl( (1-\delta_{\alpha\gamma,0}) \, \text{sgn}(\alpha-\gamma) - (1-\delta_{\beta\gamma,0})\, \text{sgn}(\beta-\gamma) \bigl) \, .
	\end{aligned}
\end{equation}
Using the $n$-tuple relation for the zeta~function\,---\,see \mbox{(23.2.14)} and \mbox{(23.10.12)} in \cite{DLMF}\,---\,we find that the residue at $u= \omega_\gamma$ is given by
\begin{align}
	\sum_{\alpha,\beta=0}^z \!\! \bigl(1-2\,\delta_{\alpha \gamma}\bigr) \bigl(1-2\,\delta_{\beta \gamma}\bigr) & \, \E^{(\rho\mspace{-1mu}(\omega_\alpha/2)-\rho\mspace{-1mu}(\omega_\beta/2)) \, \omega_\gamma \, - \, \rho\mspace{-1mu}(\omega_\gamma/2) \, ( \omega_\alpha - \omega_\beta)} \nonumber \\[-1.5ex]
	& \! \times \bigl(\rho( (\eta+\omega_\alpha)/2) - \rho(\eta) -\rho(\omega_\alpha/2) \bigr) \nonumber \\[1ex] 
	= 4 \, \rho\bigl((\eta+\omega_\gamma)/2\bigr) & - 4 \, \rho(\omega_\gamma/2) \, .
\end{align}
This means that the $(2,2\,\tau)$-periodic function 
\begin{equation}
	u \longmapsto \sum_{\gamma=0}^z \bigl( \zeta_1( (\eta+\omega_\gamma)/2) - \zeta_1(\omega_\gamma/2) \bigr) \, \wp((u+\omega_\gamma)/2)\, , 
\end{equation}
differs from $h_0$ only by an additive constant by virtue of Liouville's theorem. To find this constant, we evaluate both functions at $u=\eta$. For $h_0$ this requires some care due to the apparent pole. Remarkably, the result of the sum over $\alpha$ reduces considerably, yielding $h_0(\eta;\eta) = -4 \, \wp'(\eta)$.%
\footnote{\ This simplicity appears to be owing to the additive constant in the second line of \eqref{eq:RR'_pre}; without it the result seems more complicated.} 
Putting everything together we conclude
\begin{align}
	h_0(u;\eta) = & \sum_{\gamma=0}^z \Bigl( \rho\Bigl(\frac{\eta+\omega_\gamma}{2}\Bigr) - \rho\Bigl(\frac{\omega_\gamma}{2}\Bigr) \Bigr) \Bigl( \wp\Bigl(\frac{u+\omega_\gamma}{2}\Bigr) - \wp\Bigl(\frac{\eta+\omega_\gamma}{2}\Bigr) \Bigr) - 4 \, \wp'(\eta) \, .
\end{align}
Rescaling $u\to -u/\tau$ and specialising $\tau = -N/\omega$ yields \eqref{eq:RR_decomp}.

\subsection{Connection to Heisenberg \textsc{xyz}}

\noindent
To connect to the Heisenberg \textsc{xyz} chain we explicitly compute the local hamiltonian \eqref{eq:log_der_transf} from the expression \eqref{eq:RR_decomp}. Evaluating \eqref{eq:V^eta} at any of the special points $u=\omega_\alpha$ yields the very simple values $V\mspace{-1mu}(\omega_\alpha;\eta) = -A_\alpha(\eta)$, so in terms of the spin-spin interactions $\EE^\alpha$ defined in \eqref{eq:E^alpha} we have
\begin{equation}
	H_{\textsc{xyz}} = -\sum_{i=1}^{\mathclap{N}} 
	\vec{A}(\eta) \cdot \vec{\EE} \, .
\end{equation}
It can be rewritten in the more familiar form $\sum_{\alpha=0}^z c_\alpha \, \sigma^\alpha \otimes \sigma^\alpha$ with coefficients
\begin{equation}
\begin{aligned}
	c_0 &= \rho(\eta)\, , \\
	c_\alpha &= \rho(\eta) - \rho\Big(\frac{\eta}{2}\Big) - \rho\Big(\frac{\eta+\omega_\alpha}{2}\Big) + \rho\Big(\frac{\omega_\alpha}{2}\Big) \quad (\alpha \neq0) \, . 
\end{aligned}
\end{equation}
The coefficients with $\alpha\neq 0$ are elliptic in $\eta$ and have a fairly easy pole structure, allowing us to simplify them. The same is true for the combination $c_0(\eta) -2\,\rho(\eta/2)$. The result is
\begin{equation}
	\begin{aligned}
	c_0 & = \rho(\eta/2)-\rho(\eta) - c_1 - c_2 - c_3 \, ,  \\
	c_1 & = \frac{K}{\omega} \, \frac{1}{\sn(\eta')} \, , \ \ 
	\frac{c_2}{c_1} = \dn(\eta') \, , \ \
	\frac{c_3}{c_1} = \cn(\eta') \, , \ \
	\eta' \coloneqq \frac{2 K}{\omega} \, \eta \, . 
	\end{aligned}
\end{equation}
where the simplest form uses the complete elliptic integral of the first kind $K = \pi \, \vartheta_3(0 \mspace{1mu}|\mspace{1mu} \tau)^2/2$, with $\tau = -N/\omega$ as before, and the Jacobi elliptic functions 
\begin{equation} \label{eq:jacobis}
	\begin{aligned}
	\sn(\eta';m) & \coloneqq \frac{ \vartheta_3(0 \mspace{1mu}|\mspace{1mu} \tau) \, \vartheta_1(\eta \mspace{1mu}|\mspace{1mu} \tau) }{\vartheta_2(0 \mspace{1mu}|\mspace{1mu} \tau) \, \vartheta_4(\eta \mspace{1mu}|\mspace{1mu} \tau) }\, , \quad 
	\cn(\eta';m) \coloneqq \frac{ \vartheta_4(0 \mspace{1mu}|\mspace{1mu} \tau) \, \vartheta_2(\eta \mspace{1mu}|\mspace{1mu} \tau) }{\vartheta_2(0 \mspace{1mu}|\mspace{1mu} \tau) \,\vartheta_4(\eta \mspace{1mu}|\mspace{1mu} \tau) } \, , \\ 
	\dn(\eta';m) & \coloneqq \frac{ \vartheta_4(0 \mspace{1mu}|\mspace{1mu} \tau) \, \vartheta_3(\eta \mspace{1mu}|\mspace{1mu} \tau) }{\vartheta_3(0|\tau) \, \vartheta_4(\eta \mspace{1mu}|\mspace{1mu} \tau) }\, , 
	\end{aligned}
\end{equation}
with modulus $m=k^2 = \vartheta_2(0 \mspace{1mu}|\mspace{1mu} \tau)^4 / \vartheta_3(0 \mspace{1mu}|\mspace{1mu} \tau)^4$, here defined in terms of the theta functions \eqref{eq:theta_234}. 
We thus obtain
\begin{align} \label{eq:heisxyz}
	H_{\textsc{xyz}} = {} & N \, (c_0+c_3) + c_1 \sum_{i=1}^{N} \bigl(  \sigma^x_i \, \sigma^x_{i+1} + \Gamma \, \sigma^y_i \, \sigma^y_{i+1} + \Delta \, (\sigma^z_i \, \sigma^z_{i+1} -1)\bigr) \, ,
\end{align}
with anisotropy parameters $\Gamma \coloneqq \text{dn}(\nncrossing)$ and $\Delta \coloneqq \text{cn}(\nncrossing)$.

\section{Original Matushko--Zotov chain} \label{app:MZ_chain}

\noindent
For easy comparison we describe the SZ and MZ chains in the more standard, but for our purposes unsuitable, conventions of \cite{sechin2018r,MZ_23b}. All differences can be traced back to our use of $\theta(x)$ instead of the more common $\stheta(x)$ from \eqref{eq:theta_def_typed}.

The chiral hamiltonians $\mathsf{H}_{\textsc{mz}}^{\textsc{l},\textsc{r}}$ of \cite{MZ_23b} are written as \eqref{eq:MZ_ham_pre} in terms of the eight-vertex \textit{R}-matrix $\check{\mathsf{R}}(u;\eta)$ defined in \eqref{eq:R8v_typed}. Equivalently, they can be written in the form \eqref{eq:MZ_ham_decomp} with potential $\mathsf{V}(u;\eta)$, which is as in \eqref{eq:V^eta} except that $\mathsf{A}_\beta(\eta)$ is given by \eqref{eq:A_beta} with $\srho$ instead of $\rho$, and $\vec{\somega} \coloneqq (0,\omega, N+\omega, N)$. These hamiltonians commute with the translation operator 
\begin{equation}
	G_\textsc{mz} \coloneqq \sigma^z_N \!\!\! \ordprod_{N\geqslant i> 1} \!\!\!\! \check{\mathsf{R}}_{i-1,i}(1-i;\eta)\, . 
\end{equation}
Note that it has diagonal twist, unlike \eqref{eq:MZ_trans}. 

From the relation \eqref{eq:8v1_as-2} between the \textit{R}-matrices it follows that the two versions of the chiral hamiltonians are related as
\begin{equation} \label{eq:MZ_ham_1vs2}
	\mathsf{H}_{\textsc{mz}}^{\textsc{l},\textsc{r}} = U^{\otimes N} \, H_{\textsc{mz}'}^{\textsc{l},\textsc{r}} \; \bigl( U^{\otimes N} \bigr)^{-1} + (N-1)\,\kappa\,\eta/2 \, . 
\end{equation}
This shift is due to the following remarkably simple relation between the two potentials:
\begin{equation} \label{eq:V1_vs_V2}
	\mathsf{V}(u;\eta) = V(u;\eta) + \frac{\kappa\, \eta}{2N} \, . 
\end{equation}
We see that the potentials have the same long-range ($\kappa\to 0$) and macroscopic ($N\to\infty$) limits. 
Since $\check{\mathsf{R}}(u;N\,\gamma) \to \check{R}^\textrm{6v}(u;\gamma)$, the long-range limit of $\mathsf{H}_{\textsc{mz}}^{\textsc{l},\textsc{r}}$ is \eqref{eq:MZ_trig}.

Upon rescaling by $n_{\kappa,\eta} \sim 1/\eta$ as in \eqref{eq:normalisation_choice} the undeformed limits ($\eta\to0$) of the potential are \eqref{eq:pot_typed}. Since $\check{\mathsf{R}}(u;\eta) \to P$ this gives the SZ chain as introduced in \cite{sechin2018r}; this is called the `non-relativistic' limit in \cite{MZ_23b}. Comparing with \mbox{(2.5)} in \cite{klabbers2022coordinate} shows that to obtain the potential of this chain from the trigonometric case by wrapping as in \textsection\ref{sec:wrapping} one needs to embed the physical chain space along the imaginary direction, which can be achieved by rescaling arguments. 

In the short-range limit ($\kappa\to \infty$) the difference in \eqref{eq:V1_vs_V2} \emph{diverges}. In fact, \eqref{eq:V1_vs_V2} cannot be regularised by rescaling in this limit. Indeed, due to the shift in \eqref{eq:V1_vs_V2} the analogue of the expansion \eqref{eq:sr_limit} starts with a constant, and the $u$-dependence is subleading. This means that there is no  normalisation that would select nearest-neighbour terms only. This is why the MZ chain as given in \cite{MZ_23b} does not have a short-range limit.

\bibliography{bibliography}

\end{document}